\documentclass[aps,preprint,nofootinbib,eqsecnum]{revtex4}%
\usepackage{amsmath}
\usepackage{amsfonts}
\usepackage{amssymb}
\usepackage{graphicx}
\usepackage{amscd}%
\providecommand{\U}[1]{\protect\rule{.1in}{.1in}}

\allowdisplaybreaks
\begin{document}
\leftline {USC-06/HEP-B3 \hfill hep-th/0606045}{\small \
}{\vskip-1cm}

{\vskip2cm}

\begin{center}
{\Large \textbf{The Standard Model of Particles and Forces }}

{\Large \textbf{in the Framework of 2T-physics}}\footnote{This work was
partially supported by the US Department of Energy under grant number
DE-FG03-84ER40168.}{\Large \textbf{\ }}

{\vskip0.8cm}

\textbf{Itzhak Bars}

{\vskip0.8cm}

\textsl{Department of Physics and Astronomy}

\textsl{University of Southern California,\ Los Angeles, CA 90089-2535 USA}

{\vskip1.5cm} \textbf{Abstract}
\end{center}

In this paper it will be shown that the Standard Model in $3+1$ dimensions is
a gauge fixed version of a 2T-physics field theory in $4+2$ dimensions, thus
establishing that 2T-physics provides a correct description of Nature from the
point of view of $4+2$ dimensions. The 2T formulation leads to
phenomenological consequences of considerable significance. In particular, the
higher structure in $4+2$ dimensions prevents the problematic $F\ast F$ term
in QCD. This resolves the strong CP problem without a need for the
Peccei-Quinn symmetry or the corresponding elusive axion. Mass generation with
the Higgs mechanism is less straightforward in the new formulation of the
Standard Model, but its resolution leads to an appealing deeper physical basis
for mass, coupled with phenomena that could be measurable. In addition, there
are some brand new mechanisms of mass generation related to the higher
dimensions that deserve further study. The technical progress is based on the
construction of a new field theoretic version of 2T-physics including
interactions in an action formalism in $d+2$ dimensions. The action is
invariant under a new type of gauge symmetry which we call 2Tgauge-symmetry in
field theory. This opens the way for investigations of the Standard Model
directly in $4+2$ dimensions, or from the point of view of various embeddings
of $3+1$ dimensions, by using the duality, holography, symmetry and unifying
features of 2T-physics.

\newpage

\section{The Sp$\left(  2,R\right)  $ gauge symmetry}

The essential ingredient in two-time physics (2T-physics) is the basic gauge
symmetry Sp$(2,R)$ acting on phase space $X^{M},P_{M}$ \cite{2treviews}, or
its extensions with spin \cite{spin2t}\cite{2tbacgrounds} and/or supersymmetry
\cite{super2t}-\cite{twistorLect}. Under this gauge symmetry, momentum and
position are locally indistinguishable at any instant. This principle
inevitably leads to deep consequences, one of which is the two-time structure
of spacetime in which ordinary one-time (1T) spacetime is embedded. Some of
the 1T-physics phenomena that emerge from 2T-physics include certain types of
dualities, holography, emergent spacetimes, and a unification of certain
1T-physics systems into a single parent theory in 2T-physics.

In the present paper a field theoretic formulation of 2T-physics is given in
$d+2$ dimensions. To construct the 2T field theory, first the free field
equations are determined from the covariant quantization of the 2T particle on
the worldline, subject to the Sp$\left(  2,R\right)  $ gauge symmetry and its
extensions with spin. Next, an action is constructed from which the 2T free
field equations are derived, and then interactions are included consistently
with certain new symmetries of the action. The resulting action principle for
2T-physics in $d+2$ dimensions is then applied to construct the 2T Standard
Model in $4+2$ dimensions. It is shown that the usual Standard Model in $3+1$
dimensions is a holographic image of this $4+2$ dimensional theory. The
underlying $4+2$ structure provides some additional restrictions on the
Standard Model, with significant phenomenological consequences, as outlined in
the abstract. The $4+2$ dimensional theory suggests new non-perturbative
approaches for investigating $3+1$ dimensional field theories, including QCD.

Prior to this development, 2T-physics had been best understood for particles
in the worldline formalism interacting with all background fields
\cite{2tbacgrounds}, including gauge fields, gravitational field and all high
spin fields, and subject to the Sp$\left(  2,R\right)  $ gauge symmetry, or
its extensions with spin. For the spinless particle, the three Sp$\left(
2,R\right)  $ gauge symmetry generators $Q_{ij}\left(  X,P\right)  ,$
$i,j=1,2,$ are functions of phase space and depend on background fields
$\phi^{M_{1}M_{2}\cdots M_{s}}\left(  X\right)  $ of any integer spin $s$. The
simplest case of 2T-physics corresponds to a spinless particle moving in the
trivial constant background field $\eta^{MN}$ that corresponds to the metric
in a flat spacetime. In this case the Sp$\left(  2,R\right)  $ gauge symmetry
is generated by the operators
\begin{equation}
Q_{11}=\frac{1}{2}X\cdot X,\;\;Q_{22}=\frac{1}{2}P\cdot P,\;\;Q_{12}%
=Q_{21}=\frac{1}{2}\left(  X\cdot P+P\cdot X\right)  , \label{Qij}%
\end{equation}
where the dot product involves the the flat metric $\eta_{MN}$. Similarly, for
spinning particles of spin $s,$ phase space $\left(  \psi_{i}^{M},X^{M}%
,P^{M}\right)  ,$ includes the fermions $\psi_{i}^{M}$, $i=1,2,\cdots,2s,$ so
the gauge symmetry is enlarged to the worldline supersymmetry OSp$\left(
2s|2\right)  $ which includes Sp$\left(  2,R\right)  .$ In flat spacetime, the
generators of the gauge symmetry correspond to all the spacetime dot products
among the $\psi_{i}^{M},X^{M},P^{M}.$ These generators are first class
constraints that vanish, thus restricting the phase space $\left(  \psi
_{i}^{M},X^{M},P^{M}\right)  $ to a OSp$\left(  2s|2\right)  $ gauge invariant subspace.

To have non-trivial solutions for the constraints $Q_{ij}=0,$ etc., the flat
metric $\eta_{MN},$ which is used to form the dot products in the constraints,
must have a two-time signature. So, in the absence of backgrounds, the 2T
particle action is automatically invariant under a global SO$\left(
d,2\right)  $ symmetry in $d+2$ dimensions, where the 2T signature emerges
from the requirement of the local gauge invariance of the physical sector. In
the presence of backgrounds the 2T signature in $d+2$ dimensions is still
required by the gauge symmetry. However, the nature of the space-time global
symmetry, if any, is determined by the Killing vectors of the background in
$d+2$ dimensions, and it may be smaller or larger than SO$\left(  d,2\right)
$.

It is well understood \cite{2treviews}-\cite{2tHandAdS} that the gauge
symmetry compensates for extra dimensions in phase space $\left(  X^{M}%
,P^{M},\psi_{i}^{M}\right)  ,$ and effectively reduces the $d+2$ dimensional
space by one time and one space dimensions, thus establishing causality and
guaranteeing a ghost free 2T-physics theory. The subtlety is that there are
many ways of embedding the remaining \textquotedblleft time\textquotedblright%
\ and \textquotedblleft Hamiltonian\textquotedblright\ in the higher
space-time. Therefore there are many 1T systems that emerge in $\left(
d-1\right)  +1$ dimensions as solutions of the constraints with various gauge
choices. Some examples are given in Fig. 1.%
\begin{center}
\fbox{\includegraphics[
height=4.4944in,
width=5.9819in
]%
{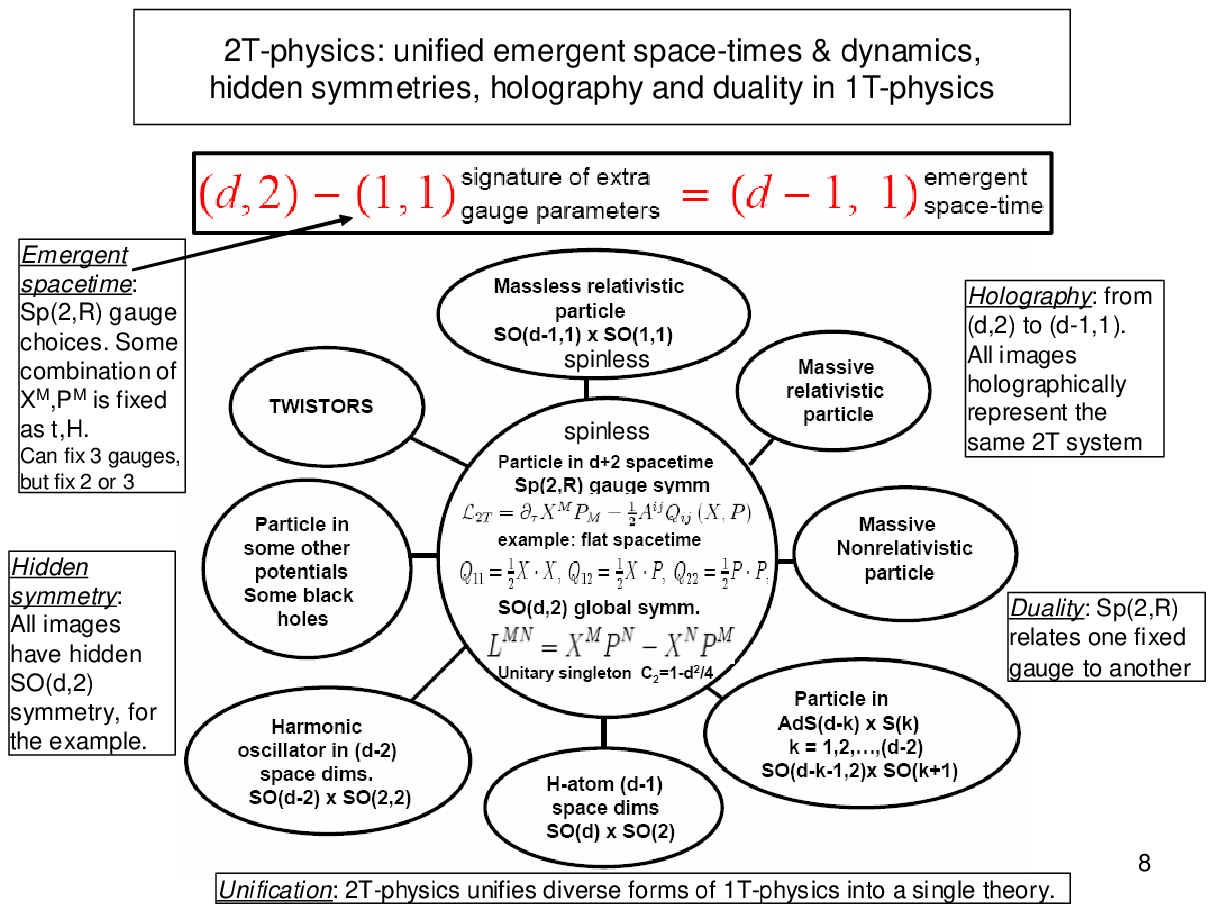}%
}\\
Fig.1 - Some 1T-physics systems that emerge from the solutions of Q$_{ij}=0.$%
\end{center}
In these emergent space-times the Hamiltonian for each 1T system is different,
hence the dynamics appears different from the point of view of 1T-physics.
However, each 1T system holographically represents the original 2T system in
$d+2$ dimensions. Of course, due to the original gauge symmetry the various 1T
systems are in some sense equivalent. This equivalence corresponds to
dualities among the various 1T systems \cite{2tHandAdS}-\cite{twistorBP1}.

Hence 2T-physics may be recognized as a unifying structure for many
1T-systems. The unification occurs through the presence of the higher
dimensions, but in a way that is very different than the Kaluza-Klein
mechanism since there are no Kaluza-Klein excitations, but instead there are
hidden symmetries that reflect $d+2$ dimensions, and also a web of dualities
among 1T systems that are holographic images of the 2T parent theory in $d+2$ dimensions.

As shown in Fig.1, simple examples of such 1T systems in $\left(  d-1\right)
+1$ dimensions, that are known to be unified by the free 2T particle in flat
$d+2$ dimensions, and have non-linear realizations of SO$\left(  d,2\right)  $
symmetry with the same Casimir eigenvalues, include the following systems for
spinless particles: free massless relativistic particle \cite{2treviews}, free
massive relativistic particle \cite{2tHandAdS}, free massive non-relativistic
particle \cite{2tHandAdS}, hydrogen atom (particle in $1/r$ potential in $d-1$
space dimensions) \cite{2tHandAdS}, harmonic oscillator in $d-2$ space
dimensions \cite{2tHandAdS}, particle on a sphere $S^{d-1}\times R$
\cite{twistorBP1}, particle on AdS$_{d-k}\times$S$^{k}$ for $k=0,1,\cdots
,\left(  d-2\right)  $ \cite{2tHandAdS}$,$ particle on maximally symmetric
curved spaces in $d$ dimensions \cite{ibquelin}, particle on BTZ black hole
(special for $d=3$ only) \cite{btzbhole}, and twistor equivalents
\cite{2ttwistor}\cite{twistorBP1} of all of these in $d$ dimensions. There are
also generalizations of these for particles with spin \cite{spin2t}, with
supersymmetry \cite{super2t}, with various background fields
\cite{2tbacgrounds}, and the twistor superstring \cite{2tsuperstring},
although details and interpretation of the 1T physics for gauges other than
the massless particle gauge remain to be developed for most of the generalizations.

The established existence of the hidden symmetries and the duality
relationships among such well known simple systems at the classical and
quantum levels provide part of the evidence for the existence of the higher
dimensions. This already validates 2T-physics as the theory that predicted
them and provided the description of the underlying deeper structure that
explain these phenomena.

Next comes the question of how to express these properties of 2T-physics in
the language of field theory, and how to include interactions. This was
partially understood \cite{2tfield} in the form of field equations, including
interactions, as reviewed in section 2. But this treatment missed an action
principle from which all the equations of motion should be derived. The field
equations were classified as those that determine \textquotedblleft
kinematics\textquotedblright\ and those that determine \textquotedblleft
dynamics\textquotedblright. The dynamical equations including interactions
could be obtained from an action as usual, but the kinematical equations,
\ which determine how $3+1$ dimensional spacetime is embedded in $4+2$
dimensional spacetime, needed to be imposed from outside as additional
constraints. This was considered incomplete in \cite{2tfield}, because a full
action principle that yields all the equations is needed to be able to study
consistently the quantum theory and other properties of the theory.

The new principles for constructing the 2T-physics action in $d+2$ dimensions
are developed in section 2. These emerge from basic properties of Sp$\left(
2,R\right)  $ and its extensions as discussed in \cite{2tfield}, supplemented
with a hint on the overall structure of the action that followed from a recent
construction of BRST field theory for 2T-physics \cite{2tbrst2006}. The new
action principle does not use the BRST formalism, but has a new type of gauge
symmetry which we name as the \textquotedblleft%
2Tgauge-symmetry\textquotedblright\ in field theory. From this action we
derive both the kinematic and the dynamical equations through the usual
variational principle.

The Standard Model in $4+2$ dimensions is constructed in section 3 by
introducing the matter and gauge field content analogous to the usual Standard
Model, but now derivatives and vector bosons are SO$\left(  4,2\right)  $
vectors, fermions are SO$\left(  4,2\right)  $=SU$\left(  2,2\right)  $
quartet spinors, while all fields are functions in $4+2$ dimensions. The
2Tgauge-symmetry dictates the overall structure and the form of the terms that
can be included.

Next, in section 4, the action for the Standard Model in $3+1$ dimensions is
derived from the action in $4+2$ dimensions by solving the subset of equations
of motion that determine the \textquotedblleft kinematics\textquotedblright.
This step is equivalent to choosing a gauge for the underlying Sp$\left(
2,R\right)  $ gauge symmetry in the worldline formalism, as illustrated in
Fig.1. In particular solving the kinematics in a particular parametrization
given in Eq.(\ref{massless}) corresponds to the \textquotedblleft massless
relativistic particle\textquotedblright\ gauge denoted in Fig.1. The solution
of the kinematic equations in this way provides a holographic image of the
$4+2$ dimensional theory in the $3+1$ dimensional spacetime. The degrees of
freedom in all the fields are thinned out from $4+2$ dimensions to $3+1$
dimensions. Both the 2Tgauge-symmetry, and solving the kinematical equations,
play a role in reducing the degrees of freedom from $4+2$ dimensions to the
proper ones in $3+1$ dimensions. The remaining dynamics in $3+1$ dimensions is
determined by the emergent $3+1$ dimensional action. In the chosen gauge the
emergent theory is the usual Standard Model action, however the emergent
Standard Model comes with some interesting restrictions on certain terms. The
additional restrictions are effects of the overall $4+2$ structure and are not
dictated by working directly in $3+1$ dimensions.

Having obtained the Standard Model in $3+1$ dimensions, one may ask, what is
new in $3+1$ dimensions? Part of the answer includes the constraints inherited
from $4+2$ dimensions that get reflected on the overall structure of the
emergent Standard Model in $3+1$ dimensions. It is fascinating that this has
phenomenological consequences. First we emphasize that all the basic
interactions that are known to work in Nature among the quarks, leptons, gauge
bosons and Yukawa couplings are permitted. The forbidden terms seem to
coincide with unobserved interactions in Nature. In particular, a forbidden
term in the $3+1$ dimensional emergent action is the problematic $F_{\mu\nu
}F_{\lambda\sigma}\varepsilon^{\mu\nu\lambda\sigma}$ term in QCD, or similar
anomalous terms in the weak interactions that cause unobserved small
violations of $B+L$. This is because there are no corresponding terms in the
$4+2$ dimensional action that would filter down to $3+1$ dimensions. As
discussed in section \ref{cp}, this provides a nice resolution of the strong
CP problem in QCD without the need for the Peccei-Quinn symmetry or the
corresponding elusive axion. This had remained as one of the unresolved issues
of the usual Standard Model (for a recent review see \cite{wittenaxion}). So,
the $4+2$ dimensional theory seems to explain more as compared to the usual
$3+1$ dimensional theory.

Mass generation is less straightforward in the emergent model than the usual
Standard Model, because a quadratic mass term for the Higgs boson is not
permitted by the underlying $4+2$ structure. This is discussed in section
\ref{massgen}. To obtain a non-trivial vacuum, one may either introduce
interactions of the Higgs with a dilaton, or invoke dynamical breakdown of the
SU$\left(  2\right)  \times$U$\left(  1\right)  $ gauge symmetry through
mechanisms such as extended technicolor. The dilaton scenario offers an
appealing deeper physical basis for mass, and generates new phenomena that
could be measurable. There are also new possibilities for mass in the emergent
theory that involve the higher dimensions. For this we only need to recall
that a massive relativistic particle can also come out from the $4+2$ theory
as illustrated in Fig.1. The mass in this case is analogous to a modulus that
comes from the higher dimensions in a non-trivial embedding of $3+1$ in $4+2$
dimensions. Since mass generation is the obscure part of the Standard Model,
and there are new mechanisms in the 2T action, the mass generation deserves
further study of what the 2T approach has to offer.

Since our proposal is that the fundamental theory is formulated in $4+2$
dimensions, one wonders if one can one test the effect of the extra
dimensions. This can be explored by studying other gauges of Sp$\left(
2,R\right)  $ (equivalent to different forms of solutions of the kinematic
equations) that lead to $3+1$ dimensional dual versions that are also
holographic images of the same $4+2$ dimensional Standard Model in the sense
of Fig.1. The exploration of these dual theories corresponds to exploring the
$4+2$ dimensional space, and is left to future work. Other remaining issues
and future directions will be discussed in section \ref{directions}.

\section{Principles for Interacting 2T Field Theory Action}

The construction of the proper action principle for 2T-physics has remained an
open problem for some time. Equations of motion for each spin, including
interactions were available, and even the Standard Model in $4+2$ dimensions
in equation of motion form was outlined \cite{2tfield}. The main stumbling
block has been the fact that there are more equations to be satisfied by each
field than the number of equations which can be derived from a standard
action. The solution given in this paper will involve some subtle properties
of the delta function $\delta\left(  X^{2}\right)  $ that imposes the
$Q_{11}\sim X^{2}=0$ constraint, and could have already been attained in
\cite{2tfield}, but it was missed. A crucial hint came from a recent BRST
field theory formulation of the problem, akin to string field theory, as
discussed in \cite{2tbrst2006}. In the present paper we by-pass the BRST
construction and use only the tip as a spring-board to construct a simpler
action for only the relevant fields of any spin, although the full BRST field
theory may also be useful to consider for more general purposes. The 2T action
principle given in this section provides the proper minimal framework to
consistently discuss new symmetries, include interactions, and perform
quantization in 2T field theory. This principle is applied to construct the
Standard Model in $4+2$ dimensions in section (\ref{SM}).

In this section we first review the derivation \cite{2tfield} of the 2T free
field equations in $d+2$ dimensions for scalars, fermions, vectors, and the
graviton from the \textit{worldline} properties of the Sp$\left(  2,R\right)
, $ OSp$\left(  1|2\right)  $, OSp$\left(  2|2\right)  $ and OSp$\left(
4|2\right)  $ gauge symmetry respectively. Then we introduce an action
principle in $d+2$ dimensional \textit{field theory} from which these field
equations are derived through the variational principle. The action has a new
kind of gauge symmetry that we call the 2Tgauge-symmetry.

The free field equations are obtained as follows. The OSp$\left(  2s|2\right)
$ are the gauge symmetry groups in the worldline formulation of the 2T
particles of spin $s$ in flat $d+2$ dimensions. In SO$\left(  d,2\right)  $
covariant first quantization, physical states are identified as those that are
gauge invariant by satisfying the first class constraints which form the
OSp$\left(  2s|2\right)  $ Lie superalgebra. In position space the constraint
equations turn into field equations in $d+2$ dimensions. So, in principle the
number of field equations for each spinning particle of spin $s$ is equal to
the number of generators of the gauge groups OSp$\left(  2s|2\right)  $, since
all generators must vanish on the physical gauge invariant field. By some
manipulation the number of equations in $d+2$ dimensions can be brought down
to a smaller set, but still there are more field equations as compared to the
familiar field equation for corresponding spinning fields in $d$ dimensions.
This must be so, because only with the additional equations it is possible to
have an equivalence between the $d+2$ dimensional field equations and their
corresponding ones in $d$ dimensions. The familiar looking $d+2$ field
equation, that is similar to the $d$ dimensional equation, is interpreted as
the dynamical equation, while the additional equations can be interpreted as
subsidiary kinematical conditions on the field in $d+2$ dimensions.

The same subsidiary kinematic equations for fields of any integer spin $s$
were also obtained by considering only the spinless particle propagating in
background fields $\phi^{M_{1}M_{2}\cdots M_{s}}\left(  X\right)  $ of any
integer spin $s$. In this case the Sp$\left(  2,R\right)  $ generators
$Q_{ij}\left(  X,P\right)  ,$ $i,j=1,2,$ are functions of the background
fields $\phi^{M_{1}M_{2}\cdots M_{s}}\left(  X\right)  $ \cite{2tbacgrounds}.
Requiring closure of the $Q_{ij}\left(  X,P\right)  $ under Poisson brackets
into Sp$\left(  2,R\right)  $ demands conditions on the background fields.
These conditions are identical to the kinematic equations. In this case the
background fields are off shell and are not required to satisfy the dynamical
equation. The kinematic equations by themselves are sufficient to reduce the
degrees of freedom in the fields $\phi^{M_{1}M_{2}\cdots M_{s}}\left(
X\right)  $ from $d+2$ dimensions to $d$ dimensions both in the spacetime
$X^{M}$ dependence of the field, and in the components of spinning fields
labeled by $M_{1},M_{2},\cdots,M_{s}$.

The 2T-physics equations that emerge from OSp$\left(  2s|2\right)  $
constraints in $d+2$ dimensions, or from the backgrounds with Sp$\left(
2,R\right)  $ gauge symmetry, coincide with Dirac's equations \cite{Dirac} in
the case of $s=0,1/2,1$ in $d=4,$ and their generalizations to spin $2$ and
higher \cite{2tfield}\cite{2tbacgrounds}\cite{vasiliev2} in any dimension.
While there were attempts in the past to write down a field theory action, the
subsidiary kinematic conditions have been treated as external conditions not
derived from the same action. The proper action principle will be given in
this paper.

In any case, it has been known \cite{Dirac}-\cite{vasiliev},\cite{2tfield} at
the level of equations of motion, that the ensemble of the $d+2$ dimensional
equations correctly reproduce the massless Klein-Gordon, Dirac, Maxwell,
Einstein, and higher spin field equations of motion in $d$ dimensions. In
2T-physics this is interpreted as an example of a more general holography from
$d+2$ dimensions to $\left(  d-1\right)  +1$ dimensions that emerges from
gauge fixing Sp$\left(  2,R\right)  $.

A particular parametrization given in Eq.(\ref{massless}) that corresponds to
the Sp$\left(  2,R\right)  $ gauge indicated as the \textquotedblleft massless
particle gauge\textquotedblright\ in Fig.1 will be used to derive the Standard
Model in $3+1$ dimensions from the $4+2$ dimensional theory. The massless
Klein-Gordon, Dirac, Maxwell, Einstein field equations in $d$ dimensions
emerge in the parametrization of Eq.(\ref{massless}). The novelty in 2T
physics is its more general property that all the other 1T interpretations
(massless, massive, curved spaces, etc.), outlined in Fig.1 also emerge from
the same 2T field equations, as different holographic images in different
Sp$\left(  2,R\right)  $ gauges, as explained before both in particle theory
\cite{2tHandAdS} and field theory \cite{2tfield}. In this paper we will use
only the massless particle interpretation related to the parametrization of
Eq.(\ref{massless}) in order to connect to the usual form of the Standard
Model. However, it should be evident that dual versions of the Standard Model,
including interactions, will emerge by taking advantage of the more general
properties of these equations\footnote{Dirac and followers regarded the
equations of motion in $4+2$ dimensions as a formulation of the hidden
conformal symmetry SO$\left(  4,2\right)  $ of massless field equations in
$d=4.$ The $4+2$ dimensional spacetime was not emphasized as being anything
other than a trick. 2T-physics developed independently from the opposite end
without awareness of Dirac's approach to conformal symmetry, and deliberately
focusing on signature $\left(  d,2\right)  $ with two times. The $\left(
d,2\right)  $ signature specifically started with a hunch \cite{hunch}
developed from $M$-theory that there are higher dimensions with signature
$\left(  10,2\right)  $ (see also \cite{vafa} for an independent idea). This
path developed through various papers \cite{2tbeginers}\cite{2tbeginers2},
gathering hints on how to correctly formulate a theory with signature $\left(
d,2\right)  $ consistent with M-theory, causality and dynamics without ghosts
in $d+2$ dimensions. The multi-particle symmetries in \cite{2tbeginers2}
provided the dynamical setup that eventually led to the introduction of the
Sp$\left(  2,R\right)  $ gauge symmetry \cite{2treviews} for a single
particle. The new features that emerged in 2T-physics include the underlying
fundamental role of the Sp$\left(  2,R\right)  $ gauge symmetry, the existence
of the other 1T interpretations as holographic images in $\left(  d-1\right)
+1$ dimensions of the same 2T system, with the same original SO$\left(
d,2\right)  $ symmetry that is not interpreted as conformal symmetry in the
various holographic images, the duality among the multiple 1T solutions, and
the corresponding interpretation of the hidden symmetries and dualities as
evidence for the higher dimensions. Only after these properties were
discovered, it was recognized in \cite{2tfield} that Dirac's approach to
conformal symmetry, that had been forgotten, could be seen as part of
2T-physics in a particular gauge.}.

\subsection{Scalar Field}

\subsubsection{Free field equations for scalars}

For the spinless 2T particle, the vanishing of the Sp$\left(  2,R\right)  $
generators implies that the physical phase space must be gauge invariant. In
covariant first quantization, physical states $|\Phi\rangle$ are identified as
those on which the generators $Q_{ij}$ vanish $X^{2}|\Phi\rangle
=0,\;P^{2}|\Phi\rangle=0,\;\left(  X\cdot P+P\cdot X\right)  |\Phi\rangle=0.$
This means that the physical states form the subset of states that are gauge
invariant under Sp$\left(  2,R\right)  .$ The field $\hat{\Phi}\left(
X\right)  $ is defined as the probability amplitude of a physical state in
position space $\hat{\Phi}\left(  X\right)  =\langle X|\Phi\rangle.$ {Since
momentum is represented as a derivative in position space }$\langle
X|P_{M}=-i\partial_{M}\langle X|${, the gauge invariance conditions }applied
on physical states $|\Phi\rangle$ give the free field equations for the field
$\hat{\Phi}\left(  X\right)  $ as
\begin{equation}
X^{2}\hat{\Phi}\left(  X\right)  =0,\;\partial_{M}\partial^{M}\hat{\Phi
}\left(  X\right)  =0,\;X^{M}\partial_{M}\hat{\Phi}\left(  X\right)
+\partial_{M}\left(  X^{M}\hat{\Phi}\left(  X\right)  \right)  =0.
\label{phieoms}%
\end{equation}
The general solution of the first equation is
\begin{equation}
\hat{\Phi}\left(  X\right)  =\delta\left(  X^{2}\right)  \Phi\left(  X\right)
, \label{phihat}%
\end{equation}
where $\Phi\left(  X\right)  $ (without the hat $\symbol{94}$)\footnote{We
distinguish between the symbols $\hat{\Phi},\hat{\Psi}_{\alpha},\hat{A}_{M}$
and $\Phi,\Psi_{\alpha},A_{M}$ to emphasize that $\hat{\Phi},\hat{\Psi
}_{\alpha},\hat{A}_{M}$ include the delta function factor. In comparing notes
with ref.\cite{2tfield} one should compare the $\Phi,\Psi_{\alpha},A_{M}$ in
that paper to the $\Phi,\Psi_{\alpha},A_{M}$ in this paper, not to the
$\hat{\Phi},\hat{\Psi}_{\alpha},\hat{A}_{M}.$\label{hats}} is any function of
$X^{M}$ which is not singular at $X^{2}=0.$ We have used the property
$X^{2}\delta\left(  X^{2}\right)  =0$ of the delta function. We also note the
following additional properties of the delta function that we use repeatedly
below
\begin{align}
\frac{\partial}{\partial X^{M}}\delta\left(  X^{2}\right)   &  =2X_{M}%
\delta^{\prime}\left(  X^{2}\right)  ,\;X\cdot\frac{\partial}{\partial
X}\delta\left(  X^{2}\right)  =2X^{2}\delta^{\prime}\left(  X^{2}\right)
=-2\delta\left(  X^{2}\right)  ,\label{delt1}\\
\partial^{2}\delta\left(  X^{2}\right)   &  =2\left(  d+2\right)
\delta^{\prime}\left(  X^{2}\right)  +4X^{2}\delta^{\prime\prime}\left(
X^{2}\right)  =2\left(  d-2\right)  \delta^{\prime}\left(  X^{2}\right)  .
\label{delt2}%
\end{align}
Here $\delta^{\prime}\left(  u\right)  ,\delta^{\prime\prime}\left(  u\right)
$ are the derivatives of the delta function with respect to its argument
$u=X^{2}.$ So we have used $u\delta^{\prime}\left(  u\right)  =-\delta\left(
u\right)  $ and $u\delta^{\prime\prime}\left(  u\right)  =-2\delta^{\prime
}\left(  u\right)  $ as the properties of the delta function of a single
variable $u$ to arrive at the above expressions. These are to be understood in
the sense of distributions under integration with smooth functions.

Inserting the solution $\hat{\Phi}\left(  X\right)  =\delta\left(
X^{2}\right)  \Phi\left(  X\right)  $ into the other two equations in
(\ref{phieoms}), and using Eqs.(\ref{delt1},\ref{delt2}), gives%
\begin{gather}
\delta\left(  X^{2}\right)  \left(  X\cdot\partial+\frac{d-2}{2}\right)
\Phi=0,\;\;\label{sp2reoms1}\\
\partial^{2}\left[  \delta\left(  X^{2}\right)  \Phi\right]  =\delta\left(
X^{2}\right)  \partial^{2}\Phi+4\delta^{\prime}\left(  X^{2}\right)  \left(
X\cdot\partial+\frac{d-2}{2}\right)  \Phi=0. \label{sp2reoms}%
\end{gather}
Here the derivatives must first be taken in the full space $X^{M}$ before the
condition $X^{2}=0$ is imposed. It is easy to see that these equations are
invariant under the following gauge transformations%
\begin{equation}
\delta_{\Lambda}\Phi=X^{2}\Lambda\left(  X\right)  \label{dellambda1}%
\end{equation}
for any function $\Lambda\left(  X\right)  .$ The $4\delta^{\prime}\left(
X^{2}\right)  $ part of the second equation in Eq.(\ref{sp2reoms}) with the
given coefficient $4$ is crucial for this invariance. If we define
\begin{equation}
\Phi\left(  X\right)  =\Phi_{0}\left(  X\right)  +X^{2}\tilde{\Phi}\left(
X\right)  , \label{phi0ph1}%
\end{equation}
where $\Phi_{0}\equiv\left[  \Phi\left(  X\right)  \right]  _{X^{2}=0},$ and
$X^{2}\tilde{\Phi}$ is the remainder, the gauge symmetry implies that
$\Phi_{0}$ is gauge invariant while $\tilde{\Phi}$ is pure gauge freedom and
completely drops out. Hence, the non-singular gauge invariant function
$\Phi_{0}\left(  X\right)  $ satisfies the following equations in $d+2$
dimensions
\begin{equation}
\left[  \left(  X\cdot\partial+\frac{d-2}{2}\right)  \Phi_{0}\right]
_{X^{2}=0}=0,\;\;\left[  \partial^{2}\Phi_{0}\right]  _{X^{2}=0}=0.
\label{2Tspin0eom}%
\end{equation}
where we have substituted the gauge invariant part $\Phi_{0}$ instead of the
full $\Phi$. This is verified directly by substituting Eq.(\ref{phi0ph1}) in
Eq.(\ref{sp2reoms1},\ref{sp2reoms}) and noting that $\tilde{\Phi}$ drops out.

The first equation in Eq.(\ref{2Tspin0eom}), together with the $X^{2}=0$
condition, are the kinematic equations, and the second one is the dynamical
equation. The kinematic equation is solved by any homogeneous function of
degree $-\left(  d-2\right)  /2.$ Namely, its general solution must satisfy
the scaling property $\Phi_{0}\left(  tX\right)  =t^{-\left(  d-2\right)
/2}\Phi_{0}\left(  X\right)  .$ Note that this homogeneity condition is much
more than assigning a scaling dimension to a field in usual field theory
because it is a restriction on the spacetime dependence of the
field\footnote{The scaling dimension alone does not require homogeneity. For
example a Klein-Gordon field in four dimensional usual field theory has
scaling dimension $-1,$ but it is not homogeneous. This is because the
dimension operator is not only the part $X\cdot\partial$ that acts on
coordinates, but also includes a part that acts on canonical field degrees of
freedom. In particular, note that the usual plane wave solutions with definite
momentum $\exp\left(  ik\cdot X\right)  $ are not homogeneous.}. With a
particular parametrization of $X^{M}$ that satisfies the other kinematic
constraint $X^{2}=0,$ as given in Eq.(\ref{massless}), plus the homogeneity
condition, one can show \cite{Dirac}\cite{2tfield} that the dynamical equation
for $\Phi_{0}\left(  X\right)  $ in $d+2$ dimensions reduces to the massless
Klein-Gordon equation $\frac{\partial^{2}\phi\left(  x\right)  }{\partial
x^{\mu}\partial x_{\mu}}=0$ in $d$ dimensions, with a definite relationship
between $\Phi_{0}\left(  X\right)  $ and $\phi\left(  x\right)  .$ We will
return to this detail of holography in the following section.

\subsubsection{Action for scalars with interactions}

We now propose the following \textit{interacting} field theory action that
reproduces both the kinematical and dynamical equations of motion
Eqs.(\ref{2Tspin0eom}). The construction of a proper action principle in
2T-physics field theory has eluded all efforts before, even though one could
write equations of motion as shown above and \cite{Dirac}-\cite{2tfield},
including interactions. We will argue that we obtain the interactions uniquely
through a gauge principle directly connected to the underlying Sp$\left(
2,R\right)  $ symmetry.

The inspiration for the following form came from a BRST formulation for
2T-physics field theory including interactions \cite{2tbrst2006}. Here we do
not use the BRST version but only extract from it a partially gauged fixed
version which has just sufficient leftover gauge symmetry for our purposes
here. Thus, the key ingredients that go into the proposed action below are
first that it should possess the gauge symmetry in Eq.(\ref{dellambda1}), and
second that it should have additional gauge symmetry to reduce the theory to
only the $\Phi_{0}$ degree of freedom, including interaction. The action is%
\begin{equation}
S\left(  \Phi\right)  =\int d^{d+2}X~\left\{  B\left(  X\right)  \partial
^{2}\left[  \Phi\delta\left(  X^{2}\right)  \right]  -\delta\left(
X^{2}\right)  \left[  B\left(  X\right)  V^{\prime}\left(  \Phi\right)
+aV\left(  \Phi\right)  \right]  \right\}  .\label{phiAaction}%
\end{equation}
Notice the delta function that imposes the vanishing Sp$\left(  2,R\right)  $
generator $X^{2}=0$ condition. The function $V\left(  \Phi\right)  $ will be
the potential energy for the field's self interactions, and its derivative is
$V^{\prime}\left(  \Phi\right)  =\partial V/\partial\Phi$.  The role of the
constant coefficient $a$ will become evident in the discussion below. We will
see that $V\left(  \Phi\right)  $ will be uniquely determined by the gauge
symmetries of the field $B\left(  X\right)  $. The field $B\left(  X\right)  $
emerged from the BRST point of view as a combination of auxiliary fields
associated with the kinematical and dynamical Sp$\left(  2,R\right)  $
generators $X\cdot P$ and $P^{2}$.

Let us first discuss the gauge symmetries of this action. The $\delta\left(
X^{2}\right)  $ structure makes it evident that we have the gauge symmetry of
Eq.(\ref{dellambda1}) $\delta_{\Lambda}\Phi=X^{2}\Lambda\left(  X\right)  , $
hence if $\Phi$ is written in the form of Eq.(\ref{phi0ph1}) $\Phi=\Phi
_{0}+X^{2}\tilde{\Phi},$ the remainder $\tilde{\Phi}$ automatically drops out.
Therefore this action really depends only on $\Phi_{0}$ automatically. We will
continue to write $\Phi$ everywhere, but it should be understood that any mode
of $\Phi$ proportional to $X^{2}$ is decoupled, and this fact will be used below.

Next we show that there is nontrivial gauge symmetry associated with the field
$B\left(  X\right)  $ under the following transformation with gauge parameter
$b\left(  X\right)  $ \footnote{This transformation was extracted from the
BRST formalism in \cite{2tbrst2006}.}
\begin{equation}
\delta_{b}B=\left(  X\cdot\partial+\frac{d-2}{2}\right)  b-\frac{1}{4}%
X^{2}\left(  \partial^{2}b-bV^{\prime\prime}\left(  \Phi\right)  \right)
,\;\text{any }b\left(  X\right)  . \label{delcA}%
\end{equation}
The transformation of the action under this $b$-symmetry gives%
\begin{align}
\delta_{b}S\left(  \Phi\right)   &  =\int d^{d+2}X~\left(
\begin{array}
[c]{c}%
\left\{  \left(  X\cdot\partial+\frac{d-2}{2}\right)  b-\frac{1}{4}%
X^{2}\left(  \partial^{2}b-bV^{\prime\prime}\left(  \Phi\right)  \right)
\right\} \\
\times\left\{  \partial^{2}\left[  \Phi\delta\left(  X^{2}\right)  \right]
-\delta\left(  X^{2}\right)  V^{\prime}\left(  \Phi\right)  \right\}
\end{array}
\right) \label{dc}\\
&  =\int d^{d+2}X~\left(
\begin{array}
[c]{c}%
\delta\left(  X^{2}\right)  \left[
\begin{array}
[c]{c}%
\left(  X\cdot\partial+\frac{d-2}{2}\right)  b\left[  \partial^{2}%
\Phi-V^{\prime}\left(  \Phi\right)  \right] \\
+\left(  \partial^{2}b-bV^{\prime\prime}\left(  \Phi\right)  \right)  \left(
X\cdot\partial+\frac{d-2}{2}\right)  \Phi
\end{array}
\right] \\
+4\delta^{\prime}\left(  X^{2}\right)  \left(  X\cdot\partial+\frac{d-2}%
{2}\right)  b~\left(  X\cdot\partial+\frac{d-2}{2}\right)  \Phi
\end{array}
\right) \label{dc2}\\
&  =\int d^{d+2}X~\left(
\begin{array}
[c]{c}%
\partial_{M}\left\{  X^{M}\left[  4\delta^{\prime}\left(  X^{2}\right)
\Phi\left(  X\cdot\partial b+\frac{d-2}{2}b\right)  -\delta\left(
X^{2}\right)  V^{\prime}\left(  \Phi\right)  b\right]  \right\} \\
-\delta\left(  X^{2}\right)  \frac{d-2}{2}b~\left[  \Phi V^{\prime\prime
}\left(  \Phi\right)  -\frac{d+2}{d-2}V^{\prime}\left(  \Phi\right)  \right]
\end{array}
\right)  \label{dc3}%
\end{align}
In going from Eq.(\ref{dc}) to Eq.(\ref{dc2}) we have used Eq.(\ref{sp2reoms})
to evaluate $\partial^{2}\left[  \Phi\delta\left(  X^{2}\right)  \right]  $
and then set $X^{2}\delta\left(  X^{2}\right)  =0$ and $X^{2}\delta^{\prime
}\left(  X^{2}\right)  =-\delta\left(  X^{2}\right)  .$ To reach
Eq.(\ref{dc3}) we do integrations by parts taking into account the delta
functions and noting the identity%
\begin{align}
&  \partial^{2}\left[  \delta\left(  X^{2}\right)  \left(  X\cdot
\partial+\frac{d-2}{2}\right)  b\right] \\
&  =\delta\left(  X^{2}\right)  \left(  X\cdot\partial+\frac{d+2}{2}\right)
\partial^{2}b+4\delta^{\prime}\left(  X^{2}\right)  \left(  X\cdot
\partial+\frac{d-2}{2}\right)  ^{2}b.
\end{align}
The total derivative in Eq.(\ref{dc3}) can be dropped, and then we see there
is a gauge symmetry $\delta_{b}S\left(  \Phi\right)  =0$ provided the
potential energy $V\left(  \Phi\right)  $ satisfies%
\begin{equation}
\Phi V^{\prime\prime}\left(  \Phi\right)  =\frac{d+2}{d-2}V^{\prime}\left(
\Phi\right)  \;\rightarrow V^{\prime}\left(  \Phi\right)  =\lambda\Phi
^{\frac{d+2}{d-2}}~. \label{Vprime}%
\end{equation}
Thus, except for the overall constant $\lambda$, the $V^{\prime}\left(
\Phi\right)  $ is uniquely determined as the given monomial, as a consequence
of imposing the $b$-gauge symmetry$.$This gauge symmetry is required to reduce
the $\Phi,B$ degrees of freedom to only $\Phi_{0}.$ But interestingly, it also
fixes the interaction uniquely.

Now let us verify that this action gives the equations of motion that we
require. The general variation of the action is obtained from
Eq.(\ref{phiAaction})%
\[
\delta S\left(  \Phi\right)  =-\int d^{d+2}X~\left[
\begin{array}
[c]{c}%
\delta\left(  X^{2}\right)  \delta B\left\{  \partial^{2}\Phi+V^{\prime
}\left(  \Phi\right)  \right\}  \\
+4\delta^{\prime}\left(  X^{2}\right)  \delta B\left(  X\cdot\partial
+\frac{d-2}{2}\right)  \Phi\\
+\delta\left(  X^{2}\right)  \delta\Phi\left[  \partial^{2}B+V^{\prime\prime
}\left(  \Phi\right)  B+aV^{\prime}\left(  \Phi\right)  \right]
\end{array}
\right]
\]
The $4\delta^{\prime}\left(  X^{2}\right)  $ comes from evaluating
$\partial^{2}\left[  \Phi\delta\left(  X^{2}\right)  \right]  $ as in
Eq.(\ref{sp2reoms}). Since the distributions $\delta\left(  X^{2}\right)
,\delta^{\prime}\left(  X^{2}\right)  $ are linearly independent the
coefficients of $\delta\left(  X^{2}\right)  \delta B$ and $\delta^{\prime
}\left(  X^{2}\right)  \delta B$ should vanish independently\footnote{In this
paper we are very careful when we make such statements. The equation
$\delta\left(  X^{2}\right)  F\left(  X\right)  +\delta^{\prime}\left(
X^{2}\right)  G\left(  X\right)  =0$ has the more general solution $\left(
G\right)  _{X^{2}=0}=0$ and $\left(  F-\tilde{G}\right)  _{X^{2}=0}=0,$ rather
than merely $\left(  F\right)  _{X^{2}=0}=0$ for the second equation. Here
$\tilde{G}$ is the remainder when one writes $G=G_{0}+X^{2}\tilde{G}.$ So
generally $F$ need not vanish on its own. However, in the present case we have
already argued that $\Phi=\Phi_{0}$ since the remainder drops out. Therefore
the two terms do vanish separately.\label{careful}}%
\begin{equation}
\left(  X\cdot\partial\Phi+\frac{d-2}{2}\Phi\right)  _{X^{2}=0}=0,\;\;\left[
\partial^{2}\Phi-V^{\prime}\left(  \Phi\right)  \right]  _{X^{2}=0}=0.
\end{equation}
In these equations we really have $\Phi=\Phi_{0}$ with no remainder
$\tilde{\Phi}$ due to the gauge symmetry $\delta_{\Lambda}\Phi=X^{2}\Lambda$
as discussed above. But we may allow any remainder as long as it is
homogeneous since this does not change the equations of motion. Thus, our
action did provide the two desired equations of motion for $\Phi_{0}$, while
the remainder $\tilde{\Phi}$ is gauge freedom, and can still be taken as
non-zero as long as it is homogeneous. In addition, the equation of motion for
$B$ is
\begin{equation}
\left[  \partial^{2}B-V^{\prime\prime}\left(  \Phi\right)  B-aV^{\prime
}\left(  \Phi\right)  \right]  _{X^{2}=0}=0.
\end{equation}
This can be understood in a bit more detail by exhibiting the remainder of $B$
in the form $B=B_{0}+X^{2}\tilde{B}.$ Then the $B$ equation really is%
\begin{equation}
\left[  \partial^{2}B_{0}-V^{\prime\prime}\left(  \Phi_{0}\right)
B_{0}-aV^{\prime}\left(  \Phi_{0}\right)  +4\left(  X\cdot\partial+\frac
{d+2}{2}\right)  \tilde{B}\right]  _{X^{2}=0}=0.\label{Atilde}%
\end{equation}
We see that this is an equation that determines the remainder $\tilde{B}$ in
terms of $B_{0},\Phi_{0},$ without fixing the dynamics of $B_{0}$ at all. So
there remains one fully undetermined function among the $B_{0},\tilde{B}$.
This is of course related to the $b$-symmetry given in Eq.(\ref{delcA}). Using
the $b$-symmetry we can choose the function $B_{0}$ at will. As in
\cite{2tbrst2006} we make the convenient gauge choice $B_{0}=\gamma\Phi_{0}$
where $\gamma$ is an overall constant to be determined consistently. In that
case the remainder $\tilde{B}$ must be determined from Eq.(\ref{Atilde}) after
inputting $B_{0}=\gamma\Phi_{0}$ and recalling the dynamical equation for
$\Phi_{0}$ up to the proportionality constant $\gamma\left[  \partial^{2}%
\Phi_{0}-V^{\prime}\left(  \Phi_{0}\right)  \right]  _{X^{2}=0}=0.$ After
using $\Phi V^{\prime\prime}\left(  \Phi\right)  =\frac{d+2}{d-2}V^{\prime
}\left(  \Phi\right)  $ as given in Eq.(\ref{Vprime}) we obtain $\tilde{B}$%
\begin{equation}
B_{0}=\gamma\Phi_{0},\;4\left(  X\cdot\partial+\frac{d+2}{2}\right)  \tilde
{B}=\left(  \frac{4\gamma}{d-2}+a\right)  V^{\prime}\left(  \Phi_{0}\right)  .
\end{equation}
Finally, we can insert the fully fixed $B_{0},\tilde{B}$ as well as $\Phi
=\Phi_{0}$ into the action and obtain an action purely in terms of $\Phi_{0}$%
\begin{align}
S\left(  \Phi_{0}\right)   &  =\int d^{d+2}X~\left[  \left(  B_{0}+X^{2}%
\tilde{B}\right)  \left\{
\begin{array}
[c]{c}%
\delta\left(  X^{2}\right)  \left(  \partial^{2}\Phi_{0}-V^{\prime}\left(
\Phi_{0}\right)  \right)  \\
+4\delta^{\prime}\left(  X^{2}\right)  \left(  X\cdot\partial+\frac{d-2}%
{2}\right)  \Phi_{0}%
\end{array}
\right\}  -\delta\left(  X^{2}\right)  aV\left(  \Phi_{0}\right)  \right]
\label{ss1}\\
&  =\int d^{d+2}X~\left[
\begin{array}
[c]{c}%
\delta\left(  X^{2}\right)  \left\{
\begin{array}
[c]{c}%
B_{0}\left(  \partial^{2}\Phi_{0}-V^{\prime}\left(  \Phi_{0}\right)  \right)
-aV\left(  \Phi_{0}\right)  \\
+4\Phi_{0}\left(  X\cdot\partial+\frac{d+2}{2}\right)  \tilde{B}%
\end{array}
\right\}  \\
+4\delta^{\prime}\left(  X^{2}\right)  B_{0}\left(  X\cdot\partial+\frac
{d-2}{2}\right)  \Phi_{0}%
\end{array}
\right]  \label{ss2}\\
&  =\int d^{d+2}X~\delta\left(  X^{2}\right)  \left[  \gamma\Phi_{0}%
\partial^{2}\Phi_{0}+\left(  a-\gamma\frac{d-6}{d-2}\right)  \Phi_{0}%
V^{\prime}\left(  \Phi_{0}\right)  -aV\left(  \Phi_{0}\right)  \right]
\label{ss5}%
\end{align}
Eq.(\ref{ss1}) is the original action Eq.(\ref{phiAaction}) rewritten in terms
of the components $\Phi_{0},B_{0},\tilde{B}$. The form in Eq.(\ref{ss2})
follows after using $X^{2}\delta\left(  X^{2}\right)  =0$ and $X^{2}%
\delta^{\prime}\left(  X^{2}\right)  =-\delta\left(  X^{2}\right)  ,$ and
performing an integration by parts in the middle term to get the structure
$\left(  X\cdot\partial+\frac{d+2}{2}\right)  \tilde{B}$. Inserting the gauge
fixed $B_{0},\tilde{B}$ we get Eq.(\ref{ss5}), where the $\delta^{\prime
}\left(  X^{2}\right)  $ term of Eq.(\ref{ss2}) becomes the total derivative
$2\gamma\partial_{M}\left(  X^{M}\delta^{\prime}\left(  X^{2}\right)  \Phi
_{0}^{2}\right)  $ and drops out in this gauge.

\bigskip We must require that the equation of motion for $\Phi_{0}$ that
follows from this gauge fixed action be the same as $\partial^{2}\Phi
_{0}-V^{\prime}\left(  \Phi_{0}\right)  =0$ as given by the original action.
For this to be the case, the coefficient $\gamma$ that had appeared  in the
gauge fixing $B_{0}=\gamma\Phi_{0}$ must be determined self-consistently in
terms of the constant $a$ as $\gamma=-\frac{d-2}{4}a,$ so that the potential
energy terms in Eq.(\ref{ss5}) sum up to being $-2\gamma V\left(  \Phi\right)
$ and match the normalization of the kinetic terms as follows%
\begin{equation}
S\left(  \Phi\right)  =2\gamma\int d^{d+2}X~\delta\left(  X^{2}\right)
\left[  \frac{1}{2}\Phi\partial^{2}\Phi-\lambda\frac{d-2}{2d}\Phi^{\frac
{2d}{d-2}}\right]  .\label{phiactionFixed}%
\end{equation}
The potential energy $V\left(  \Phi\right)  $ has now been fixed uniquely up
to an overall coupling constant as the monomial%
\begin{equation}
V\left(  \Phi\right)  =\lambda\frac{d-2}{2d}\Phi^{\frac{2d}{d-2}}.
\end{equation}
The constant $2\gamma$ is an overall normalization factor that will be
absorbed later into the normalization of the volume in $\left(  d-1\right)
+1$ dimensions.

In our derivation the $\Phi$ in the action of Eq.(\ref{phiactionFixed}) was
strictly $\Phi_{0}.$ But we can add a remainder to $\Phi_{0}+X^{2}\tilde{\Phi
}$ without changing the physics, as long as the remainder is homogeneous and
satisfies $\left(  X\cdot\partial+\frac{d+2}{2}\right)  \tilde{\Phi}=0.$ Then
the full $\Phi$ satisfies $\left(  X\cdot\partial+\frac{d-2}{2}\right)
\Phi=0$ when it is on shell. We have the freedom to add a homogeneous
remainder because the action in Eq.(\ref{phiactionFixed}), including the
homogeneous remainder but off shell $\Phi_{0},$ has a leftover gauge symmetry
$\delta\Phi=X^{2}\Lambda$ as long as $\Lambda$ is homogeneous $\left(
X\cdot\partial+\frac{d+2}{2}\right)  \Lambda=0.$ To demonstrate this symmetry
observe the general variation of the action in Eq.(\ref{phiactionFixed})
\begin{align}
\delta S\left(  \Phi\right)   &  =2\gamma\int d^{d+2}X~\delta\left(
X^{2}\right)  \left\{  \delta\Phi\left(  \frac{1}{2}\partial^{2}\Phi
-V^{\prime}\left(  \Phi\right)  \right)  +\frac{1}{2}\Phi\partial^{2}\left(
\delta\Phi\right)  \right\}  \\
&  =2\gamma\int d^{d+2}X~\delta\Phi\left\{  \delta\left(  X^{2}\right)
\left(  \frac{1}{2}\partial^{2}\Phi-V^{\prime}\left(  \Phi\right)  \right)
+\frac{1}{2}\partial^{2}\left(  \delta\left(  X^{2}\right)  \Phi\right)
\right\}  \\
&  =2\gamma\int d^{d+2}X~\delta\Phi\left\{
\begin{array}
[c]{c}%
\delta\left(  X^{2}\right)  \left[  \partial^{2}\Phi-V^{\prime}\left(
\Phi\right)  \right]  \\
+2\delta^{\prime}\left(  X^{2}\right)  \left[  X\cdot\partial\Phi+\frac
{d-2}{2}\Phi\right]
\end{array}
\right\}  ,\label{phivary}%
\end{align}
If we substitute the gauge transformation $\delta\Phi=X^{2}\Lambda$ in
Eq.(\ref{phivary}) we get the gauge variation $\delta_{\Lambda}S\left(
\Phi\right)  $ which becomes
\begin{align}
\delta_{\Lambda}S\left(  \Phi\right)   &  =2\gamma\int d^{d+2}X~\Lambda
\delta\left(  X^{2}\right)  \left[  X\cdot\partial\Phi+\frac{d-2}{2}%
\Phi\right]  \\
&  =2\gamma\int d^{d+2}X~\partial_{M}\left(  X^{M}\Lambda\Phi\delta\left(
X^{2}\right)  \right)  =0.
\end{align}
In the first line we have already dropped a term due to $X^{2}\delta\left(
X^{2}\right)  =0$ and used $X^{2}\delta^{\prime}\left(  X^{2}\right)
=-\delta\left(  X^{2}\right)  .$ The resulting form is a total divergence as
given in the second line as long as $\Lambda$ is homogeneous $\left(
X\cdot\partial+\frac{d+2}{2}\right)  \Lambda=0.$

Furthermore, the action in Eq.(\ref{phiactionFixed}), including the
homogeneous remainder, has a leftover $b$-symmetry $\delta_{b}\Phi$ given in
Eq.(\ref{delcA}) as long as the $b$ parameter is homogeneous $\left(
X\cdot\partial+\frac{d-2}{2}\right)  b=0$. In that case, from
Eq.(\ref{phivary}) we derive
\begin{equation}
\delta_{b}S\left(  \Phi\right)  =\gamma\left(  d-2\right)  \int d^{d+2}%
X~\delta\left(  X^{2}\right)  ~b~\left[  \Phi V^{\prime\prime}\left(
\Phi\right)  -\frac{d+2}{d-2}V^{\prime}\left(  \Phi\right)  \right]
\label{bleftover}%
\end{equation}
So, requiring the symmetry $\delta_{b}S\left(  \Phi\right)  =0$ fixes the
potential uniquely.

In conclusion, in the gauge fixed action in Eq.(\ref{phiactionFixed}) we can
allow any $\Phi$ whose remainder $\tilde{\Phi}$ has the homogeneity property
stated. Of course this permits gauge fixing off-shell all the way to
$\Phi=\Phi_{0}$ if so desired, but we will rather keep the homogeneous gauge
freedom as the remainder\footnote{Note that the last coefficient in
Eq.(\ref{phivary}) is $2\delta^{\prime}\left(  X^{2}\right)  $ and not
$4\delta^{\prime}\left(  X^{2}\right)  .$ If it had been $4\delta^{\prime
}\left(  X^{2}\right)  $ there would have been a greater symmetry with
arbitrary $\Lambda$ and arbitrary $b$ rather than homogeneous $\Lambda,$ and
homogeneous $b.$ Of course, the original action in Eq.(\ref{phiAaction}) has
the greater symmetry before gauge fixing.} of the $\Lambda$ and $b$ symmetries.

Now we show that the action in Eq.(\ref{phiactionFixed}) is adequate to
generate both the kinematic and dynamical equations of motion. Using
Eq.(\ref{phivary}) we impose the variational principle $\delta S\left(
\Phi\right)  =0$ which gives
\begin{equation}
\delta\left(  X^{2}\right)  \left[  \partial^{2}\Phi-V^{\prime}\left(
\Phi\right)  \right]  +2\delta^{\prime}\left(  X^{2}\right)  \left[
X\cdot\partial\Phi+\frac{d-2}{2}\Phi\right]  =0.
\end{equation}
This results in \textit{two different equations on }$\Phi$\textit{, not just
one} because the coefficients of both $\delta\left(  X^{2}\right)  $ and
$\delta^{\prime}\left(  X^{2}\right)  $ must vanish separately. By contrast in
an ordinary field theory the variation of a single field would result in a
single equation. This is one of the crucial observations that was not
appreciated in our previous attempts to construct an action principle that
gave both equations of motion. 

Being careful as explained in footnote (\ref{careful}), the coefficients of
$\delta\left(  X^{2}\right)  $ and $\delta^{\prime}\left(  X^{2}\right)  $
that vanish are%
\begin{equation}
\left[  \partial^{2}\Phi_{0}-V^{\prime}\left(  \Phi_{0}\right)  +\left(
X\cdot\partial+\frac{d+2}{2}\right)  \tilde{\Phi}\right]  _{X^{2}}=0,\;\left(
X\cdot\partial+\frac{d-2}{2}\right)  \Phi_{0}=0.
\end{equation}
The resulting equations are precisely the desired ones, provided $\tilde{\Phi
}$ is homogeneous, $\left(  X\cdot\partial+\frac{d+2}{2}\right)  \tilde{\Phi
}=0$ as is the case in our gauge fixed action as explained above. In that case
we can write the equations of motion without splitting $\Phi$ into components
in the form
\begin{equation}
\left(  X\cdot\partial+\frac{d-2}{2}\right)  \Phi=0\;\text{and\ \ }%
\partial^{2}\Phi-V^{\prime}\left(  \Phi\right)  =0.\; \label{variational}%
\end{equation}
We have made the point that it is crucially important that it is understood
that the action $S\left(  \Phi\right)  $ in Eq.(\ref{phiactionFixed}) is a
gauge fixed version of the original action Eq.(\ref{phiAaction} that contains
only the gauge fixed $\Phi$ up to an arbitrary homogeneous remainder, rather
than the most general remainder $\tilde{\Phi}.$ For the most general remainder
$\tilde{\Phi}$ the action in Eq.(\ref{phiactionFixed}) would give the wrong
dynamical equation.

Therefore, the correct action is either the simplified form
Eq.(\ref{phiactionFixed}) with the gauge fixed $\Phi$ up to a homogeneous
remainder that corresponds to remaining gauge freedom, or it is the more
general gauge invariant form in Eq.(\ref{phiAaction}) that includes all the
degrees of freedom in $\Phi$ as well as those of $B$. Recall that the
$b$-gauge symmetry uniquely determined the interaction $V\left(  \Phi\right)
.$

The advantage of the gauge fixed version in Eq.(\ref{phiactionFixed}) is its
simplicity in terms of a single field $\Phi$, but we must point out a subtle
feature. To arrive at the two equations of motion from this gauge fixed
action, we note that we have applied a slightly unconventional variational
approach. Specifically, note that two equations follow from the fact that the
general variation $\delta\Phi=\delta\Phi_{0}+X^{2}\delta\tilde{\Phi}$ contains
two general variational parameters $\delta\Phi_{0},\delta\tilde{\Phi},$ in
which neither $\delta\Phi_{0}$ nor $\delta\tilde{\Phi}$ are homogeneous,
although the remainder $\tilde{\Phi}$ is required to be homogeneous after the
variation. The unconvetional part is the requirement of a homogeneous
$\tilde{\Phi}$ (as determined in our gauge fixing discussion), but a general
$\delta\tilde{\Phi}$ to yield the second equation for $\Phi_{0},$ namely
$\left(  X\cdot\partial+\frac{d-2}{2}\right)  \Phi_{0}=0.$ By taking a
homogeneous $\tilde{\Phi}$ but a general $\delta\tilde{\Phi}$ we have devised
a tool to keep track of the effects of the 2T gauge symmetry off-shell, whose
utility is demonstrated in Eq.(\ref{bleftover}), and which will come in handy
in later investigations. The alternative to the above is to take from the
beginning a gauge fixed action instead of Eq.(\ref{phiactionFixed}) with a
homogeneous $\Phi_{0}$ as well as homogeneous $\delta\Phi_{0},$ and have no
$\tilde{\Phi},\delta\tilde{\Phi}$ at all, however in so doing we completely
lose track of the 2T gauge symmetry.

There is another physically equivalent gauge fixed form of the action that
makes the underlying Sp$\left(  2,R\right)  $ symmetry more evident. This is
given in Appendix A.

We have now shown that the field $\Phi\left(  X\right)  $ described by our
action satisfies precisely the same free field equations of motion in
Eqs.(\ref{2Tspin0eom}) that follow from the Sp$\left(  2,R\right)  $
constraints, plus consistent interactions. So the physical degrees of freedom
and gauge symmetries of the 2T spinless free particle are correctly described
by our action principle. In addition we have introduced a gauge principle that
leads to unique self interactions.

In this process we have also discovered a new gauge symmetry that we will call
the 2Tgauge-symmetry. This includes both the $\Lambda$ and the $b$ gauge
symmetries. These gauge symmetries are responsible for removing gauge degrees
of freedom, and identify the physical field as $\Phi_{0}\left(  X\right)
=\left[  \Phi\left(  X\right)  \right]  _{X^{2}=0}.$ We will see that the
2Tgauge-symmetry persists in the presence of all interactions of the field.
Furthermore, for each field in the theory there is an extension of this
symmetry, so it is a rather general symmetry that dictates the structure of
the action.

\subsubsection{Interactions among several scalars}

Let us now describe interactions among several scalar fields. For convenience
we will do this in the gauge fixed version\footnote{The fully gauge invariant
version must have a $B^{i}\left(  X\right)  $ field corresponding to each
$\Phi^{i}\left(  X\right)  $. Furthermore, there is a separate $\Lambda^{i}$
and $b^{i}$ gauge parameter for each $i.$ After the gauge fixing procedure
described in the previous section for each $i$ we end up in the physical
sector without the $B^{i}\left(  X\right)  ,$ and only with the $\Phi
^{i}\left(  X\right)  $ whose remainder $\tilde{\Phi}^{i}\left(  X\right)  $
is gauge fixed to be homogeneous. \label{bgauge}} by using directly $\Phi
^{i}=\Phi_{0}^{i}$ for all the fields labeled by $i=1,2,\cdots$. However, for
brevity we will omit the zero subscript in $\Phi_{0}^{i}.$ We identify as
$S_{0}\left(  \Phi^{i}\right)  =-\frac{1}{2}Z\int d^{d+2}X~\delta\left(
X^{2}\right)  \Phi^{i}\partial^{2}\Phi^{i}$ the quadratic part of the action
in Eq.(\ref{phiactionFixed}) at zero coupling constant. The interaction term
is then identified as
\begin{equation}
S_{1}\left(  \Phi^{i}\right)  =-\int d^{d+2}X~\delta\left(  X^{2}\right)
V\left(  \Phi^{i}\right)  .
\end{equation}
The $b$-symmetry requires $V\left(  \Phi^{i}\right)  $ to be overall
homogeneous of degree $\frac{2d}{d-2}.$ For example, if there are two scalar
fields, say $\Phi\left(  X\right)  $ and $H\left(  X\right)  ,$ the total
action must be taken in the form%
\begin{equation}
S\left(  \Phi,H\right)  =S_{0}\left(  \Phi\right)  +S_{0}\left(  H\right)
-\int d^{d+2}X~\delta\left(  X^{2}\right)  V\left(  \Phi,H\right)  .
\label{SphiH}%
\end{equation}
where the allowed potential energy can only be of the form
\begin{equation}
V\left(  \Phi,H\right)  =\lambda_{\phi}\Phi^{\frac{2d}{d-2}}+\lambda
_{H}H^{\frac{2d}{d-2}}+\sum_{k,l}\lambda_{kl}\Phi^{k}H^{l}\delta
_{k+l-\frac{2d}{d-2}}. \label{VphiH}%
\end{equation}
The coupling constants are all dimensionless in any dimension. Note that in
four dimensions $d=4$ only quartic interactions are allowed $V\left(
\Phi,H\right)  =\lambda_{\phi}\Phi^{4}+\lambda_{H}H^{4}+\sum_{k,l}\lambda
_{kl}\Phi^{k}H^{l}\delta_{k+l-4}.$ No quadratic mass terms with dimensionful
constants can be included. This will impact our discussion of mass generation
as will be seen below.

It is possible to modify the rigid result on the form of the potential
discussed above by modifying the action with another type of term that
includes $\delta^{\prime}\left(  X^{2}\right)  $ instead of only
$\delta\left(  X^{2}\right)  $. To illustrate this consider again the single
field case and include an additional term in the action of the form
\begin{equation}
S_{2}\left(  \Phi\right)  =-\int d^{d+2}X~\delta^{\prime}\left(  X^{2}\right)
W\left(  \Phi\right)  . \label{s2W}%
\end{equation}
The equations of motion as well as the gauge symmetries are altered with the
total action
\begin{equation}
S_{tot}\left(  \Phi\right)  =S_{0}\left(  \Phi\right)  +S_{1}\left(
\Phi\right)  +S_{2}\left(  \Phi\right)  .
\end{equation}
Therefore the final potential $V\left(  \Phi\right)  $ could be different.
After a few trials with a few such functions it becomes clear that the
resulting equations of motion may have only trivial solutions except for
special combinations of $V$ and $W$ that must be chosen consistently to avoid
a trivial system. So far we have found only one very simple non-trivial case,
given by a quadratic $W=\frac{1}{2}a\Phi^{2}$. This changes the kinematic
equation for $\Phi,$ and requires $\Phi$ to be homogeneous $\Phi\left(
tX\right)  =t^{k\left(  a\right)  }\Phi\left(  X\right)  $ with a degree
$k\left(  a\right)  $ that depends on the constant $a.$ Consistent with the
new homogeneity degree of $\Phi$, the potential energy is again a monomial
$V\left(  \Phi\right)  \sim\Phi^{p\left(  a\right)  }$ with a new power
$p\left(  a\right)  $ so that $V\left(  \Phi\right)  $ has total homogeneity
degree $-d$ just as before. Thus, the power $p\left(  a\right)  $ of the
monomial in the potential $V\left(  \Phi\right)  $ can be changed arbitrarily
as a function of the coefficient $a$ in $W=\frac{1}{2}a\Phi^{2}.$ Currently we
do not know of other examples for which the coupled kinematic and dynamical
equations have non-trivial solutions for other functionals $W\left(
\Phi\right)  $.

It has now become clear that $V\left(  \Phi\right)  $ could be altered by
tinkering with the additional term $W\left(  \Phi\right)  $. However, it is
not a priori clear what forms of $V\left(  \Phi\right)  $ exist consistently
with the coupled differential equations$.$

If there is more than one scalar, such as $\Phi,H,$ then the system of
equations derived from $W\left(  \Phi,H\right)  $ and $V\left(  \Phi,H\right)
$ gets more complicated. A general study of which forms of $V\left(
\Phi\right)  $ or $V\left(  \Phi,H\right)  $ can consistently be found through
this procedure is not currently available. We will return to this topic when
we discuss the mass generation mechanism in the Standard Model.

\subsection{Spinor Field}

\subsubsection{Free field equations for fermions}

For particles of spin $1/2$ on the worldline the phase space $\left(
X^{M},P^{M},\psi^{M}\right)  $ includes the anticommuting fermions $\psi^{M}$.
The worldline gauge symmetry acting on phase space is enlarged from Sp$\left(
2,R\right)  $ to the supergroup OSp$\left(  1|2\right)  $ \cite{spin2t}. The
generators of this symmetry in the flat background are proportional to $X^{2}%
$, $P^{2},\left(  X\cdot P+P\cdot X\right)  $, $X\cdot\psi,$ $P\cdot\psi.$ The
quantized fermions $\psi^{M}$ form a Clifford algebra and therefore are
represented by gamma matrices\footnote{An explicit form of SO$\left(
d,2\right)  $ gamma matrices in even dimensions labelled by $M=0^{\prime
},1^{\prime},\mu$ and $\mu=0,i$, is given by $\Gamma^{0^{\prime}}=-i\tau
_{1}\times1,$ $\Gamma^{1^{\prime}}=\tau_{2}\times1$, $\Gamma^{0}=1\times
1$,\ $\Gamma^{i}=\tau_{3}\times\gamma^{i},$ \ where $\gamma^{i}$ are the
SO$\left(  d-1\right)  $ gamma matrices. It is convenient to use a lightcone
type basis by defining\ $\Gamma^{\pm^{\prime}}=\frac{1}{\sqrt{2}}\left(
\Gamma^{0^{\prime}}\pm\Gamma^{1^{\prime}}\right)  =-i\sqrt{2}\tau^{\pm}%
\times1$. The $\bar{\Gamma}^{M}$ are the same as the $\Gamma^{M}$ for
$M=\pm^{\prime},i,$ but for $M=0$ we have $\bar{\Gamma}^{0}=-\Gamma
^{0}=-1\times1.$ From these we construct the traceless $\Gamma^{MN}$ as
$\Gamma^{+^{\prime}-^{\prime}}=\left(
\genfrac{}{}{0pt}{}{-1}{0}%
\genfrac{}{}{0pt}{}{0}{1}%
\right)  $,\ $\Gamma^{+^{\prime}\mu}=i\sqrt{2}\left(
\genfrac{}{}{0pt}{}{0}{0}%
\genfrac{}{}{0pt}{}{\bar{\gamma}^{\mu}}{0}%
\right)  $,\ $\Gamma^{-^{\prime}\mu}=-i\sqrt{2}\left(
\genfrac{}{}{0pt}{}{0}{\gamma^{\mu}}%
\genfrac{}{}{0pt}{}{0}{0}%
\right)  $,\ $\Gamma^{\mu\nu}=\left(
\genfrac{}{}{0pt}{}{\bar{\gamma}^{\mu\nu}}{0}%
\genfrac{}{}{0pt}{}{0}{\gamma^{\mu\nu}}%
\right)  $, where $\gamma_{\mu}=\left(  1,\gamma^{i}\right)  $, $\bar{\gamma
}_{\mu}=\left(  -1,\gamma^{i}\right)  $, noting the lower $\mu$ indices. Then
$\frac{1}{2}\Gamma_{MN}J^{MN}=-\Gamma^{+^{\prime}-^{\prime}}J^{+^{\prime
}-^{\prime}}$+~ $\frac{1}{2}J_{\mu\nu}\Gamma^{\mu\nu}-$ $\Gamma_{~\mu
}^{+^{\prime}}J^{-^{\prime}\mu}-$ $\Gamma_{~\mu}^{-^{\prime}}J^{+^{\prime}\mu
}$ takes an explicit matrix form. We can further write $\gamma^{1}=\sigma
^{1}\times1,$ $\gamma^{2}=\sigma^{2}\times1$ and $\gamma^{r}=\sigma^{3}%
\times\rho^{r},$ where the $\rho^{r}$ are the gamma matrices for SO$\left(
d-3\right)  $. It is possible to choose hermitian $\rho^{r}.$ In $d=4$ the
$\rho^{r}$ are replaced just by the number $1$ and then the $\gamma_{\mu}%
,\bar{\gamma}_{\mu}$ are just $2\times2$ Pauli matrices. These gamma matrices
are consistent with the metric in spinor space $\eta=-i\tau_{1}\times
1\times1=\Gamma^{0^{\prime}}=\bar{\Gamma}^{0^{\prime}}$ that is used to
construct the contravariant spinor $\overline{\Psi_{L,R}}\equiv\Psi
_{L,R}^{\dagger}\eta.$ The metric $\eta$ has the properties $\eta\left(
\Gamma^{M}\right)  \eta^{-1}=-\left(  \bar{\Gamma}^{M}\right)  ^{\dagger}$ and
$\eta\Gamma^{MN}\eta^{-1}=-\left(  \Gamma^{MN}\right)  ^{\dagger}$ or
equivalently $\left(  \eta\Gamma^{M}\right)  =\left(  \eta\bar{\Gamma}%
^{M}\right)  ^{\dagger}$ and $\left(  \eta\Gamma^{MN}\right)  =\left(
\eta\Gamma^{MN}\right)  ^{\dagger}.$ Therefore we have the hermiticity
properties $\left(  \overline{\psi_{1L}}\Gamma^{M}\psi_{2R}\right)  ^{\dagger
}=\overline{\psi_{2R}}\bar{\Gamma}^{M}\psi_{1L}$ and $\left(  \overline
{\psi_{1L}}\Gamma^{MN}\psi_{2L}\right)  ^{\dagger}=\overline{\psi_{2L}}%
\Gamma^{MN}\psi_{1L}$. We can also define the charge conjugation matrix $C$ by
$C=\tau_{1}\times\sigma_{2}=-\tilde{C}\bar{\Gamma}^{0^{\prime}},\;$%
with\ $\tilde{C}\equiv C\Gamma^{0^{\prime}}=-1\times i\sigma_{2}$. The
property of $C$ is such that $C\Gamma^{M}C^{-1}=\left(  \Gamma^{M}\right)
^{T},\;\;C\bar{\Gamma}^{M}C^{-1}=\left(  \bar{\Gamma}^{M}\right)  ^{T}$ and
$C\Gamma^{MN}C^{-1}=-\left(  \bar{\Gamma}^{MN}\right)  ^{T},\;\;C\bar{\Gamma
}^{MN}C^{-1}=-\left(  \Gamma^{MN}\right)  ^{T}$. Then $C\Gamma^{M}\;$are
antisymmetric matrices and group theoretically corresponds to $\left(
4\times4\right)  _{antisymmetric}=6$ for SU$\left(  2,2\right)  $
representations. \label{gamms}} $\Gamma^{M},$ $\bar{\Gamma}^{M}$ acting on the
two chiral spinors of SO$\left(  d,2\right)  $ (assuming even $d$). The gamma
matrices satisfy $\Gamma^{M}\bar{\Gamma}^{N}+\Gamma^{N}\bar{\Gamma}^{M}%
=2\eta^{MN}$ which is equivalent to the quantization rules for $\psi^{M}$.

The physical states $|\Psi\rangle$ correspond to the gauge invariant subset of
states on which all of the OSp$\left(  1|2\right)  $ generators vanish. It is
sufficient to impose $X\cdot\psi|\Psi\rangle=P\cdot\psi|\Psi\rangle=0$ because
the remaining constraints follow from these. The chiral field $\hat{\Psi
}_{\dot{\alpha}}^{L}\left(  X\right)  $ is defined as the probability
amplitude of a physical state in position and spinor space $\hat{\Psi}%
_{\dot{\alpha}}^{L}\left(  X\right)  =\langle X,\dot{\alpha}|\Psi\rangle.$ The
second chiral spinor field $\hat{\Psi}_{\alpha}^{R}\left(  X\right)  =\langle
X,\alpha|\Psi\rangle$ is associated with the second spinor labeled with
$\alpha$ instead of $\dot{\alpha}$. The number of components of each chiral
spinor of SO$\left(  d,2\right)  $ is $2^{\frac{d}{2}}.$ In the case of
$d+2=6$ these 4-component spinors form the two fundamental representations of
SU$\left(  2,2\right)  =$SO$\left(  4,2\right)  .$

Both chiral spinors must satisfy the physical state conditions. Thus, defining
$\hat{\Psi}_{\dot{\alpha}}\equiv\hat{\Psi}_{\dot{\alpha}}^{L}$, it satisfies
\begin{equation}
\left(  \not X  \hat{\Psi}\right)  _{\alpha}=0,\;\left(  \not \partial
\hat{\Psi}\right)  _{\alpha}=0,\;\hat{\Psi}_{\alpha}\left(  X\right)
=\text{chiral spinor of SO}\left(  d,2\right)  . \label{osp12}%
\end{equation}
Note that the gamma matrices $\left(  \Gamma^{M}\right)  _{\alpha}%
^{~\dot{\beta}}$ have the labels $\alpha,\dot{\beta}$ of both chiral spinors
of SO$\left(  d,2\right)  .$ We use the notation $\not \partial \equiv
\Gamma^{M}\partial_{M}$ , $\not X  \equiv\Gamma^{M}X_{M},$ and similarly
$\overline{\not \partial }\equiv\bar{\Gamma}^{M}\partial_{M}$ , $\overline
{\not X  }\equiv\bar{\Gamma}^{M}X_{M}.$

The general solution of $\not X  \hat{\Psi}=0$ is
\begin{equation}
\hat{\Psi}\left(  X\right)  =\delta\left(  X^{2}\right)  \overline{\not X
}\Psi\left(  X\right)  , \label{psihat}%
\end{equation}
where we have used $\not X  \overline{\not X  }=X^{2}$ and $X^{2}\delta\left(
X^{2}\right)  =0.$ Here $\Psi_{\alpha}\left(  X\right)  $ (without the hat
$^{\symbol{94}}$ and with $\alpha$ instead of $\dot{\alpha}$) is labeled like
the other chiral spinor of SO$\left(  d,2\right)  .$

Next we examine the second equation $\not \partial \hat{\Psi}=0$ which can be
put into several forms as follows
\begin{align}
0  &  =\not \partial \hat{\Psi}=\not \partial \left[  \delta\left(
X^{2}\right)  \overline{\not X  }\Psi\right]  =\delta\left(  X^{2}\right)
\left[  \not \partial \left(  \overline{\not X  }\Psi\right)  -2\Psi\right]
\label{delpsi}\\
&  =\delta\left(  X^{2}\right)  \left[  -\not X  \overline{\not \partial }%
\Psi+2\left(  X\cdot\partial+\frac{d}{2}\right)  \Psi\right] \label{Lpsi1}\\
&  =\delta\left(  X^{2}\right)  \left[  \left(  \frac{1}{2i}\Gamma^{MN}%
L_{MN}+\frac{d}{2}\right)  \Psi+\left(  X\cdot\partial+\frac{d}{2}\right)
\Psi\right]  . \label{Lpsi}%
\end{align}
In the first line we have used $\partial_{M}\delta\left(  X^{2}\right)
=2X_{M}\delta^{\prime}\left(  X^{2}\right)  $ and $2\delta^{\prime}\left(
X^{2}\right)  \not X  \overline{\not X  }=2\delta^{\prime}\left(
X^{2}\right)  X^{2}=-2\delta\left(  X^{2}\right)  $ to obtain an overall delta
function. In the second line we changed the order $\overline{\not \partial
}\not X  =-\overline{\not X  }\not \partial +2X\cdot\partial+d+2$. In the
third line we have used the definition $\Gamma^{MN}=\frac{1}{2}\left(
\Gamma^{M}\bar{\Gamma}^{N}-\Gamma^{N}\bar{\Gamma}^{M}\right)  ,$ while
$L^{MN}=-i\left(  X^{M}\partial^{N}-X^{N}\partial^{M}\right)  $ is the
SO$\left(  d,2\right)  $ orbital angular momentum. The structure $\frac{1}%
{2i}\Gamma^{MN}L_{MN}=-\Gamma^{MN}X_{M}\partial_{N}=-\not X  \overline
{\not \partial }+X\cdot\partial$ can be regarded as the analog of spin-orbit
coupling, where the SO$\left(  d,2\right)  $ spin angular momentum is given by
$S^{MN}=\frac{1}{2i}\Gamma^{MN}.$ As we saw in the case of the scalar in
Appendix A, the appearance of $L^{MN}$ is naturally expected from the point of
view of Sp$\left(  2,R\right)  $ symmetry.

By applying $\overline{\not X  }$ on Eq.(\ref{Lpsi1}) we can derive a
homogeneity condition for $\Psi$ as follows
\begin{align}
0  &  =\delta\left(  X^{2}\right)  \left[  -\overline{\not X  }\not X
\overline{\not \partial }\Psi+2\overline{\not X  }\left(  X\cdot\partial
+\frac{d}{2}\right)  \Psi\right] \\
&  =2\delta\left(  X^{2}\right)  \left(  X\cdot\partial+\frac{d+2}{2}\right)
\left(  \overline{\not X  }\Psi\right)  ,
\end{align}
where we have used $\overline{\not X  }\not X  \delta\left(  X^{2}\right)
=X^{2}\delta\left(  X^{2}\right)  =0.$ According to the last line $\left(
\overline{\not X  }\Psi\right)  $ is homogeneous of degree $-\frac{d+2}{2}.$
Then from Eq.(\ref{delpsi}) we learn that $\Psi_{\dot{\alpha}}=\frac{1}%
{2}\not \partial \left(  \overline{\not X  }\Psi\right)  $ is homogeneous of
degree $-\frac{d}{2}$ since the right hand side has this homogeneity degree.
This requires the second terms in Eqs.(\ref{Lpsi1},\ref{Lpsi}) to vanish,
hence the two terms of Eqs.(\ref{Lpsi1},\ref{Lpsi}) vanish
independently\footnote{If the expression $2\left(  X\cdot\partial+\frac{d}%
{2}\right)  \Psi$ that appears in Eq.(\ref{Lpsi1}) had occured with a
different coefficient than $2$, the same arguments can still be given to prove
that the two terms in Eqs.(\ref{Lpsi1},\ref{Lpsi}) separately vanish. This
remark allows us to drop the parameter $\alpha$ mentioned after Eq.() below,
which seems to play no role.\label{alpha}}. Therefore we learn that the
OSp$\left(  1|2\right)  $ gauge invariance conditions of Eq.(\ref{osp12})
require $\Psi_{\dot{\alpha}}$ to satisfy the following equations
\begin{equation}
\left[  \left(  X\cdot\partial+\frac{d}{2}\right)  \Psi_{\alpha}\right]
_{X^{2}=0}=0,\;\;\left[  \not X  \overline{\not \partial }\Psi\right]
_{X^{2}=0}=0\text{ ~or~ }\left[  \left(  \frac{1}{2i}\Gamma^{MN}L_{MN}%
+\frac{d}{2}\right)  \Psi\right]  _{X^{2}=0}=0. \label{2Tspin1/2eom}%
\end{equation}
The derivatives must be taken before the condition $X^{2}=0$ is imposed.

If the expression $2\left(  X\cdot\partial+\frac{d}{2}\right)  \Psi$ that
appears in Eq.(\ref{Lpsi1}) had occured with a different coefficient than $2$,
the same arguments can still be given to prove that the two terms in
Eqs.(\ref{Lpsi1},\ref{Lpsi}) separately vanish. This remark allows us to drop
the parameter $\alpha$ mentioned after Eq.(\ref{fermiS0}) below, which seems
to play no role.

This analysis is performed independently for the two spinors $\hat{\Psi}%
_{\dot{\alpha}}^{L}=\delta\left(  X^{2}\right)  \left(  \overline{\not X
}\Psi^{R}\right)  _{\dot{\alpha}}$ and $\hat{\Psi}_{\alpha}^{R}=\delta\left(
X^{2}\right)  \left(  \overline{\not X  }\Psi^{L}\right)  _{\alpha}.$ So, the
free field equations for the two chiral fermions of SO$\left(  d,2\right)  $
are of the form of Eq.(\ref{2Tspin1/2eom}), except for interchanging
$\Gamma^{M}\leftrightarrow\bar{\Gamma}^{M}$ to describe the $L,R$ sectors.

The first equation in (\ref{2Tspin1/2eom}) is the kinematical equation and the
second is the dynamical equation. Both will be derived from an action in the
next subsection, and consistent interactions will be introduced after that.

These 2T-physics equations for chiral fermions in $d+2$ dimensions
\cite{2tfield} coincide with the equations proposed by Dirac \cite{Dirac} for
$d+2=6$. So it has been known for a long time that they reproduce the massless
Dirac equation in $d=4.$ More precisely, from $\Psi^{L}\left(  X\right)  $ we
reproduce the massless Weyl equation for a left handed SL$\left(  2,C\right)
$ spinor and from $\Psi^{R}\left(  X\right)  $ we reproduce the massless Weyl
equation for a right handed SL$\left(  2,C\right)  $ spinor. We will return to
this detail explicitly when we derive the Standard Model in $3+1$ dimensions
from the one in $4+2$ dimensions.

For the application to the Standard Model it is essential to classify the
left/right handed chiral fermions differently under SU$\left(  3\right)
\times$SU$\left(  2\right)  \times$U$\left(  1\right)  .$ This is why we were
careful in our analysis above to distinguish between left/right spinors
directly in $d+2$ dimensions.

One may ask how do we do away with the extra spinor components in going from
the 4-component SU$\left(  2,2\right)  $ spinors $\Psi_{\alpha}^{L},\Psi
_{\dot{\alpha}}^{R}$ in six dimensions, to the 2-component SL$\left(
2,C\right)  $ chiral spinors in four dimensions. The explanation is rooted in
the 2Tgauge-symmetry extended to fermions. The following fermionic gauge
symmetry was first noted in \cite{2tfield}
\begin{equation}
\delta_{\zeta}\Psi^{L}=X^{2}\zeta_{1}^{L}+\not X  \zeta_{2}^{R},\;\delta
_{\zeta}\Psi^{R}=X^{2}\zeta_{1}^{R}+\overline{\not X  }\zeta_{2}^{L}%
,\;\zeta_{1,2}^{L,R}\text{=SO}\left(  d,2\right)  \text{ fermionic spinors.}
\label{fermiGaugetransf}%
\end{equation}
The easiest way to notice this symmetry is through Eq.(\ref{psihat}), where it
is evident that the transformations above leave the physical states $\hat
{\Psi}^{L}=\delta\left(  X^{2}\right)  \overline{\not X  }\Psi^{R}$ or
$\hat{\Psi}^{R}=\delta\left(  X^{2}\right)  \overline{\not X  }\Psi^{L}$
invariant$. $ One may follow the symmetry down to the equations
(\ref{2Tspin1/2eom}) written in terms of $\Psi$ (rather than $\hat{\Psi}$)
where it is sufficient to have homogeneous $\zeta^{L,R}.$ This symmetry will
be an automatic fundamental symmetry in the fermion action proposed below.

Note that in the discussion above the gauge parameters $\zeta_{1,2}^{L,R}$ are
independent. When we introduce all the fermions in the Standard Model, each
chiral fermion will naturally have its own independent local fermionic
symmetry parameters $\zeta_{1,2}^{L,R}$ with their SU$\left(  3\right)
\times$SU$\left(  2\right)  \times$U$\left(  1\right)  $ charges identical to
those of the corresponding fermion. So the fermionic 2Tgauge-symmetry will be
just large enough to remove all ferminoic gauge degrees of freedom. This will
be a symmetry that is elegantly built in the structure of the action for fermions.

The fermionic 2Tgauge-symmetry can eliminate only half of the components in
each $\Psi^{L,R}$ because, despite appearances, $\not X  \zeta_{2}%
^{R},\overline{\not X  }\zeta_{2}^{L}$ contain only half as many independent
degrees of freedom as $\Psi^{L,R}.$ This fact, that will become more evident
below by constructing $\not X  $,$\overline{\not X  }$ as explicit matrices,
is due to the condition $X^{2}=0$. So, half of the components in each
$\Psi^{L,R}\left(  X\right)  $ can be gauge fixed, while their dependence on
the $d+2$ dimensions $X^{M}$ can be reduced to $\left(  d-1\right)  +1$
dimensions $x^{\mu}$ by solving the kinematic conditions in
(\ref{2Tspin1/2eom}). This leaves the correct physical degrees of freedom for
chiral fermions in $d$ dimensions.

Thus, in the application to the Standard Model in the following sections,
where we use a specific embedding of $d$ dimensions as given in
Eq.(\ref{massless}), four component SO$\left(  4,2\right)  $ spinors
$\Psi^{L,R}\left(  X\right)  $ in six dimensions will be equivalent to two
component chiral fermions $\psi^{L,R}\left(  x\right)  $ in four dimensions
(SL$\left(  2,C\right)  $ doublets) after eliminating the gauge components of
each quark and lepton via fermionic-gauge fixing.

We emphasize that in 2T-physics this reduction to $d$ dimensions is understood
as one of the possible gauge choices that provides a holographic image of the
$d+2$ dimensional theory in the sense of Fig.1.

\subsubsection{Free field action for fermions}

We now propose the following action whose minimization gives the fermion
equations (\ref{2Tspin1/2eom})%
\begin{equation}
S_{0}\left(  \Psi\right)  =\frac{i}{2}\int\left(  d^{d+2}X\right)
\delta\left(  X^{2}\right)  \left(  \bar{\Psi}\not X  \overline{\not \partial
}\Psi+\bar{\Psi}\overleftarrow{\not \partial }\overline{\not X  }\Psi\right)
. \label{fermiS0}%
\end{equation}
This action is manifestly hermitian. It is possible to add to $S_{0}\left(
\Psi\right)  $ the term $\frac{\alpha}{2}\int\left(  d^{d+2}X\right)
\delta\left(  X^{2}\right)  \bar{\Psi}\left(  -\overleftarrow{\partial}\cdot
X+X\cdot\partial\right)  \Psi$ with an arbitrary real coefficient $\alpha.$
However, as remarked after Eq.(\ref{2Tspin1/2eom}) above, and after
Eq.(\ref{homogZeta}) below regarding gauge symmetries, this term does not
change any part of the discussion, hence we will suppress $\alpha$ in this
paper for simplicity, but it should be kept in mind in future investigations.
This action can be rewritten only in terms of $L^{MN}$ as in Eq.(\ref{Lpsi}),
so we can easily argue that this action is invariant under Sp$\left(
2,R\right)  $. Here the contravariant $\bar{\Psi}^{\dot{\alpha}}$ is defined
as $\bar{\Psi}=\Psi^{\dagger}\eta$ by using the SU$\left(  2,2\right)  $
metric $\eta$ given in footnote (\ref{gamms}), and we have used the notation
$\bar{\Psi}\overleftarrow{\not \partial }\equiv\partial_{M}\bar{\Psi}%
\Gamma^{M}$. Upon general variation, this action gives%
\begin{align}
\delta S_{0}\left(  \Psi\right)   &  =i\int\left(  d^{d+2}X\right)  \delta
\bar{\Psi}\left\{  \frac{1}{2}\delta\left(  X^{2}\right)  \not X
\overline{\not \partial }\Psi-\frac{1}{2}\partial_{M}\left[  \delta\left(
X^{2}\right)  \Gamma^{M}\overline{\not X  }\Psi\right]  \right\}  +h.c.\\
&  =i\int\left(  d^{d+2}X\right)  \delta\bar{\Psi}\left\{  \delta\left(
X^{2}\right)  \left[  \frac{1}{2}\left(  \not X  \overline{\not \partial
}-\not \partial \overline{\not X  }\right)  \Psi+\Psi\right]  \right\}
+h.c.\\
&  =i\int\left(  d^{d+2}X\right)  ~\delta\left(  X^{2}\right)  ~\delta
\bar{\Psi}\left[  \not X  \overline{\not \partial }\Psi-\left(  X\cdot
\partial+\frac{d}{2}\right)  \Psi\right]  +h.c.
\end{align}
In the first line an integration by parts has been performed to collect the
coefficient of $\delta\bar{\Psi},$ while $h.c.$ stands for the Hermitian
conjugate term that contains $\delta\Psi$. The last term in the second line
comes from taking the derivative of the delta function $-\frac{1}{2}\left(
\partial_{M}\delta\left(  X^{2}\right)  \right)  \Gamma^{M}\overline{\not X
}\Psi=-\frac{1}{2}\delta^{\prime}\left(  X^{2}\right)  \not X  \overline
{\not X  }\Psi=-X^{2}\delta^{\prime}\left(  X^{2}\right)  \Psi=\delta\left(
X^{2}\right)  \Psi,$ thus obtaining an overall factor of $\delta\left(
X^{2}\right)  .$ To derive the third line we interchange orders $\overline
{\not \partial }\not X  =-\overline{\not X  }\not \partial +2X\cdot
\partial+d+2.$

Next we point out the fermionic 2Tgauge-symmetry when we substitute
\begin{equation}
\delta_{\zeta}\bar{\Psi}=X^{2}\bar{\zeta}_{1}+\bar{\zeta}_{2}\overline
{\not X } \label{psisymmetry}%
\end{equation}
instead of the general variation $\delta\bar{\Psi}$
\begin{align}
\delta_{\zeta}S_{0}\left(  \Psi\right)   &  =-i\int\left(  d^{d+2}X\right)
\delta\left(  X^{2}\right)  \left(  X^{2}\bar{\zeta}_{1}+\bar{\zeta}%
_{2}\overline{\not X }\right)  \left[  \not X \overline{\not \partial }%
\Psi-\left(  X\cdot\partial+\frac{d}{2}\right)  \Psi\right]  +h.c.\\
&  =-i\int\left(  d^{d+2}X\right)  \partial_{M}\left[  X^{M}\bar{\zeta}%
_{2}\overline{\not X }\Psi\delta\left(  X^{2}\right)  \right]  +h.c.=0
\end{align}
In the first line the $\delta\left(  X^{2}\right)  X^{2}\bar{\zeta}_{1}$ terms
and the $\delta\left(  X^{2}\right)  \bar{\zeta}_{2}\overline{\not X }%
\not X \overline{\not \partial }\Psi=\delta\left(  X^{2}\right)  X^{2}%
\bar{\zeta}_{2}\overline{\not \partial }\Psi$ vanish trivially, while the
remaining term $\delta\left(  X^{2}\right)  \bar{\zeta}_{2}\overline{\not X
}\left(  X\cdot\partial+\frac{d}{2}\right)  \Psi$ can be written as the total
divergence in the second line provided $\bar{\zeta}_{2}$ satisfies the
homogeneity condition
\begin{equation}
\left(  X\cdot\partial\bar{\zeta}_{2}+\frac{d+2}{2}\bar{\zeta}_{2}\right)
_{X^{2}=0}=0. \label{homogZeta}%
\end{equation}
If the term proportional to $\alpha$ noted following Eq.(\ref{fermiS0}) had
been included, the same arguments still hold for any $\alpha$. Note that the
symmetry holds off-shell without requiring $\Psi$ or $\bar{\zeta}_{1}$ to be
restricted in any way. By applying the operator $\left(  X\cdot\partial
+\frac{d}{2}\right)  $ on both sides of Eq.(\ref{psisymmetry}) and using the
homogeneity of $\bar{\zeta}_{2}$ we note that $\left(  X\cdot\partial+\frac
{d}{2}\right)  \delta_{\zeta}\bar{\Psi}=X^{2}\left(  X\cdot\partial+\frac
{d+4}{2}\right)  \bar{\zeta}_{1}$ is proportional to $X^{2}.$ This implies
that, while the off-shell $\Psi$ in general has only pure gauge degrees of
freedom in its parts proportional to $X^{2}$ (symmetry $X^{2}\zeta_{1}),$ only
homogeneous parts of the off-shell $\Psi$ with homogeneity degree of
$-\frac{d}{2}$ contain gauge degrees of freedom of the type $\overline
{\not X }\zeta_{2}$.

We now return to the general $\delta\bar{\Psi}$ and require the variational
principle $\delta S_{0}\left(  \Psi\right)  =0.$ The resulting equation of
motion $\delta\left(  X^{2}\right)  \left[  \not X  \overline{\not \partial
}\Psi-\left(  X\cdot\partial+\frac{d}{2}\right)  \Psi\right]  =0$ is identical
to Eq.(\ref{Lpsi1}), which is also equivalent to Eqs.(\ref{delpsi},\ref{Lpsi})
by the same arguments supplied when studying those equations. As we have
argued before, the two terms must vanish separately, and therefore the action
principle gives the correct equations of motion derived in
Eq.(\ref{2Tspin1/2eom}).

Since the on-shell free field is already homogeneous we can eliminate half of
its degrees of freedom by fixing the fermionic 2Tgauge-symmetry $\zeta_{2}$.
Hence our action reproduces the correct degrees of freedom, and the field
equations required by the 2T-physics constraints that follow from OSp$\left(
1|2\right)  $ gauge symmetry, namely $\not \partial \hat{\Psi}=0,\;\not X
\hat{\Psi}=0$ with $\hat{\Psi}\left(  X\right)  =\delta\left(  X^{2}\right)
\overline{\not X  }\Psi\left(  X\right)  .$

\subsubsection{Interactions for fermions}

The interaction terms for fermions must respect the fermionic 2Tgauge-symmetry
as well as the SO$\left(  d,2\right)  $ invariance. Interaction with gauge
fields constructed by replacing ordinary derivatives $\partial_{M}$ with
covariant derivatives $D_{M}=\partial_{M}+A_{M}$ are automatically consistent
with these symmetries. The $A_{M}$ are associated with the adjoint
representation of an internal gauge symmetry, and the fermions are in some
representation of the gauge group, as usual in non-Abelian gauge theories.

In the application to the Standard Model the $A_{M}$ are the SU$\left(
3\right)  \times$SU$\left(  2\right)  \times$U$\left(  1\right)  $ gauge
fields, and the SU$\left(  2,2\right)  =$SO$\left(  4,2\right)  $ fermions
$\Psi^{L,R}$ are classified under this internal group exactly like left/right
handed quarks and leptons are classified in the usual Standard Model.

The analysis in Eqs.(\ref{fermiS0}-\ref{homogZeta}) goes through unchanged in
every step except for replacing the covariant derivative $D_{M}$ in all steps
instead of $\partial_{M}$. Note that this modifies the homogeneity (or
kinematic) conditions on the fermionic gauge parameters in Eq.(\ref{homogZeta}%
) to include the gauge field $\left(  X\cdot D+\frac{d+2}{2}\right)
\bar{\zeta}_{2}=0,$ since the $\bar{\zeta}_{1,2}$ are classified just like
$\bar{\Psi}$ under the gauge group. However, as seen below, it is possible to
choose a gauge $X\cdot A=0$ in which $X\cdot D=X\cdot\partial$ so that in this
gauge the kinematic equations revert to the ordinary homogeneity condition.

We now consider interactions with scalars. Due to the fermionic, as well as
the SO$\left(  d,2\right)  $ symmetry, the interactions must have the
fermionic structures $\bar{\Psi}^{L}\not X  \Psi^{R}$, $\bar{\Psi}%
^{R}\overline{\not X  }\Psi^{L}$ coupled to the scalars and multiplied with
the delta function%
\begin{equation}
S_{int}\left(  \psi,scalars\right)  =\int\left(  d^{d+2}X\right)
~\delta\left(  X^{2}\right)  ~\left[  \bar{\Psi}^{L}\not X  \Psi^{R}%
\times(scalars)+h.c.\right]  . \label{fermiscalarint}%
\end{equation}
The group theoretical explanation is as follows. First, these structures are
SO$\left(  d,2\right)  $ invariant because the product of a left spinor and an
anti-right spinor makes an invariant when dotted with the vector $X^{M}$. For
example, for SO$\left(  4,2\right)  =$SU$\left(  2,2\right)  $ which is our
main interest, the left spinor $\Psi^{L}$ is a $4$ of SU$\left(  2,2\right)
,$ and the right spinor $\Psi^{R}$ is a $4^{\ast},$ while the anti-left spinor
$\bar{\Psi}^{L}$ is again a $4^{\ast}.$ The SU$\left(  2,2\right)  $ product
$4^{\ast}\times4^{\ast}$ antisymmetrized is exactly the SO$\left(  4,2\right)
$ vector in six dimensions. So the structure$\bar{\Psi}^{L}\Gamma_{M}\Psi^{R}$
must be dotted with the vector $X^{M}$ to make the SO$\left(  d,2\right)  $
invariant $\bar{\Psi}^{L}\not X  \Psi^{R}$. Second, under the fermionic gauge
transformations of Eq.(\ref{fermiGaugetransf}) the variation of $\bar{\Psi
}^{L}\not X  \Psi^{R}$ produces a factor of $X^{2},$ as in
\begin{align}
\delta_{\zeta}\left(  \bar{\Psi}^{L}\not X  \Psi^{R}\right)   &  =\left(
\bar{\zeta}_{1}^{L}X^{2}+\bar{\zeta}_{2}^{R}\overline{\not X  }\right)
\not X  \Psi^{R}+\bar{\Psi}^{L}\not X  \left(  X^{2}\zeta_{1}^{R}%
+\overline{\not X  }\zeta_{2}^{L}\right) \\
&  =X^{2}\left[  \left(  \bar{\zeta}_{1}^{L}\not X  \Psi^{R}+\bar{\Psi}%
^{L}\not X  \zeta_{1}^{R}\right)  +\left(  \bar{\zeta}_{2}^{R}\Psi^{R}%
+\bar{\Psi}^{L}\zeta_{2}^{L}\right)  \right]  .
\end{align}
When multiplied with the delta function this vanishes $X^{2}\delta\left(
X^{2}\right)  =0.$ Therefore the interaction $S_{int}\left(  \psi
,scalars\right)  $ is invariant under the fermionic 2Tgauge-symmetry.

Now we turn to the internal gauge group to see how to couple the scalars in
Eq.(\ref{fermiscalarint}). Let us assume that $\bar{\Psi}^{L}\not X  \Psi^{R}
$ is not invariant under the internal gauge group. For example, if we classify
$\Psi^{L,R}$ under SU$\left(  3\right)  \times$SU$\left(  2\right)  \times
$U$\left(  1\right)  $ as in the Standard Model, then $\bar{\Psi}^{L}\not X
\Psi^{R}$ is invariant under SU$\left(  3\right)  ,$ but transforms as a
doublet under SU$\left(  2\right)  $ and has a non-trivial phase under
U$\left(  1\right)  .$ Therefore to make an invariant we must introduce a
complex scalar $H$ that has just the opposite classification under the gauge
group to be able to construct a gauge invariant in the form $\bar{\Psi}%
^{L}\not X  \Psi^{R}H,$ exactly like the Higgs in the Standard Model. We will
see that, in the reduction from $d+2$ to $d$ dimensions by solving the
kinematic equations, this term will become the familiar Yukawa coupling
$\bar{\psi}^{L}\psi^{R}h$ in $d$ dimensions, and will lead to the correct
Yukawa couplings of the Higgs field in the Standard Model in $d=4$ dimensions.

There is one more point to consider to obtain non-trivial interactions with
scalars, if $d\neq4.$ This is not immediately evident from the action, but
becomes clear when we examine the equations of motion. The dynamical fermion
equation of motion including interaction with scalars has the form $\not X
(\overline{\not \partial }\Psi^{L}+H\Psi^{R}+\cdots)=0$. In addition there are
the kinematic equations of motion which demand definite homogeneity for each
field. We learned above that, the on-shell, $\Psi^{L,R}$ have homogeneity
degree $-d/2$ and the scalar $H$ has homogeneity degree $-\left(  d-2\right)
/2$ (assuming the $W$ term analogous to Eq.(\ref{s2W}) is zero for the $H$
field). We see that, unless $d=4,$ the homogeneity of $\overline
{\not \partial }\Psi$ which is $-\left(  d+2\right)  /2$ does not match the
homogeneity of the term $H\Psi$ which is $-d+1.$ Then each term must vanish
separately, and the interaction becomes trivial. To avoid this we need to
multiply $H\Psi$ with a factor that establishes the same homogeneity for both
terms. Thus, in addition to $H$ we consider another scalar $\Phi$ that is
neutral under the SU$\left(  3\right)  \times$SU$\left(  2\right)  \times
$U$\left(  1\right)  $ gauge group. Assuming that $\Phi$ has homogeneity
degree $-\left(  d-2\right)  /2$, we see that $H\Psi^{R}\Phi^{-\frac{d-4}%
{d-2}}$ has degree $-\left(  d+2\right)  /2,$ which is the same degree as that
of $\overline{\not \partial }\Psi^{L}.$ Hence the action that produces
nontrivial interactions between fermions and scalars has the form
\begin{equation}
S_{int}\left(  \psi,H,\Phi\right)  =g_{H}\int\left(  d^{d+2}X\right)
~\delta\left(  X^{2}\right)  ~\left(  \bar{\Psi}^{L}\not X  \Psi^{R}%
H+\bar{\Psi}^{R}\overline{\not X  }\Psi^{L}\bar{H}\right)  \Phi^{-\frac
{d-4}{d-2}}, \label{psiHphi}%
\end{equation}
with a dimensionless coupling $g_{H}$ in any dimension. The exponent in
$\Phi^{-\frac{d-4}{d-2}}$ would be shifted to another value if $W\left(
\Phi\right)  \neq0$ in Eq.(\ref{s2W}), since in that case the homogeneity of
$\Phi$ would be different, but still some power of $\Phi$ will be needed as
long as $d\neq4.$ With this form of interaction we also find that the
interactions in the dynamical equations of motion for $H$ and $\Phi$ are also
consistent with their homogeneity. We see here that the role of $\Phi$ is
similar to the role of a dilaton. In $d=4$ the $\Phi$ disappears and the
interaction reduces to a renormalizable Yukawa interaction\footnote{There are
also other invariant couplings that occur in the symmetric product of $\left(
4^{\ast}\times4^{\ast}\right)  _{s}=10$ of SU$\left(  2,2\right)  .$ This
product takes the form $\bar{\Psi}^{L}\Gamma_{MNK}\Psi^{R}$ and it can be
coupled invariantly to the gauge field and Higgs field in the form $\bar{\Psi
}^{L}\Gamma_{MNK}F^{MN}X^{K}\Psi^{R}H$ consistently with the SU$\left(
3\right)  \times$SU$\left(  2\right)  \times$U$\left(  1\right)  $ gauge
symmetry of the Standard Model. Furthermore this term is invariant also under
the fermionic gauge symmetry after taking into account the property of the
gauge field $X_{M}F^{MN}=0$ given in the next section. So this interaction is
permitted by all the symmetries of the theory. However, this term cannot be
included in the action of the Standard Model because it leads to a
non-renormalizable interaction. Instead, it is expected to arise with a
calculable coupling from the quantum effects in the theory, and contribute to
the anomalous magnetic moment \cite{barsYoshimura}.}.

\subsection{Gauge Field}

The gauge field equations of motion can be obtained from two different
approaches in 2T-physics. We will not provide the derivation here because it
would require too much space, and instead refer to past work. The first
approach uses OSp$\left(  2|2\right)  $ as the gauge group on the worldline,
that acts linearly on the phase space superquartet $\left(  \psi_{1}^{M}%
,\psi_{2}^{M},X^{M},P^{M}\right)  $ which contains two worldline fermions
$\psi_{i}^{M}\left(  \tau\right)  $ \cite{spin2t}. In $d+2=6$ dimensional flat
spacetime, the physical states in the covariant quantization of this theory
are precisely the gauge bosons that we need in $d+2=6$ dimensions
\cite{2tfield}. The second approach is to consider a spinless particle moving
in background fields in any dimension $d+2$, and subject to the Sp$\left(
2,R\right)  $ gauge symmetry. In this case, phase space contains only $\left(
X^{M},P^{M}\right)  $ but the action of Sp$\left(  2,R\right)  $ is non-linear
in a way that depends on the background fields $\phi^{M_{1}M_{2}\cdots M_{s}%
}\left(  X\right)  $ of all integer spins $s=0,1,2,3,\cdots$. The background
fields include gauge fields $A_{M}\left(  X\right)  .$ The requirement of the
consistent closure of the Sp$\left(  2,R\right)  $ Lie algebra generates
kinematical equations for all background fields, including the gauge fields
\cite{2tbacgrounds}. The two approaches give the same kinematic equations for
$A_{M},$ namely%
\begin{equation}
X^{M}F_{MN}=0,\;\text{where }F_{MN}=\partial_{M}A_{N}-\partial_{N}A_{M}%
-ig_{A}\left[  A_{M},A_{N}\right]  , \label{Akinematic}%
\end{equation}
while also demanding $X^{2}=0$. The dynamical equation follows from the
OSp$\left(  2|2\right)  $ approach for $4+2$ dimensions as given in
\cite{2tfield}, and can be extended\footnote{For general $d+2,$ the quantum
spectrum of the OSp$\left(  2|2\right)  $ theory gives the gauge fields
$A_{M_{1}\cdots M_{p+1}}$ for a $p$-brane with $p=\left(  d-4\right)  /2$
\cite{2tfield}. Thus, for a gauge field $A_{M}$ we can use the OSp$\left(
2|2\right)  $ setup only for $d=4.$ The background field method does not have
this limitation and applies to any $d.$ Thus, after obtaining the appropriate
equations for gauge fields, we have managed to extend the gauge field
equations to any dimension $d$ by introducing the dilaton factor as in
Eq.(\ref{Adynamic}).} to any $d+2$ dimensions by including a dilatonic factor
$\Phi^{\frac{2\left(  d-4\right)  }{d-2}}$%
\begin{equation}
\left(  D_{M}\left(  \Phi^{\frac{2\left(  d-4\right)  }{d-2}}F^{MN}\right)
\right)  _{X^{2}=0}=sources. \label{Adynamic}%
\end{equation}
These equations have been constructed to be covariant under Yang-Mills
transformations for some non-Abelian gauge group $G$, so $D_{M}$ is the usual
covariant derivative. The dilaton factor $\Phi^{\frac{2\left(  d-4\right)
}{d-2}}$ is a singlet under the gauge group $G$. It disappears in four
dimensions, but is non-trivial in dimensions other than $d=4.$ Its role is to
provide consistency with the kinematical conditions for all the fields in the
theory. Recall that the kinematical equations follows from gauge invariance
under the Sp$\left(  2,R\right)  $ generator $\left(  X\cdot P+P\cdot
X\right)  $. The exponent $2\frac{d-4}{d-2}$ is determined by homogeneity
which comes from this condition, with a reasoning similar to the one that led
to the dilaton factor in Eq.(\ref{psiHphi}). This exponent would be shifted to
another value if $W\left(  \Phi\right)  \neq0$ in Eq.(\ref{s2W}), since in
that case the homogeneity degree of the dilaton $\Phi$ would be different, but
the exponent still vanishes if $d=4.$

In the rest of this section we discuss the new 2Tgauge-symmetry beyond the
usual Yang-Mills gauge symmetry $G.$ To do so we will first use $G$ to choose
some gauges for the Yang-Mills field, and then discuss the 2Tgauge-symmetry in
the fixed Yang-Mills gauge. We could discuss the 2Tgauge-symmetry of the
equations of motion without choosing any Yang-Mills gauge, as will be done for
the action in the next subsection. However, here we want to take the
opportunity to point out some useful gauge choices.

Using the Yang-Mills type gauge invariance we may impose various gauge
conditions. In the gauge $X^{M}A_{M}=0$ the kinematic constraint
(\ref{Akinematic}) reduces to a homogeneity constraint as follows%
\begin{equation}
X\cdot A=0:\;X^{M}F_{MN}=\left(  X\cdot\partial+1\right)  A_{N}=0.
\label{Aaxial}%
\end{equation}
Note that the homogeneity degree of $A_{M}$ is $-1,$ which is the same
homogeneity as the derivative $\partial_{M},$ consistent with the covariant
derivative $D_{M}=\partial_{M}+A_{M},$ in all dimensions. In this gauge the
equations above, taken for $d=4,$ agree with Dirac's equations \cite{Dirac} as
generalized to non-Abelian gauge fields \cite{salam}, and so it has been known
for a long time that after solving the kinematic equation, the dynamical
equation reproduces the correct equations of motion for non-Abelian gauge
fields in $3+1$ dimensions. We will return to this detail when we discuss the
derivation of the Standard Model in $3+1$ dimensions from the one in $4+2$ dimensions.

Now we turn to the new 2Tgauge-symmetry with the transformation law%
\begin{equation}
\delta_{a}A_{M}=X^{2}a_{M}\Phi^{-\frac{2\left(  d-4\right)  }{d-2}%
},\;\;\text{with\ }\left[  \left(  X\cdot D+d-1\right)  a_{N}-X_{N}D\cdot
a\right]  _{X^{2}=0}=0,\;\text{and }X\cdot a=0. \label{Aa}%
\end{equation}
The local parameters $a_{M}$ are Lie algebra valued and must be in the adjoint
representation of the Yang-Mills gauge group $G$ to be consistent with the
Yang-Mills classification of $A_{M}.$ We examine the dynamical equation to
verify the 2Tgauge-symmetry for on shell fields $A_{M}$ as follows%
\begin{align}
&  \delta_{a}\left\{  D_{M}\left(  \Phi^{\frac{2\left(  d-4\right)  }{d-2}%
}F^{MN}\right)  \right\}  _{X^{2}=0}\\
&  =\left\{  -i\Phi^{\frac{2\left(  d-4\right)  }{d-2}}\left[  \left(
\delta_{a}A_{M}\right)  ,F^{MN}\right]  +D_{M}\left(  \Phi^{\frac{2\left(
d-4\right)  }{d-2}}D^{[M}\delta_{a}A^{N]}\right)  \right\}  _{X^{2}=0}\\
&  =\left\{  -iX^{2}\left[  a_{M},F^{MN}\right]  +D_{M}\left(  \Phi
^{\frac{2\left(  d-4\right)  }{d-2}}D^{[M}\left(  X^{2}a^{N]}\Phi
^{-\frac{2\left(  d-4\right)  }{d-2}}\right)  \right)  \right\}  _{X^{2}=0}\\
&  =2\left\{  D_{M}\left[  \left(  X^{M}a^{N}-X^{N}a^{M}\right)  \right]
+X_{M}\left(  D^{M}a^{N}-D^{N}a^{M}\right)  +\left(  d-4\right)
a^{N}\right\}  _{X^{2}=0}\\
&  =4\left\{  \left(  d-1\right)  a^{N}+X\cdot Da^{N}-X^{N}D\cdot a\right\}
_{X^{2}=0}=0
\end{align}
In this computation all terms proportional to $X^{2}$ have been dropped after
the derivatives are evaluated. In going from the third line to the fourth we
have used $X\cdot a=0,$ and $\left[  X\cdot\partial\Phi+\frac{d-2}{2}%
\Phi\right]  _{X^{2}=0}=0$ as in Eq.(\ref{variational}), to obtain the term
proportional to $d-4.$ This shows that the dilaton factor is required in the
dynamical equation of the gauge field Eq.(\ref{Adynamic}) in order to have the
2Tgauge-symmetry in all dimensions $d$. Finally the last line vanishes due to
the property of the gauge parameter $a_{M}$ given in Eq.(\ref{Aa}).

The new gauge transformation says that the part of $A_{M}$ proportional to
$X^{2}$ contains gauge degrees of freedom. If we first identify different
parts of the gauge field as
\begin{equation}
A_{M}\left(  X\right)  =A_{M}^{0}\left(  X\right)  +X^{2}\tilde{A}_{M}\left(
X\right)  \label{A0A1}%
\end{equation}
such that $A_{M}^{0}\left(  X\right)  \equiv\left[  A_{M}\left(  X\right)
\right]  _{X^{2}=0},$ then the remainder $\tilde{A}_{M}\left(  X\right)  $ can
be completely removed from the equations of motion by using this gauge
symmetry, provided the on-shell $\tilde{A}_{M}$ satisfies the same conditions
as $a_{M}$ as given in Eq.(\ref{Aa}).

Note that the Yang-Mills field strength $F_{MN}$ is not invariant under the
new gauge symmetry. But the equation of motion is invariant as seen above, and
also the action will be shown to be gauge invariant. The non-invariance of
$F_{MN}$ is welcome because this is how the $d+2$ dimensional theory will
contain the same physical information that resides in the field strength
$F_{\mu\nu}$ in $\left(  d-1\right)  +1$ dimensions.

\subsubsection{Non-Abelian gauge field action}

We now propose the action principle that gives both the kinematical and
dynamical equations of motion in Eqs.(\ref{Akinematic},\ref{Adynamic}). In the
following we will assume that the physical gauge for the 2Tgauge-symmetry
discussed in the previous section is already chosen, and build the action
starting directly with the physical component $A_{M}^{0}\left(  X\right)  $.
By comparison to the scalar action, this is analogous to building the action
directly for $\Phi_{0},$ skipping the fully gauge invariant treatment for
$A_{M}\left(  X\right)  $ that we gave for both the scalar and the fermions.
Thus everywhere we write $A_{M}$ below should be understood as being the gauge
invariant part $A_{M}^{0}\left(  X\right)  =\left[  A_{M}\left(  X\right)
\right]  _{X^{2}=0}.$

The consistent 2T-physics Yang-Mills type action in any dimension
is\footnote{It is also possible to construct a fully Sp$\left(  2,R\right)  $
invariant action for gauge fields. This is given in the second part of
Appendix A. The point is that every derivative can appear in the form $L^{MN}$
to display fully the invariance under the underlying Sp$\left(  2,R\right)  .$
The physical sector of either treatment is identical.}%
\begin{equation}
S\left(  A\right)  =-\frac{1}{4}\int\left(  d^{d+2}X\right)  ~\delta\left(
X^{2}\right)  ~\Phi^{\frac{2\left(  d-4\right)  }{d-2}}~Tr\left(  F_{MN}%
F^{MN}\right)  . \label{Aaction}%
\end{equation}
The dilaton factor $\Phi^{\frac{2\left(  d-4\right)  }{d-2}}$ is necessary for
the consistency of homogeneous terms in the equations of motion. Evidently,
this factor disappears for $d=4,$ which is the case of interest for the
application to the Standard Model. The gauge coupling $g_{A}$ as defined in
Eq.(\ref{Akinematic}) is dimensionless in any dimension.

The general variation with respect to the gauge field gives%
\begin{align}
\delta S\left(  A\right)   &  =-\int\left(  d^{d+2}X\right)  ~\delta\left(
X^{2}\right)  ~\Phi^{\frac{2\left(  d-4\right)  }{d-2}}~Tr\left(  F^{MN}%
D_{M}\left(  \delta A_{N}\right)  \right) \label{AdeltaS1}\\
&  =\int\left(  d^{d+2}X\right)  Tr\left\{  \delta A_{N}D_{M}\left[
\Phi^{\frac{2\left(  d-4\right)  }{d-2}}\delta\left(  X^{2}\right)
F^{MN}\right]  \right\} \\
&  =\int\left(  d^{d+2}X\right)  Tr\left\{  \delta A_{N}\left[
\begin{array}
[c]{c}%
\delta\left(  X^{2}\right)  D_{M}\left(  \Phi^{\frac{2\left(  d-4\right)
}{d-2}}F^{MN}\right) \\
+2\Phi^{\frac{2\left(  d-4\right)  }{d-2}}\delta^{\prime}\left(  X^{2}\right)
X_{M}F^{MN}%
\end{array}
\right]  \right\}  . \label{AdeltaS}%
\end{align}
In $d\neq4,$ there is also a contribution to the equations of motion of the
\textquotedblleft dilaton\textquotedblright\ $\Phi,$ through the variation
$\delta\Phi$ which is not shown. The equations of motion that follow from this
action include both the kinematical and dynamical equations since the
coefficients of $\delta\left(  X^{2}\right)  ,\delta^{\prime}\left(
X^{2}\right)  $ must vanish separately. There are however subtleties in the
delta functions that need to be taken into account as in footnote
(\ref{careful}). For this reason the $A\left(  X\right)  $ that appears in
this action is already gauge fixed $A_{M}\left(  X\right)  =A_{M}^{0}\left(
X\right)  ,$ excluding the remainder $\tilde{A}_{M}\left(  X\right)  ,$ as
emphasized in the beginning of this section. After taking this point into
consideration, we see that this action yields precisely the correct equations
given in Eqs.(\ref{Akinematic},\ref{Adynamic}), so we have the correct action
principle for the physical sector $A_{M}\left(  X\right)  =A_{M}^{0}\left(
X\right)  $. Of course, $S\left(  A\right)  $ has already been built to be
gauge invariant under some Yang-Mills type gauge symmetry group $G$. For
example $G=$ SU$\left(  3\right)  \times$SU$\left(  2\right)  \times$U$\left(
1\right)  $ in the application to the Standard Model.

Next we show that we can add to $A_{M}^{0}\left(  X\right)  $ a remainder of
the form $A_{M}\left(  X\right)  =A_{M}^{0}\left(  X\right)  +X^{2}\tilde
{A}_{M}\left(  X\right)  $ without changing the physics, provided $\tilde
{A}_{M}\left(  X\right)  $ satisfies the same equations that $\tilde{a}_{M}$
satisfies as given in Eq.(\ref{Aa}). For this we first assume that the action
(\ref{Aaction}) is already written for the more general $A$ that includes the
special $\tilde{A}_{M}.$ We will then show that this action has the
2Tgauge-symmetry in which $\tilde{A}_{M}\left(  X\right)  $ can be changed by
arbitrary amounts $a_{M}.$ Therefore one can choose the gauge $\tilde{A}%
_{M}=0$ if so desired,

Thus consider the transformation $\delta_{a}A_{M}$ of Eq.(\ref{Aa}), and
insert it in Eq.(\ref{AdeltaS1}) to show there is a gauge symmetry $\delta
_{a}S\left(  A\right)  =0$ as follows%
\begin{align}
\delta_{a}S\left(  A\right)   &  =\int\left(  d^{d+2}X\right)  Tr\left\{
\Phi^{-\frac{2\left(  d-4\right)  }{d-2}}X^{2}a_{N}\left[
\begin{array}
[c]{c}%
\delta\left(  X^{2}\right)  D_{M}\left(  \Phi^{\frac{2\left(  d-4\right)
}{d-2}}F^{MN}\right) \\
+2\Phi^{\frac{2\left(  d-4\right)  }{d-2}}\delta^{\prime}\left(  X^{2}\right)
X_{M}F^{MN}%
\end{array}
\right]  \right\} \label{dsa1}\\
&  =-2\int\left(  d^{d+2}X\right)  \delta\left(  X^{2}\right)  Tr\left\{
a_{N}\left(  X\cdot\partial+1\right)  A^{N}-a_{N}D^{N}\left(  X\cdot A\right)
\right\} \label{dsa2}\\
&  =-2\int\left(  d^{d+2}X\right)  \delta\left(  X^{2}\right)  Tr\left\{
\left[
\begin{array}
[c]{c}%
\left(  X\cdot\partial+1\right)  \left(  A\cdot a\right)  -D^{N}\left(  X\cdot
Aa_{N}\right) \\
-A^{N}X\cdot\partial a_{N}+\left(  X\cdot A\right)  \left(  D\cdot a\right)
\end{array}
\right]  \right\} \label{dsa3}\\
&  =-2\int\left(  d^{d+2}X\right)  \delta\left(  X^{2}\right)  Tr\left\{
\begin{array}
[c]{c}%
\left(  X\cdot\partial+d\right)  \left(  A\cdot a\right)  -\partial_{N}\left(
X\cdot Aa^{N}\right) \\
+A^{N}\left[  -\left(  X\cdot D+d-1\right)  a_{N}+X_{N}D\cdot a\right]
\end{array}
\right\} \label{dsa4}\\
&  =-2\int\left(  d^{d+2}X\right)  Tr\left\{
\begin{array}
[c]{c}%
\partial_{N}\left[  \left(  X^{N}A\cdot a-X\cdot Aa^{N}\right)  \delta\left(
X^{2}\right)  \right] \\
+A^{N}D^{M}\left[  \left(  X_{N}a_{M}-X_{M}a_{N}\right)  \delta\left(
X^{2}\right)  \right]
\end{array}
\right\}  =0 \label{dsa}%
\end{align}
The steps in this calculation are explained as follows. In Eq.(\ref{dsa1}) we
use $X^{2}\delta\left(  X^{2}\right)  =0$ to drop the first term, and then use
$X^{2}\delta^{\prime}\left(  X^{2}\right)  =-\delta\left(  X^{2}\right)  $ and
write out $X_{M}F^{MN}$ in the form $X_{M}F^{MN}=\left(  X\cdot\partial
+1\right)  A^{N}-D^{N}\left(  X\cdot A\right)  $ to obtain Eq.(\ref{dsa2}).
The form in Eq.(\ref{dsa3}) is equivalent to Eq.(\ref{dsa2}) after evaluating
the derivatives. To get to Eq.(\ref{dsa4}) we added and subtracted the terms
proportional to $d$, and moved the non-Abelian term in the covariant
derivative from the first line to the second line. One can show that this
takes the form of Eq.(\ref{dsa}). Indeed after evaluating the derivatives in
Eq.(\ref{dsa}) we get back Eq.(\ref{dsa4}). Then we note that the expression
$D^{M}\left[  \left(  X_{N}a_{M}-X_{M}a_{N}\right)  \delta\left(
X^{2}\right)  \right]  $ vanishes by using the conditions on the gauge
parameters $a_{M}$ given in Eq.(\ref{Aa})$.$ Finally the total divergence can
be dropped.

We see that as long as $\tilde{A}_{M}$ satisfies the same equation as $a_{M}$
this gauge symmetry can gauge fix it to zero. Thus the physics is the same for
any remainder $X^{2}\tilde{A}_{M}$ of this type. We will see below that, in a
special Yang-Mills gauge, the class of allowed remainders $\tilde{A}%
_{M}\left(  X\right)  $ are those that are homogeneous of degree $-3.$

The Yang-Mills gauge symmetry combined with the new 2Tgauge-symmetry discussed
in this section are just sufficient to reduce the degrees of freedom in
$A_{M}\left(  X\right)  $ to be physical and ghost free. The physical degrees
of freedom will then agree with the degrees of freedom of the gauge field
$A_{\mu}\left(  x\right)  $ in $\left(  d-1\right)  +1$ dimensions. This
reduction will be discussed in particular for the Standard Model in $3+1$
dimensions in the following sections.

We emphasize that the Yang-Mills field strength $F_{MN}$ is not covariant
under the 2Tgauge-symmetry transformation $\delta_{a}A_{M}$ as seen in the
computations above. Therefore, using the new gauge symmetry it is possible to
gauge fix some of the components of $F_{MN}$ at will. We will use this freedom
in the reduction from $4+2$ to $3+1$ dimensions to show that only the $3+1$
dimensional components $F_{\mu\nu}\left(  x\right)  $ survive as the physical
field strengths.

\section{The Standard Model in $4+2$ Dimensions \label{SM}}

In the previous section we have constructed the action principle for field
theories in the framework of 2T-physics in any $d+2$ dimensions. The theory
contains scalars $H^{i}$,$\Phi,$ left/right handed chiral fermions
$\Psi^{L_{a}}$, $\Psi^{R_{\beta}}$ and gauge bosons $A_{M}^{r}$ classified
according to any gauge group, and can also be extended to include
gravity\footnote{The equations of motion for the gravitational field in
2T-physics is derived in \cite{2tfield}. The 2T action that generates these
equations is constructed using the methods of the present paper. This will be
given in a separate paper.} in any $d+2$ dimensions. Among the scalars we
distinguish one of them as the dilaton $\Phi.$ Although the dilaton factors in
Eqs.(\ref{psiHphi},\ref{Aaction}) disappear for $d=4,$ the dilaton may still
couple to the other scalars $H^{i}$ as in Eq.(\ref{VphiH}) even if $d=4$, so
we keep the dilaton as one of the fields in the theory.

The 2Tgauge-symmetry was derived in a tortuous way by starting from 2T-physics
equations of motion based on Sp$\left(  2,R\right)  $ local symmetry and its
extensions on the worldline. However, it is possible to reverse the reasoning
and suggest the 2Tgauge-symmetry directly in field theory as one of the
principles for building an action. This would then lead to the action
suggested above in a unique way. This is the point of view we take in this section.

Thus, in addition to the field theoretic guiding principles for constructing
the Standard Model in four dimensions, we add a new one, namely the
2Tgauge-symmetry given by%
\begin{align}
&  \left.
\begin{array}
[c]{l}%
\delta_{\Lambda}\Phi=X^{2}\Lambda,\;\delta_{\Lambda}H^{i}=X^{2}\Lambda
^{i},~~\\
\delta_{b}B_{\Phi},\delta_{b}B_{H^{i}}\text{ as in Eq.(\ref{delcA})}%
\end{array}
\right\}  \;i\text{ spans all other scalar fields,}\label{delphi}\\
&  \left.
\begin{array}
[c]{c}%
\delta_{\zeta}\Psi^{L_{a}}=X^{2}\zeta_{1}^{L_{\alpha}}+\not X  \zeta
_{2}^{R_{\alpha}},\;\\
\delta_{\zeta}\Psi^{R_{\beta}}=X^{2}\zeta_{1}^{R_{\beta}}+\overline{\not X
}\zeta_{2}^{L_{\beta}},\;
\end{array}
\right\}  \;\alpha,\beta\text{ span all fermions,}\label{delpsiLR}\\
&  \left.
\begin{array}
[c]{l}%
\delta_{a}A_{M}^{r}=X^{2}a_{M}^{r}\Phi^{-\frac{2\left(  d-4\right)  }{d-2}%
},\;\\
\delta_{b}B_{A_{M}^{r}}\;\text{similar to Eq.(\ref{delcA})}%
\end{array}
\right\}  \;\;r\text{ spans all gauge bosons.} \label{delA}%
\end{align}
There is a separate 2Tgauge-symmetry parameter for each field\footnote{The
last one, $\delta_{b}B_{A_{M}^{r}},\;$is similar to Eq.(\ref{delcA}). This is
the gauge symmetry required in order to allow arbitrary remainders $\tilde
{A}_{M}$ for the gauge field $A_{M}=A_{M}^{0}+X^{2}\tilde{A}_{M}.$ In the
presence of this gauge symmetry there would be no conditions on the parameters
$a_{M}\left(  X\right)  .$ We bypassed this more general setting that would
include the additional field $B_{A_{M}^{r}}$, and considered the gauge fixed
form of the action after the extra field $B_{A_{M}^{r}}$ is eliminated by a
gauge choice. In this gauge only to a specialized subset of $a_{M}$ and
corresponding $\tilde{A}_{M}$ play a role as discussed in the text. This was
sufficient for our purposes here.}. Thus, degrees of freedom can be removed
from every field in $d+2$ dimensions, such as to make it equivalent to a field
in $\left(  d-1\right)  +1$ dimensions.

Under the requirement of the 2Tgauge-symmetry the Lagrangian must include an
overall $\delta\left(  X^{2}\right)  $ or $\delta^{\prime}\left(
X^{2}\right)  $ factor, and furthermore it acquires the form of the actions
given in the previous section. Then the theory must be constructed in $d+2$
dimensions. The action will not have translation symmetry in $d+2$ dimensions
since $X^{M}$ appears explicitly through the delta function $\delta\left(
X^{2}\right)  $ and in the fermion terms, but the action will have SO$\left(
d,2\right)  $ symmetry. When the SO$\left(  d,2\right)  $ is interpreted as
the conformal symmetry in $\left(  d-1\right)  +1$ dimensions, which is the
case in one of the embeddings of $\left(  d-1\right)  +1$ in $d+2, $ then
Poincar\'{e} invariance, that includes translation invariance and Lorentz
invariance in $\left(  d-1\right)  +1$ dimensions emerges as part of conformal
symmetry. Thus, we take the point of view that the added 2Tgauge-symmetry
principle requires the two-time structure in field theory, and this is fully
consistent with everything we know in $\left(  d-1\right)  +1 $ dimensions.

In addition to the 2Tgauge-symmetry there is of course the principles of
Yang-Mills gauge symmetry and renormalizability requirements for $d=4$. The
Yang-Mills gauge symmetry is straightforward as discussed in the previous
section. By renormalizability, in the present paper, we mean that the emergent
$3+1$ dimensional field theory should be renormalizable. This amounts to
requiring that the emergent theory in four spacetime dimensions, at the
classical level, should not contain any terms of dimension larger than four.
In turn, this becomes a principle for restricting the types of terms that can
be included in the classical 2T-physics field theory in $4+2$ dimensions. For
example we cannot include high powers of fields in the classical 2T-physics
action. Eventually, when we develop the techniques of computation with the
quantum theory directly in $4+2$ dimensions, we need to replace the
requirement of renormalizability to mean the same directly in $4+2$ dimensions
at the quantum level.

Given the principles stated above we now construct the Standard Model in $4+2
$ dimensions. The internal Yang-Mills group structure is identical to the
usual Standard Model, but the spacetime structure is different, thus all the
fields are 6-dimensional fields instead of 4-dimensional fields. The 6
dimensional structure will impose certain restrictions on the emergent
Standard Model in $3+1$ dimensions as outlined in the Abstract. To be
completely explicit we write out the details below.

The Yang-Mills gauge group is $G=$SU$\left(  3\right)  \times$SU$\left(
2\right)  \times$U$\left(  1\right)  ,$ so we have the corresponding
Yang-Mills fields $A_{M}=\left(  G_{M},W_{M},B_{M}\right)  ,$ namely gluons
$G_{M}$ and electroweak gauge bosons $\left(  W_{M},B_{M}\right)  .$ These are
in the usual adjoint representations denoted by the dimensions for SU$\left(
3\right)  \times$SU$\left(  2\right)  $ and by the charge for U$\left(
1\right)  $ written as a subscript, as follows%
\begin{equation}
\text{vectors of SO}\left(  4,2\right)  :\;G_{M}=\left(  8,1\right)
_{0},\;W_{M}=\left(  1,3\right)  _{0},\;B_{M}=\left(  1,1\right)  _{0}%
\end{equation}
The SO$\left(  4,2\right)  $ scalar fields include the dilaton $\Phi$ which is
neutral under SU$\left(  3\right)  \times$SU$\left(  2\right)  \times
$U$\left(  1\right)  $ and the Higgs doublet $H$ classified as usual%
\begin{equation}
\text{scalars of SO}\left(  4,2\right)  :\Phi=\left(  1,1\right)
_{0},\;\;H^{i}=\left(
\genfrac{}{}{0pt}{}{H^{+}}{H^{0}}%
\right)  _{\frac{1}{2}}=\left(  1,2\right)  _{\frac{1}{2}}.
\end{equation}
Of course more scalars can be included, but for now we will assume a minimal
number as above. The fermionic matter fields are the three generations of
quarks and leptons $\Psi^{L_{a}}\left(  X\right)  ,$ $\Psi^{R_{\beta}}\left(
X\right)  $ taken as the left/right quartet spinors of SU$\left(  2,2\right)
=$SO$\left(  4,2\right)  ,$ and in in the usual representations of the
Yang-Mills gauge group $G$, as follows
\begin{align}
4\text{ of SU}\left(  2,2\right)   &  :\left(
\genfrac{}{}{0pt}{}{u^{L}}{d^{L}}%
\right)  _{\frac{1}{6}},\left(
\genfrac{}{}{0pt}{}{\nu_{e}^{L}}{e^{L}}%
\right)  _{-\frac{1}{2}};\left(
\genfrac{}{}{0pt}{}{c^{L}}{s^{L}}%
\right)  _{\frac{1}{6}},\left(
\genfrac{}{}{0pt}{}{\nu_{\mu}^{L}}{\mu^{L}}%
\right)  _{-\frac{1}{2}};\left(
\genfrac{}{}{0pt}{}{t^{L}}{b^{L}}%
\right)  _{\frac{1}{6}},\left(
\genfrac{}{}{0pt}{}{\nu_{\tau}^{L}}{\tau^{L}}%
\right)  _{-\frac{1}{2}}\\
4^{\ast}\text{ of SU}\left(  2,2\right)   &  :%
\genfrac{}{}{0pt}{}{\left(  u^{R}\right)  _{\frac{2}{3}}}{\left(
d^{R}\right)  _{-\frac{1}{3}}}%
,%
\genfrac{}{}{0pt}{}{\left(  \nu_{e}^{R}\right)  _{0}}{\left(  e^{R}\right)
_{-1}}%
;%
\genfrac{}{}{0pt}{}{\left(  c^{R}\right)  _{\frac{2}{3}}}{\left(
s^{R}\right)  _{-\frac{1}{3}}}%
,%
\genfrac{}{}{0pt}{}{\left(  \nu_{\mu}^{R}\right)  _{0}}{\left(  \mu
^{R}\right)  _{-1}}%
;%
\genfrac{}{}{0pt}{}{\left(  t^{R}\right)  _{\frac{2}{3}}}{\left(
b^{R}\right)  _{-\frac{1}{3}}}%
,%
\genfrac{}{}{0pt}{}{\left(  \nu_{\tau}^{R}\right)  _{0}}{\left(  \tau
^{R}\right)  _{-1}}%
\end{align}
We have included the right handed neutrinos assuming these particles develop
Dirac or Majorana-type masses. To describe the fermions in a more compact
notation we further introduce the following definitions. The three left handed
quark and lepton doublets are defined as $\left(  Q^{L_{i}}\right)  _{\frac
{1}{6}}$ , $\left(  L^{L_{i}}\right)  _{-\frac{1}{2}}$ respectively with
$i=1,2,3$ denoting the three families. Similarly, we define the family
labeling $j=1,2,3$ for the right handed quarks and leptons as $u^{R_{j}%
}=\left(  u^{R},c^{R},t^{R}\right)  _{\frac{2}{3}},$ $d^{R_{j}}=\left(
d^{R},s^{R},b^{R}\right)  _{-\frac{1}{3}}$, $e^{R_{j}}=\left(  e^{R},\mu
^{R},\tau^{R}\right)  _{-1}$ and $\nu^{R_{j}}=\left(  \nu_{e}^{R},\nu_{\mu
}^{R},\nu_{\tau}^{R}\right)  _{0}.$ All quarks are triplets and all leptons
are singlets under color SU$\left(  3\right)  $. The left handed quarks and
leptons $Q^{L_{i}}$ , $L^{L_{i}}$ are doublets and the right handed quarks and
lepton $u^{R_{j}}$, $d^{R_{j}}$, $e^{R_{j}}$, $\nu^{R_{j}}$ are singlets under
SU$\left(  2\right)  $ as listed above. Furthermore each field is charged
under U$\left(  1\right)  $ with the charges marked as subscripts above. The
electric charge of each field is then given by $Q=I_{3}+Y,$ where $Y$ is the
U$\left(  1\right)  $ charge and $I_{3}$ is the third generator of SU$\left(
2\right)  $ represented as $\frac{1}{2}\sigma_{3}$ on all doublets.

The covariant derivatives for each field is then straightforward, as usual in
gauge theories. Then the action for the Standard Model in $4+2$ dimensions is
given by%
\begin{align}
S\left(  A,\Psi^{L,R},H,\Phi\right)   &  =Z\int\left(  d^{6}X\right)
~\delta\left(  X^{2}\right)  ~L\left(  A,\Psi^{L,R},H,\Phi\right)
\label{2TSMaction}\\
L\left(  A,\Psi^{L,R},H,\Phi\right)   &  =L\left(  A\right)  +L\left(
A,\Psi^{L,R}\right)  +L\left(  \Psi^{L,R},H\right)  +L\left(  A,\Phi,H\right)
\end{align}
where $Z$ is an overall normalization factor that will be chosen below, and
$A=\left(  G,W,B\right)  $ is a short hand notation for the gauge fields.
Thus, after peeling off the overall volume element $Z\int\left(
d^{6}X\right)  ~\delta\left(  X^{2}\right)  $ the various parts of the
Lagrangian are given as follows. The factor of $Z$ will be fixed later to
normalize the emergent volume element in $3+1$ dimensions.

To get to the physics as simply as possible we assume the 2Tgauge-symmetry has
already been gauge fixed to simplify the action as discussed in the previous
sections. This means that the simplified actions given below for the Standard
Model contains fields whose remainders (i.e. parts proportional to $X^{2})$
are not the most general. For example, all remainders can be fixed to be zero.
More generally, they are gauge freedoms that have appropriate homogeneity
properties to be consistently removable by the remaining 2Tgauge-symmetry. The
latter version is not only more general, but can also be simpler for
understanding the reduction from $4+2$ to $3+1$ dimensions discussed in the
next section.

The Lagrangian for the gauge bosons $L\left(  A\right)  $ is then
\begin{equation}
L\left(  A\right)  =-\frac{1}{4}Tr_{3}\left(  G_{MN}G^{MN}\right)  -\frac
{1}{4}Tr_{2}\left(  W_{MN}W^{MN}\right)  -\frac{1}{4}B_{MN}B^{MN}.
\label{Alagrangian}%
\end{equation}
Note that there is no dilaton factor in the Yang-Mills action since $d=4.$
Each field strength is of the form $A_{MN}=\partial_{M}A_{N}-\partial_{N}%
A_{M}-ig_{A}\left[  A_{M},A_{N}\right]  $, for $A=\left(  G,W,B\right)  $ with
corresponding gauge groups SU$\left(  3\right)  \times$SU$\left(  2\right)
\times$U$\left(  1\right)  ,$ and with different dimensionless coupling
constants $g_{3},g_{2},g_{1}$ appearing instead of the $g_{A}$. Of course for
the Abelian $B_{MN}$ there is no quadratic term proportional to the U$\left(
1\right)  $ coupling $g_{1}.$

The Lagrangian $L\left(  A,\Phi,H\right)  $ is of the form Eq.(\ref{SphiH})%
\begin{equation}
L\left(  A,\Phi,H\right)  =\frac{1}{2}\Phi\partial^{2}\Phi+\frac{1}{2}\left(
H^{\dagger}D^{2}H+\left(  D^{2}H\right)  ^{\dagger}H\right)  -V\left(
\Phi,H\right)  \label{phiHlagrangian}%
\end{equation}
where the covariant derivative $D_{M}H$ is given by%
\begin{equation}
D_{M}H=\left(  \partial_{M}-ig_{2}\vec{W}_{M}\cdot\frac{\vec{\tau}}{2}%
-i\frac{g_{1}}{2}B_{M}\right)  H.
\end{equation}
The potential energy can be written in the following gauge invariant form%
\begin{equation}
V\left(  \Phi,H\right)  =\frac{\lambda}{4}\left(  H^{\dagger}H-\alpha^{2}%
\Phi^{2}\right)  ^{2}+V\left(  \Phi\right)  \label{VHphi}%
\end{equation}
where the couplings $\lambda,\alpha$ are dimensionless. Recall that quadratic
mass terms for $H$ are not allowed in the potential by the consistency of the
kinematic equations of motion that require homogeneity as discussed in
Eq.(\ref{VphiH}). For this reason we needed to introduce a coupling to the
dilaton field $\Phi.$ Now $V\left(  \Phi,H\right)  $ has a nontrivial minimum
for the Higgs field $H$ where the minimum occurs at $H^{\dagger}H=\alpha
^{2}\Phi^{2}$ or
\begin{equation}
H=\left(
\genfrac{}{}{0pt}{}{H^{+}}{H^{0}}%
\right)  =\alpha\Phi\left(
\genfrac{}{}{0pt}{}{0}{1}%
\right)  . \label{vacuum}%
\end{equation}
This breaks the SU$\left(  2\right)  \times$U$\left(  1\right)  $ gauge
symmetry down to the electro-magnetic U$\left(  1\right)  $ subgroup. Next we
need to discuss $V\left(  \Phi\right)  $ that stabilizes the dilaton $\Phi$ at
some constant expectation value $\langle\Phi\rangle\neq0$ that sets the scale
for the weak interactions as $\langle H^{0}\rangle=\alpha\langle\Phi\rangle=v$
in the range of 100 $GeV.$ For this we refer to the later discussion on the
topic of mass generation in section (\ref{massgen}).

The Lagrangian $L\left(  A,\Psi^{L,R}\right)  $ for fermions is of the form
given in Eq.(\ref{fermiS0}) but otherwise has all the terms in parallel to the
Standard Model in $3+1$ dimensions as follows%
\begin{align}
L\left(  A,\Psi^{L,R}\right)   &  =\frac{i}{2}\left(  \bar{Q}^{L_{i}}\not X
\overline{\not D  }Q^{L_{i}}+\bar{Q}^{L_{i}}\overleftarrow{\not D  }%
\overline{\not X  }Q^{L_{i}}\right)  +\frac{i}{2}\left(  \bar{L}^{L_{i}%
}\not X  \overline{\not D  }L^{L_{i}}+\bar{L}^{L_{i}}\overleftarrow{\not D
}\overline{\not X  }L^{L_{i}}\right) \label{psilagrangian}\\
&  -\frac{i}{2}\left(  \bar{d}^{R_{j}}\overline{\not X  }\not D  d^{R_{j}%
}+\bar{d}^{R_{j}}\overleftarrow{\overline{\not D  }}\not X  d^{R_{j}}\right)
-\frac{i}{2}\left(  \bar{e}^{R_{j}}\overline{\not X  }\not D  e^{R_{j}}%
+\bar{e}^{R_{j}}\overleftarrow{\overline{\not D  }}\not X  e^{R_{j}}\right) \\
&  -\frac{i}{2}\left(  \bar{u}^{R_{j}}\overline{\not X  }\not D  u^{R_{j}%
}+\bar{u}^{R_{j}}\overleftarrow{\overline{\not D  }}\not X  u^{R_{j}}\right)
-\frac{i}{2}\left(  \bar{\nu}^{R_{j}}\overline{\not X  }\not D  \nu^{R_{j}%
}+\bar{\nu}^{R_{j}}\overleftarrow{\overline{\not D  }}\not X  \nu^{R_{j}%
}\right)
\end{align}
Note that $\Gamma^{M}\leftrightarrow\bar{\Gamma}^{M}$ are interchanged in
comparing the left/right sectors, while the sign patterns in the L/R sectors
are chosen so that the emergent theory in $3+1$ dimensions has the correct
normalization for the kinetic terms (see Eq.(\ref{norm})). Also, we have
replaced the ordinary derivative $\partial_{M}\Psi^{L,R}$ by the Yang-Mills
covariant derivatives $D_{M}$ as follows, again in parallel to the usual
Standard Model in $3+1$ dimensions%
\begin{align}
D_{M}Q^{L_{i}}  &  =\left(  \partial_{M}-ig_{3}G_{M}^{a}\frac{\lambda}{2}%
^{a}-ig_{2}\vec{W}_{M}\cdot\frac{\vec{\tau}}{2}-i\frac{g_{1}}{6}B_{M}\right)
Q^{L_{i}},\\
D_{M}u^{R_{j}}  &  =\left(  \partial_{M}-ig_{3}G_{M}^{a}\frac{\lambda}{2}%
^{a}-i\frac{2g_{1}}{3}B_{M}\right)  u^{R_{j}}\\
D_{M}d^{R_{j}}  &  =\left(  \partial_{M}-ig_{3}G_{M}^{a}\frac{\lambda}{2}%
^{a}+i\frac{g_{1}}{3}B_{M}\right)  d^{R_{j}}\\
D_{M}L^{L_{i}}  &  =\left(  \partial_{M}-ig_{2}\vec{W}_{M}\cdot\frac{\vec
{\tau}}{2}+i\frac{g_{1}}{2}B_{M}\right)  L^{L_{i}}\\
D_{M}\nu^{R_{j}}  &  =\partial_{M}\nu^{R_{j}}\\
D_{M}e^{R_{j}}  &  =\left(  \partial_{M}+ig_{1}B_{M}\right)  e^{R_{j}}%
\end{align}
where $\frac{\lambda}{2}^{a}$ are the $3\times3$ matrices that represent the
generators of SU$\left(  3\right)  $ and $\frac{\vec{\tau}}{2}$ are the
$2\times2$ Pauli matrices that represent the generators of SU$\left(
2\right)  .$ This part of the Lagrangian is invariant under a global family
symmetry group $F$ that transforms only the fermions indicated as subscripts
\begin{equation}
F=U\left(  3\right)  _{Q^{L}}\times U\left(  3\right)  _{u^{R}}\times U\left(
3\right)  _{d^{R}}\times U\left(  3\right)  _{L^{L}}\times U\left(  3\right)
_{e^{R}}\times U\left(  3\right)  _{\nu^{R}}.
\end{equation}

The Lagrangian $S\left(  \Psi^{L,R},H\right)  $ for Yukawa couplings is of the
form in Eq.(\ref{psiHphi}), but without the dilaton factor since $d=4.$ This
couples the three families as follows, again in parallel to the usual Standard
Model in $3+1$ dimensions
\begin{equation}
L\left(  \Psi^{L,R},H\right)  =\left(
\begin{array}
[c]{c}%
\left(  g_{u}\right)  _{ij}\bar{Q}^{L_{i}}\not X  u^{R_{j}}H^{c})+\left(
g_{u}^{\dagger}\right)  _{ji}\bar{H}^{c}\bar{u}^{R_{j}}\overline{\not X
}Q^{L_{i}}\\
+\left(  g_{d}\right)  _{ij}\bar{Q}^{L_{i}}\not X  d^{R_{j}}H+\left(
g_{u}^{\dagger}\right)  _{ji}\bar{H}\bar{d}^{R_{j}}\overline{\not X  }%
Q^{L_{i}}\\
+\left(  g_{\nu}\right)  _{ij}L^{L_{i}}\not X  \nu^{R_{j}}H^{c}+\left(
g_{\nu}^{\dagger}\right)  _{ji}\bar{H}^{c}\bar{\nu}^{R_{j}}\overline{\not X
}L^{L_{i}}\\
+\left(  g_{e}\right)  _{ij}L^{L_{i}}\not X  e^{R_{j}}H+-\left(
g_{e}^{\dagger}\right)  _{ji}\bar{H}\bar{e}^{R_{j}}\overline{\not X  }%
L^{L_{i}}%
\end{array}
\right)  . \label{yukawa}%
\end{equation}
Here $H^{c}=i\tau_{2}H^{\ast}=\left(
\genfrac{}{}{0pt}{}{\bar{H}^{0}}{-H^{-}}%
\right)  _{-\frac{1}{2}}$ is the SU$\left(  2\right)  $ charge conjugate of
$H$, which transforms as an SU$\left(  2\right)  $ doublet and has opposite
U$\left(  1\right)  $ charge. The dimensionless Yukawa couplings $\left(
g_{u}\right)  _{ij}$, $\left(  g_{d}\right)  _{ij}$, $\left(  g_{\nu}\right)
_{ij}$, $\left(  g_{e}\right)  _{ij}$ are complex 3$\times3$ constant matrices
since this is the most general permitted by the gauge symmetry SU$\left(
3\right)  \times$SU$\left(  2\right)  \times$U$\left(  1\right)  .$ These
couplings break the global family symmetry $F$ of the lagrangian $\,L\left(
A,\Psi^{L,R}\right)  $ mentioned above. As is well known, by using the freedom
of the global family symmetry $F$ it is possible to choose a basis for the
quarks and leptons such that $g_{u}$ and $g_{e}$ are real and diagonal, while
$g_{d}$ and $g_{\nu}$ become Hermitian but non-diagonal. This relates to the
Kobayashi-Maskawa matrices for the quarks and for the neutrinos. The off
diagonal entries mix families so that separate family number is not conserved.
This leads to the explanation of how the more massive families decay to the
less massive ones, and how neutrino mixing occurs. Using the symmetry $F$ to
its maximum to eliminate phases in the remaining off diagonal entries, the
mixing between quark families $\left(  1,2\right)  $ can be chosen real, but
the mixing between families $\left(  1,3\right)  $ and $\left(  2,3\right)  $
remain complex. The complex phases in $g_{d}$ and $g_{\nu}$ violate the
discrete CP symmetry.

Based on the principles of 2Tgauge-symmetry in field theory, that emerged from
the underlying Sp$\left(  2,R\right)  $ gauge symmetry in the worldline
formalism, we have constructed the Standard Model in $4+2$ dimensions. This
action generates consistently both the kinematic and dynamical equations of
motion for every on shell field in the theory. The kinematic equations,
together with the 2Tgauge-symmetry, are just the necessary ingredients to make
the $4+2$ dimensional theory equivalent to the $3+1$ dimensional Standard
Model. However the $4+2$ structure imposes some restrictions on the emergent
$3+1$ dimensional theory that relate to unresolved issues in the Standard
Model in $3+1$ dimensions, including the issues of the strong CP problem and
the mass generation mechanism. These are discussed in the following sections.

\section{The emergent Standard Model in $3+1$}

In this section we demonstrate how the $3+1$ dimensional Standard Model
emerges from $4+2$ dimensions. The new 2Tgauge-symmetry in field theory is
essential to show that every field in the theory can be gauge fixed so that it
becomes independent of $X^{2},$ as already assumed in the simplified gauge
fixed form of the Lagrangian given in the previous section. Then by using the
condition $X^{2}=0$ imposed by the delta function we can eliminate one of the
components of $X^{M}$ from every field in the theory. A second component of
$X^{M}$ will also be eliminated from every field by putting every field
partially on-shell by satisfying the kinematic equations that follow from the
action. These two conditions are precisely the two Sp$\left(  2,R\right)  $
generators in the worldline formalism $X^{2}=\left(  X\cdot P+P\cdot X\right)
=0$ that are solved explicitly in a fixed gauge to obtain a holographic image
of the $4+2$ dimensional system in $3+1$ dimensions, as in Fig.1. We now
discuss how this happens in field theory for the Standard Model. We emphasize
that the third Sp$\left(  2,R\right)  $ generator is modified by the
interactions, and will be left off-shell in the discussion below.

We start by choosing a lightcone type basis in $4+2$ dimensions so that the
flat metric takes the form $ds^{2}=dX^{M}dX^{N}\eta_{MN}=-2dX^{+^{\prime}%
}dX^{-^{\prime}}+dX^{\mu}dX^{\nu}\eta_{\mu\nu},$ where $\eta_{\mu\nu},$ with
$\mu,\nu=0,1,2,3$ is the Minkowski metric and $X^{\pm^{\prime}}=\frac{1}%
{\sqrt{2}}\left(  X^{0^{\prime}}\pm X^{1^{\prime}}\right)  $ are the lightcone
coordinates for the extra one space and one time dimensions. Furthermore we
choose the following parametrization which defines the emergent $3+1$
dimensional spacetime $x^{\mu}$
\begin{align}
X^{+^{\prime}}  &  =\kappa,\;X^{-^{\prime}}=\kappa\lambda,\;X^{\mu}=\kappa
x^{\mu},\label{massless}\\
\kappa &  =X^{+^{\prime}},\;\;\lambda=\frac{X^{-^{\prime}}}{X^{+^{\prime}}%
},\;x^{\mu}=\frac{X^{\mu}}{X^{+^{\prime}}}.
\end{align}
This provides one of the many possible embedding of $3+1$ dimensions in $4+2$
dimensions. Each such embedding corresponds to a Sp$\left(  2,R\right)  $
gauge choice in the underlying 2T-physics worldline theory. The present one
corresponds to the one labeled as the \textquotedblleft relativistic massless
particle\textquotedblright\ in Fig.1. In other embeddings, we will obtain a
different $3+1$ dimensional view of the $4+2$ dimensional theory, as in the
examples of Fig.1.

The fields are parameterized as $\Phi\left(  X\right)  =\Phi\left(
\kappa,\lambda,x^{\mu}\right)  ,$ and similarly for the others, where
$\lambda,x^{\mu}$ are homogeneous coordinates which do not change under
rescaling $\Phi\left(  tX\right)  =\Phi\left(  t\kappa,\lambda,x^{\mu}\right)
.$ The kinematic equations will be solved in this parametrization to reduce
the theory from fields in the spacetime $X^{M}$ to fields in the smaller
spacetime $x^{\mu}.$ After this, there remains to satisfy the dynamical
equations, including interactions, only in terms of the fields in the smaller
spacetime $x^{\mu}$. The dynamics in the reduced space is described by the
reduced action which holographically captures all of the information in the
$4+2$ dimensional theory.

With the parametrization of Eq.(\ref{massless}) we get $X^{2}=-2X^{+^{\prime}%
}X^{-^{\prime}}+X^{\mu}X_{\mu}=\kappa^{2}\left(  -2\lambda+x^{2}\right)  .$ So
the volume element that appears in the action Eq.(\ref{2TSMaction}) takes the
form%
\begin{equation}
\left(  d^{6}X\right)  ~\delta\left(  X^{2}\right)  =\kappa^{5}d\kappa
~d^{4}x~d\lambda~\delta\left(  \kappa^{2}\left(  2\lambda-x^{2}\right)
\right)  .
\end{equation}
When $\lambda=x^{2}/2$ is imposed, the $4+2$ dimensional flat metric reduces
to the conformal metric in $3+1$ dimensions $ds^{2}=dX^{M}dX_{M}=\kappa
^{2}\left(  dx\right)  ^{2}.$ This is how the SO$\left(  4,2\right)  $ in
$4+2$ dimensions becomes the conformal symmetry in $3+1$ dimensions. The
non-linear realization of conformal transformations in $3+1$ dimensions
$x^{\mu}$ is nothing but the SO$\left(  4,2\right)  $ Lorentz transformations
in the space $X^{M}.$

Recall that derivatives must be taken before the $X^{2}=0$ is imposed. Let us
now use the chain rule $\partial_{M}=\left(  \partial_{M}\kappa\right)
\frac{\partial}{\partial\kappa}+\left(  \partial_{M}\lambda\right)
\frac{\partial}{\partial\lambda}+\left(  \partial_{M}x^{\mu}\right)
\frac{\partial}{\partial x^{\mu}}$ to compute derivatives as follows
\begin{equation}
\;\frac{\partial}{\partial X^{\mu}}=\frac{1}{\kappa}\frac{\partial}{\partial
x^{\mu}},\;\;\frac{\partial}{\partial X^{-^{\prime}}}=\frac{1}{\kappa}%
\frac{\partial}{\partial\lambda},\quad\frac{\partial}{\partial X^{+^{\prime}}%
}=\frac{1}{\kappa}\left(  \kappa\frac{\partial}{\partial\kappa}-\lambda
\frac{\partial}{\partial\lambda}-x^{\mu}\frac{\partial}{\partial x^{\mu}%
}\right)  . \label{chain}%
\end{equation}
Using these we further compute $X^{M}\partial_{M}=\kappa\frac{\partial
}{\partial\kappa},$ and the Laplace operator in $4+2$ dimensions
\begin{equation}
\partial^{M}\partial_{M}=\frac{1}{\kappa^{2}}\left(  \frac{\partial}{\partial
x^{\mu}}+x_{\mu}\frac{\partial}{\partial\lambda}\right)  ^{2}-\frac{1}%
{\kappa^{2}}\left(  2\kappa\frac{\partial}{\partial\kappa}+d-2\right)
\frac{\partial}{\partial\lambda}+\frac{1}{\kappa^{2}}\left(  2\lambda
-x^{2}\right)  \left(  \frac{\partial}{\partial\lambda}\right)  ^{2}.
\label{laplace}%
\end{equation}
We will also need the structures $\Gamma^{M}X_{M},\bar{\Gamma}^{M}X_{M}%
,\Gamma^{M}\partial_{M},\bar{\Gamma}^{M}\partial_{M}$ that appear in the
fermion equations in $4+2$ dimensions by using explicitly the gamma matrix
representation given in footnote (\ref{gamms})%
\begin{align}
X^{M}\Gamma_{M}  &  =-\Gamma^{+^{\prime}}X^{-^{\prime}}-\Gamma^{-^{\prime}%
}X^{+^{\prime}}+\Gamma_{\mu}X^{\mu}\\
&  =\kappa\left(
\begin{array}
[c]{cc}%
x^{\mu}\sigma_{\mu} & i\sqrt{2}\lambda\\
i\sqrt{2} & -x^{\mu}\bar{\sigma}_{\mu}%
\end{array}
\right)  , \label{X}%
\end{align}
and%
\begin{align}
\bar{\Gamma}_{M}\partial^{M}  &  =\bar{\Gamma}^{+^{\prime}}\partial
_{+^{\prime}}+\bar{\Gamma}^{-^{\prime}}\partial_{-^{\prime}}+\bar{\Gamma}%
^{\mu}\partial_{\mu}\\
&  =\frac{1}{\kappa}\left(
\begin{array}
[c]{cc}%
\bar{\sigma}^{\mu}\partial_{\mu} & -i\sqrt{2}\left(  \kappa\partial_{\kappa
}-\lambda\partial_{\lambda}-x^{\mu}\partial_{\mu}\right) \\
-i\sqrt{2}\partial_{\lambda} & -\sigma^{\mu}\partial_{\mu}%
\end{array}
\right)  .\; \label{Delgamma}%
\end{align}
For $\bar{\Gamma}^{M}X_{M},\Gamma^{M}\partial_{M}$ we obtain the the same
structures as above, but replacing $\sigma_{\mu}$ by $\bar{\sigma}_{\mu}$ and
vice versa.

\subsection{Reduction of scalars}

We now proceed to solve the kinematical equations. We start with the kinematic
equations of the scalars $\left(  X\cdot\partial+\frac{d-2}{2}\right)
\Phi=\left(  \kappa\frac{\partial}{\partial\kappa}+1\right)  \Phi=0$. The
Higgs scalar $H$ also satisfies a similar equation but with the covariant
derivative replacing the ordinary derivative. We fix the Yang-Mills gauge
symmetry so that
\begin{equation}
X\cdot A=0\text{ for all YM fields }A=\left(  G,W,B\right)
\end{equation}
In this gauge $X\cdot D=X\cdot\partial$ therefore $H$ satisfies the same
kinematic condition as the singlet $\Phi.$ These homogeneity conditions
determine the kappa dependence fully as an overall factor%
\begin{equation}
\Phi\left(  X\right)  =\kappa^{-1}\underline{\Phi}\left(  x,\lambda\right)
,\text{ similarly for }H. \label{phihom}%
\end{equation}
Now recall that according to Eq.(\ref{delphi}) the part of the
\textit{homogeneous} scalar field proportional to $X^{2}$ is gauge freedom
with respect to the 2Tgauge-symmetry. We have already said that we can gauge
fix the remainder to zero, but let's look at the details of how this is done.
Thus if we define $\phi\left(  x\right)  \equiv\underline{\Phi}\left(
x,0\right)  $ and write $\underline{\Phi}\left(  x,\lambda\right)
=\phi\left(  x\right)  +\left(  \lambda-\frac{x^{2}}{2}\right)  \tilde{\phi
}\left(  x,\lambda\right)  ,$ then the remainder $\tilde{\phi}\left(
x,\lambda\right)  $ can be gauge fixed at will. It is convenient to choose the
gauge $\tilde{\phi}\left(  x,\lambda\right)  =0$ which makes the field
$\Phi\left(  X\right)  $ independent of $\lambda$
\begin{equation}
\Phi\left(  X\right)  =\kappa^{-1}\phi\left(  x\right)  ,\text{ similarly
}H\left(  X\right)  =\kappa^{-1}h\left(  x\right)  .
\end{equation}
Then it is simple to compute the $4+2$ dimensional Laplacian $\partial
^{M}\partial_{M}$ of Eq.(\ref{laplace}) since all derivatives $\frac{\partial
}{\partial\lambda}$ vanish and we obtain\footnote{If we don't choose the
simplifying gauge $\tilde{\phi}\left(  x,\lambda\right)  =0,$ we still obtain
the same result as follows. In computing $\partial^{M}\partial_{M}\Phi\left(
X\right)  $ we recall that $\lambda=\frac{x^{2}}{2}$ is imposed by the delta
function after all derivatives are computed. Then the operator $\partial
^{M}\partial_{M}$ in Eq.(\ref{laplace}) simplifies as follows. The last term
drops because of the factor $(\lambda-\frac{x^{2}}{2})=0$ even after
differentiation. The second term drops because of the form of the homogeneous
solution (\ref{phihom}). The first term simplifies because derivatives with
respect to $x^{\mu}$ appear only in the combination $\frac{\partial}{\partial
x^{\mu}}+x_{\mu}\frac{\partial}{\partial\lambda}.$ Then, setting
$\lambda=x^{2}/2$ after differentiation with the derivative operator
$\frac{\partial}{\partial x^{\mu}}+x_{\mu}\frac{\partial}{\partial\lambda},$
gives the same result as setting $\lambda=x^{2}/2$ before differentiation and
then differentiating with respect to the total $x$ dependence including the
part coming from $\lambda=x^{2}/2$
\begin{equation}
\left[  \left(  \frac{\partial}{\partial x^{\mu}}+x_{\mu}\frac{\partial
}{\partial\lambda}\right)  \tilde{\Phi}\left(  x,\lambda\right)  \right]
_{\lambda=x^{2}/2}=\frac{\partial}{\partial x^{\mu}}\tilde{\Phi}\left(
x,\frac{x^{2}}{2}\right)  .
\end{equation}
Therefore we can set $\tilde{\Phi}\left(  x,\lambda\right)  |_{\lambda
=x^{2}/2}=\tilde{\Phi}\left(  x,\frac{x^{2}}{2}\right)  =\phi\left(  x\right)
$ before differentiation. This is equivalent to dropping the term
$\frac{\partial}{\partial\lambda}$ in the derivative operator $\frac{\partial
}{\partial x^{\mu}}+x_{\mu}\frac{\partial}{\partial\lambda}.$ Hence
$\partial^{M}\partial_{M}$ in $4+2$ dimensions reduces to the Laplace operator
in $3+1$ dimensions Eq.(\ref{laplace4}), in agreement with the simpler
derivation based on the fixed gauge.\label{lambda}}%
\begin{equation}
\partial^{M}\partial_{M}\Phi\left(  X\right)  =\frac{1}{\kappa^{3}}%
\frac{\partial^{2}\phi\left(  x\right)  }{\partial x^{\mu}\partial x_{\mu}}.
\label{laplace4}%
\end{equation}
It will be argued below that the gauge fields will also get reduced to four
dimensional fields\footnote{As seen in Eq.(\ref{Afixed}) the gauge fixed form
of $A_{M}$ includes the non-zero component $A^{-^{\prime}}=\frac{1}{\kappa
}x^{\mu}A_{\mu}\left(  x\right)  =-A_{+^{\prime}},$ while $A^{+^{\prime}%
}=-A_{-^{\prime}}=0.$ Therefore the covariant derivatives in the extra
dimensions $D_{\pm^{\prime}}$ take the form $D_{-^{\prime}}=\partial
_{-^{\prime}}=\frac{1}{\kappa}\partial_{\lambda}$ and $D_{+^{\prime}}%
=\partial_{+^{\prime}}-iA_{+^{\prime}}=\partial_{\kappa}+\frac{i}{\kappa
}x^{\mu}A_{\mu}.$ These appear in $D^{M}D_{M}=-D_{+^{\prime}}D_{-^{\prime}%
}-D_{-^{\prime}}D_{+^{\prime}}+D^{\mu}D_{\mu}.$ Since $D_{-^{\prime}}=\frac
{1}{\kappa}\partial_{\lambda}$ vanishes on the $\lambda$ independent $h\left(
x\right)  $ and $A_{\mu}\left(  x\right)  $ we obtain the reduction
$D^{M}D_{M}\rightarrow D^{\mu}D_{\mu}$ which then leads to the four
dimensional theory. \label{dlambda}}, so that $D^{2}H\left(  X\right)
=\frac{1}{\kappa^{3}}D^{\mu}D_{\mu}h$ will involve only the four dimensional
component of the gauge field $A_{\mu}\left(  x\right)  $. In this way the
scalar Lagrangian $L\left(  A,\Phi,H\right)  $ is reduced to the form%
\begin{equation}
L\left(  A_{M}\left(  X\right)  ,\Phi\left(  X\right)  ,H\left(  X\right)
\right)  =\frac{1}{\kappa^{4}}L\left(  A_{\mu}\left(  x\right)  ,\phi\left(
x\right)  ,h\left(  x\right)  \right)
\end{equation}
where $L\left(  A_{\mu}\left(  x\right)  ,\phi\left(  x\right)  ,h\left(
x\right)  \right)  $ is purely a four dimensional Lagrangian that has the same
form as $L\left(  A_{M}\left(  X\right)  ,\Phi\left(  X\right)  ,H\left(
X\right)  \right)  $ except for the fact that only four dimensional fields and
only four dimensional covariant derivatives appear. Replacing this in the
action we obtain
\begin{align}
S\left(  A,\Phi,H\right)   &  =Z\int\left\vert \kappa\right\vert ^{5}%
d\kappa~d^{4}x~d\lambda~\delta\left(  \kappa^{2}\left(  2\lambda-x^{2}\right)
\right)  \times\frac{1}{\kappa^{4}}L\left(  A_{\mu}\left(  x\right)
,\phi\left(  x\right)  ,h\left(  x\right)  \right) \\
&  =\left[  Z\int d\kappa du~\delta\left(  2\left\vert \kappa\right\vert
u\right)  \right]  \int d^{4}xL\left(  A_{\mu}\left(  x\right)  ,\phi\left(
x\right)  ,h\left(  x\right)  \right)  .
\end{align}
In the last step we have changed integration variable to $u=\lambda
-\frac{x^{2}}{2}.$ We see that the action has an overall logarithmically
divergent factor which is cancelled by choosing the overall normalization $Z$
in front of the whole action in Eq.(\ref{2TSMaction}), so that
\begin{equation}
Z\int d\kappa du~\delta\left(  2\left\vert \kappa\right\vert u\right)  =1.
\label{Z}%
\end{equation}
The same factor will appear in all the terms of the action in the reduction
from $4+2$ to $3+1$ dimensions. The four dimensional action $S\left(
A,\phi,h\right)  =\int d^{4}xL\left(  A_{\mu}\left(  x\right)  ,\phi\left(
x\right)  ,h\left(  x\right)  \right)  $ is translation and Lorentz invariant,
and captures all of the information contained in the six dimensional action
$S\left(  A,\Phi,H\right)  $ without losing any information. In this sense the
four dimensional action is a holographic image of the higher dimensional one.

\subsection{Reduction of chiral fermions}

We start by writing every fermion in the form $\Psi^{L,R}\left(  X\right)
=\Psi_{0}^{L,R}\left(  X\right)  +X^{2}\widetilde{\Psi}^{L,R}\left(  X\right)
$ where $\Psi_{0}^{L,R}$ is defined as $\Psi_{0}^{L,R}\left(  X\right)
\equiv\left[  \Psi^{L,R}\left(  X\right)  \right]  _{X^{2}=0}=\Psi_{0}%
^{L,R}\left(  \kappa,x\right)  $ which is independent of $\lambda.$ We can
gauge fix $\widetilde{\Psi}^{L,R}\left(  X\right)  =0$ off shell by using the
$X^{2}\zeta_{1}$ part of the 2Tgauge-symmetry for fermions of
Eq.(\ref{delpsiLR}). In this gauge $\Psi^{L,R}\left(  X\right)  \equiv\Psi
_{0}^{L,R}\left(  \kappa,x\right)  $ becomes fully independent of $\lambda.$
Next we use the the kinematical equations $\left(  X\cdot\partial+\frac{d}%
{2}\right)  \Psi^{L,R}=\left(  \kappa\frac{\partial}{\partial\kappa}+2\right)
\Psi^{L,R}=0.$ This determines the kappa dependence fully as an overall factor
for all fermions, and we get the homogeneous form%
\begin{equation}
\Psi^{L,R}\left(  X\right)  =\kappa^{-2}\chi^{L,R}\left(  x\right)  .
\end{equation}
Since we now have homogeneous fermions, half of the degrees of freedom can be
removed by using the fermionic 2Tgauge-symmetry of Eq.(\ref{delpsiLR}) with
the parameters $\not X  \zeta_{2}^{R_{\alpha}},\;\overline{\not X  }\zeta
_{2}^{L_{\beta}}$ that have the same degree of homogeneity. It is convenient
to choose the lightcone type gauge $\Gamma^{+^{\prime}}\Psi^{L,R}=0\;$in the
extra dimensions that requires the two lower components of $\Psi^{L,R}$ to
vanish
\begin{equation}
\Psi^{L,R}\left(  X\right)  =\frac{1}{2^{1/4}\kappa^{2}}\left(
\begin{array}
[c]{c}%
\psi^{L,R}\left(  x\right) \\
0
\end{array}
\right)  ,\;\bar{\Psi}^{L,R}\left(  X\right)  =\frac{-i}{2^{1/4}\kappa^{2}%
}\left(  0,\;\bar{\psi}^{L,R}\left(  x\right)  \right)  .
\end{equation}
Note that $\bar{\Psi}^{L,R}$ is constructed by taking Hermitian conjugation
and applying the SU$\left(  2,2\right)  $ metric $\eta=-i\tau_{1}\times1$
given in footnote (\ref{gamms}). At this point we remain with only four
dimensional fields $\psi^{L,R}\left(  x\right)  $ written in the form of
SL$\left(  2,C\right)  $ doublets.

For the gauge fixed form of $\Psi^{L,R}$given above, the structures
$\overline{\not D  }\Psi^{L}$ and $\bar{\Psi}^{L}\not X  $ that appear in the
action take the following forms after using Eqs.(\ref{X},\ref{Delgamma})%
\begin{align}
\bar{\Psi}^{L}\not X   &  =\frac{-i}{2^{1/4}\kappa}\left(  0,\;\bar{\psi}%
^{L}\right)  \left(
\begin{array}
[c]{cc}%
x^{\mu}\sigma_{\mu} & i\sqrt{2}\lambda\\
i\sqrt{2} & -x^{\mu}\bar{\sigma}_{\mu}%
\end{array}
\right) \\
&  =\frac{-i}{2^{1/4}\kappa}\left(  i\sqrt{2}\bar{\psi}^{L},\;-\bar{\psi}%
^{L}x^{\mu}\bar{\sigma}_{\mu}\right)
\end{align}
and%
\begin{align}
\overline{\not D  }\Psi^{L}  &  =\frac{1}{2^{1/4}\kappa}\left(
\begin{array}
[c]{cc}%
\bar{\sigma}^{\mu}D_{\mu} & -i\sqrt{2}\left(  \kappa D_{\kappa}-\lambda
\partial_{\lambda}-x^{\mu}D_{\mu}\right) \\
-i\sqrt{2}\partial_{\lambda} & -\sigma^{\mu}D_{\mu}%
\end{array}
\right)  \left(
\begin{array}
[c]{c}%
\frac{1}{\kappa^{2}}\psi^{L}\left(  x\right) \\
0
\end{array}
\right) \\
&  =\frac{1}{2^{1/4}\kappa^{3}}\left(
\begin{array}
[c]{c}%
\bar{\sigma}^{\mu}D_{\mu}\psi^{L}\left(  x\right) \\
0
\end{array}
\right)  . \label{Dpsi}%
\end{align}
where we have used\footnote{Even if we had not chosen the simplifying gauge
$\Psi_{1}^{L,R}\left(  X\right)  =0$ which made $\Psi^{L,R}\left(  X\right)  $
independent of $\lambda,$ we can still reach the same conclusion for
Eq.(\ref{Dpsi}) for any $\Psi_{1}^{L,R}\left(  X\right)  ,$ after first
differentiating with respect to $\lambda$ and then setting $\lambda=x^{2}/2.$
The argument for this is similar to footnote (\ref{lambda}).} $\partial
_{\lambda}\psi^{L}\left(  x\right)  =0.$ Note that only $\partial_{\lambda}$
appears instead of $D_{\lambda}$ due to a Yang-Mills gauge choice
$A_{-^{\prime}}=-A^{+^{\prime}}=0$ as explained below and in footnote
(\ref{dlambda}).

There remains to show that the dynamical equations of the original fermions
reduce to the usual four dimensional massless fermion equations in $3+1$
dimensions. This can be done either for the equations of motion or more
concisely for the Lagrangian, with equivalent conclusions. Thus consider the
fermionic structures of the type $i\bar{\Psi}^{L}\not X \overline{\not D }%
\Psi^{L}$, $-i\bar{\Psi}^{R}\overline{\not X }\not D \Psi^{R}$, $g\bar{\Psi
}^{L}\not X \Psi^{R}H,$ etc. that appear in the $4+2$ dimensional action in
Eqs.(\ref{psilagrangian},\ref{yukawa}). From the above gauge fixed expressions
for $\Psi^{L,R}$, $\bar{\Psi}^{L}\not X $ and $\overline{\not D }\Psi^{L}$ we
compute%
\begin{equation}%
\begin{array}
[c]{c}%
g\bar{\Psi}^{L}\not X \Psi^{R}H=\frac{g}{\kappa^{4}}\bar{\psi}^{L}\psi
^{R}h,\;\;\;\;~i\bar{\Psi}^{L}\not X \overline{\not D }\Psi^{L}=\frac
{i}{\kappa^{4}}\bar{\psi}^{L}\bar{\sigma}^{\mu}D_{\mu}\psi^{L},\;\;\\
g\bar{\Psi}^{R}\overline{\not X }\Psi^{L}H^{\ast}=\frac{g}{\kappa^{4}}%
\bar{\psi}^{R}\psi^{L}h^{\ast},\;\;-i\bar{\Psi}^{R}\overline{\not X }%
\not D \Psi^{R}=-\frac{i}{\kappa^{4}}\bar{\psi}^{L}\sigma^{\mu}D_{\mu}\psi^{L}%
\end{array}
\; \label{norm}%
\end{equation}
Note that all explicit dependence on $X^{M}$ has disappeared. These emergent
forms are precisely the correctly normalized translation and Lorentz invariant
kinetic and Yukawa coupling terms that should appear in the four dimensional
action. Therefore, we have shown that after using the kinematical equations,
and imposing some gauge fixing, all the fermion terms in the $4+2$ dimensional
action reduce to a four dimensional theory%
\begin{align}
S\left(  fermions\right)   &  =Z\int\left\vert \kappa\right\vert ^{5}%
d\kappa~d^{4}x~d\lambda~\delta\left(  \kappa^{2}\left(  2\lambda-x^{2}\right)
\right)  \times\frac{1}{\kappa^{4}}L\left(  fermions\right) \\
&  =\int d^{4}x~L_{fermions}\left(  A_{\mu}\left(  x\right)  ,h\left(
x\right)  ,\psi^{L,R}\left(  x\right)  \right)
\end{align}
where Eq.(\ref{Z}) is used. Here $L_{fermions}\left(  A_{\mu}\left(  x\right)
,h\left(  x\right)  ,\psi^{L,R}\left(  x\right)  \right)  $ is the reduced
form of the fermion terms $L\left(  A,\Psi^{L,R}\right)  +L\left(  \Psi
^{L,R},H\right)  $ that appear in the $4+2$ theory in Eqs.(\ref{psilagrangian}%
,\ref{yukawa}). This is precisely the usual chiral fermion terms in the
Standard Model Lagrangian in $3+1$ dimensions interacting with the SU$\left(
3\right)  \times$SU$\left(  2\right)  \times$U$\left(  1\right)  $ gauge
bosons and the Higgs field.

\subsection{Reduction of gauge bosons}

We start by gauge fixing the Yang-Mills gauge symmetry in the form of
Eq.(\ref{Aaxial}) $X\cdot A=0,$ so that the kinematic equations for the
$A_{M}=\left(  G_{M},W_{M},B_{M}\right)  $ become the simple homogeneity
condition $X^{N}F_{NM}=\left(  X\cdot\partial+1\right)  A_{M}=\left(
\kappa\partial_{\kappa}+1\right)  A_{M}=0.$ Thus in this gauge we can write%
\begin{equation}
A^{M}\left(  X\right)  =\kappa^{-1}\underline{A}^{M}\left(  x^{\mu}%
,\lambda\right)
\end{equation}
There is a leftover gauge symmetry given by Yang-Mills transformations
$\delta_{\Lambda}A_{M}=D_{M}\Lambda=\partial_{M}\Lambda-i\left[  A_{M}%
,\Lambda\right]  $ that do not change the gauge condition $X\cdot A=0.$ This
leftover symmetry corresponds to homogeneous $\Lambda\left(  X\right)  $ of
degree $0$%
\begin{equation}
X\cdot\delta_{\Lambda}A=0\;\rightarrow X\cdot\partial\Lambda=0.
\end{equation}
This is just sufficient gauge symmetry to remove one full degree of freedom in
$d+2$ dimensions from a homogeneous gauge field $A^{M}\left(  X\right)  $.
Using this freedom we choose a lightcone type gauge in the extra dimensions
$A^{+^{\prime}}\left(  X\right)  =0$ (note $A_{-^{\prime}}=-\eta_{-^{\prime
}+^{\prime}}A^{+^{\prime}}=0$), and also use the fact that $X\cdot A=0$ to
solve for the other lightcone component in terms of the four Minkowski
components $A^{\mu}\left(  X\right)  .$ We find
\begin{equation}
A^{M}\left(  X\right)  :\;A^{+^{\prime}}=-A_{-^{\prime}}=0,\;\;A^{-^{\prime}%
}=-A_{+^{\prime}}=\frac{1}{\kappa}x^{\mu}\underline{A}_{\mu}\left(  x^{\mu
},\lambda\right)  ,\;\;A^{\mu}\left(  X\right)  =\frac{1}{\kappa}\underline
{A}^{\mu}\left(  x^{\mu},\lambda\right)  . \label{Afixed}%
\end{equation}

We now turn to the 2Tgauge-symmetry. In the present gauge its parameters
$a^{M}\left(  X\right)  $ of Eq.(\ref{Aa}) must be restricted to maintain the
gauge choice $A^{+^{\prime}}\left(  X\right)  =0.$ Therefore we must take
$a^{+^{\prime}}\left(  X\right)  =0.$ Furthermore due to $X\cdot a=0$ they
must have the same form as $A^{M}\left(  X\right)  ,$ namely $a^{+^{\prime}%
}=0$,\ $a^{-^{\prime}}=X^{\mu}a_{\mu}.$ The conditions that the remaining
$a^{\mu}\left(  X\right)  $ must satisfy in this gauge follow from
Eq.(\ref{Aa}) as
\begin{equation}
\left(  X\cdot\partial+3\right)  a^{\mu}\left(  X\right)  =X^{\mu}D\cdot
a,\;\left(  X\cdot\partial+3\right)  a^{\pm^{\prime}}\left(  X\right)
=X^{\pm^{\prime}}D\cdot a,
\end{equation}
Since $a^{+^{\prime}}=0$ we must have $D\cdot a=0,$ hence $a^{\mu}\left(
X\right)  $ must be homogeneous $\left(  X\cdot\partial+3\right)  a^{\mu
}\left(  X\right)  =0$, and therefore we have%
\begin{equation}
a^{\mu}\left(  X\right)  =\frac{1}{\kappa^{3}}\underline{a}^{\mu}\left(
x,\lambda\right)  .
\end{equation}
We can now use this homogeneous degree of freedom in the 2Tgauge-symmetry to
eliminate the $\lambda$ dependence from the gauge field \underline{$A$}%
$^{M}\left(  x^{\mu},\lambda\right)  .$

To proceed we first identify different parts of of the gauge field as
\begin{equation}
A_{\mu}\left(  X\right)  =A_{\mu}^{0}\left(  X\right)  +X^{2}\underline
{A}_{\mu}=\frac{1}{\kappa}\left[  A_{\mu}\left(  x\right)  +\left(
\lambda-\frac{x^{2}}{2}\right)  V_{\mu}\left(  x,\lambda\right)  \right]
\end{equation}
such that $A_{\mu}^{0}\left(  X\right)  \equiv\left[  A_{\mu}\left(  X\right)
\right]  _{X^{2}=0}=\kappa^{-1}A^{\mu}\left(  x\right)  $. Then, use the gauge
parameters $\underline{a}^{\mu}\left(  x,\lambda\right)  $ to gauge fix
$V^{\mu}\left(  x,\lambda\right)  =0.$ In this way we have arrived at a gauge
fixed field $A^{M}\left(  X\right)  $ that is independent of $\lambda$%
\begin{equation}
A^{M}\left(  X\right)  :\;A^{+^{\prime}}=-A_{-^{\prime}}=0,\;\;A^{-^{\prime}%
}=-A_{+^{\prime}}=\frac{1}{\kappa}x^{\mu}A_{\mu}\left(  x\right)  ,\;\;A^{\mu
}\left(  X\right)  =\frac{1}{\kappa}A^{\mu}\left(  x\right)
\end{equation}

We can now compute field strengths. Recall from the chain rule in
Eq.(\ref{chain}) that $\partial_{-^{\prime}}=\kappa^{-1}\partial_{\lambda}$
will vanish when applied on the $\lambda$ independent fields. Therefore we
find%
\begin{align}
F_{\mu\nu}\left(  X\right)   &  =\kappa^{-2}F_{\mu\nu}\left(  x\right)
,\;\text{with }F_{\mu\nu}\left(  x\right)  =\partial_{\mu}A_{\nu}%
-\partial_{\nu}A_{\mu}-i\left[  A_{\mu},A_{\nu}\right] \\
F_{+^{\prime}\mu}\left(  X\right)   &  =\kappa^{-2}x^{\nu}F_{\mu\nu}\left(
x\right)  ,\;\;F_{-^{\prime}\mu}\left(  X\right)  =0,\;\;F_{+^{\prime
}-^{\prime}}\left(  X\right)  =0.
\end{align}
The Lagrangian density becomes%
\begin{equation}
L\left(  A\left(  X\right)  \right)  =-\frac{1}{4}Tr\left(  F_{MN}%
F^{MN}\right)  \left(  X\right)  =-\frac{1}{4\kappa^{4}}Tr\left(  F_{\mu\nu
}F^{\mu\nu}\right)  \left(  x\right)  .
\end{equation}
We see that only the four dimensional field strength $F_{\mu\nu}$ has survived
as the only independent field. It was possible to gauge fix the components
$F_{-^{\prime}\mu}\left(  X\right)  ,F_{+^{\prime}-^{\prime}}\left(  X\right)
$ to zero, because $F_{MN}\left(  X\right)  $ is not gauge invariant under the
2Tgauge-symmetry, although the action as well as the dynamical equations
constructed from $F_{MN}$ are gauge invariant.

Using these results we can now see that after using the kinematical equations,
and imposing some gauge fixing, all the gauge theory terms in the $4+2$
dimensional action reduce to a four dimensional theory%
\begin{align}
S\left(  A\right)   &  =Z\int\left\vert \kappa\right\vert ^{5}d\kappa
~d^{4}x~d\lambda~\delta\left(  \kappa^{2}\left(  2\lambda-x^{2}\right)
\right)  \times\frac{1}{\kappa^{4}}L\left(  A^{\mu}\left(  x\right)  \right)
\\
&  =\int d^{4}x~L\left(  A_{\mu}\left(  x\right)  \right)
\end{align}
This is precisely the usual Yang-Mills terms in the Standard Model Lagrangian
in $3+1$ dimensions for the SU$\left(  3\right)  \times$SU$\left(  2\right)
\times$U$\left(  1\right)  $ gauge fields.

We have thus demonstrated that the $3+1$ dimensional Standard Model emerges
from $4+2$ dimensions.

\section{Resolution of the strong CP problem \label{cp}}

Recall that the strong CP problem in QCD is due to the fact that a term of the
form $S_{\theta}=\frac{\theta}{4!}\int dx^{4}\varepsilon_{\mu\nu\lambda\sigma
}Tr\left(  G^{\mu\nu}G^{\lambda\sigma}\right)  $ can be added to the QCD
action in $3+1$ dimensions without violating any of the gauge or global
symmetries. Unfortunately this term violates CP conservation of the strong
interactions. So, phenomenologically speaking, if it is not absolutely zero,
it must be extremely small. However there is no explanation of this fact
within the simple version of the Standard Model. This problem can be
circumvented by extending the Standard Model with an additional U$\left(
1\right)  $ symmetry, called the Peccei-Quinn symmetry\cite{pq}, by doubling
the Higgs bosons. The spontaneous breakdown of this symmetry, along with
SU$\left(  2\right)  \times$U$\left(  1\right)  $ leads to the Goldstone boson
called the axion. So far searches for the axion have limited its parameters
sufficiently to basically rule it out. This leaves us with a fundamental
problem to solve.

We will argue that there is a resolution of this problem in the $4+2$
formulation of the Standard Model. The key point is that a term similar to the
form $\varepsilon_{\mu\nu\lambda\sigma}Tr\left(  G^{\mu\nu}G^{\lambda\sigma
}\right)  $ that appears in the QCD Lagrangian in $3+1$ dimensions cannot be
written down in $4+2$ dimensions as an invariant under the symmetries. This is
because in $4+2$ dimensions the Levi-Civita symbol $\varepsilon^{M_{1}%
M_{2}M_{3}M_{4}M_{5}M_{6}}$ has six indices instead of four.

We may ask if there are any additional invariant terms that we could have
included in the $4+2$ dimensional theory that could lead to $S_{\theta}$ upon
reduction to $3+1$ dimensions? In providing an answer to this question we must
take into account the 2Tgauge-symmetry as well as the principle of
renormalizability. The latter says that we should not include any terms in
$4+2$ dimensions that would lead to non-renormalizable interactions in $3+1$
dimensions. Then one cannot find any terms that include the delta function
$\delta\left(  X^{2}\right)  $ in the volume element, except for the following
one%
\begin{equation}
\int\left(  d^{6}X\right)  ~\delta\left(  X^{2}\right)  ~X_{M_{1}}%
\partial_{M_{2}}Tr\left(  F_{M_{3}M_{4}}F_{M_{5}M_{6}}\right)  \varepsilon
^{M_{1}M_{2}M_{3}M_{4}M_{5}M_{6}}.
\end{equation}
This term vanishes identically as follows. The ordinary derivative
$\partial_{M_{2}}$ can be rewritten as the covariant derivative $D_{M_{2}}$
when applied on each of the field strengths inside the trace, since the
non-Abelian terms sum up to become the commutator of matrices that vanish due
to the trace. Then we use the Biachi identities $D_{[M_{2}}F_{M_{3}M_{4}]}=0$
to show that the term is null.

There remains the following type of term to consider without the delta
function%
\begin{equation}
\int\left(  d^{6}X\right)  ~Tr\left(  F_{M_{1}M_{2}}F_{M_{3}M_{4}}%
F_{M_{5}M_{6}}\right)  \varepsilon^{M_{1}M_{2}M_{3}M_{4}M_{5}M_{6}}.
\end{equation}
The fact that it is cubic rather than quadratic seems to already violate the
renormalizability requirements. However, this is a topological term that can
be written as a total divergence, so it cannot violate renormalizability.
Furthermore, since it is a total divergence, it is automatically invariant
under all the infinitesimal gauge symmetries we discussed before. Furthermore,
it cannot contribute to the equations of motion. For the SU$\left(  3\right)
\times$SU$\left(  2\right)  \times$U$\left(  1\right)  $ gauge group there are
several such gauge invariants, namely
\begin{equation}
\int\left(  d^{6}X\right)  ~\left[
\begin{array}
[c]{c}%
aTr\left(  G_{M_{1}M_{2}}G_{M_{3}M_{4}}G_{M_{5}M_{6}}\right) \\
+bTr\left(  W_{M_{1}M_{2}}W_{M_{3}M_{4}}W_{M_{5}M_{6}}\right) \\
+c\left(  B_{M_{1}M_{2}}B_{M_{3}M_{4}}B_{M_{5}M_{6}}\right) \\
+dB_{M_{1}M_{2}}Tr\left(  W_{M_{3}M_{4}}W_{M_{5}M_{6}}\right) \\
+eB_{M_{1}M_{2}}Tr\left(  G_{M_{3}M_{4}}G_{M_{5}M_{6}}\right)
\end{array}
\right]  \varepsilon^{M_{1}M_{2}M_{3}M_{4}M_{5}M_{6}},
\end{equation}
The one that comes closest to producing $S_{\theta}$ is the last one, since
upon reduction to $3+1$ dimensions we might get the term $\int dx^{4}%
\varepsilon_{\mu\nu\lambda\sigma}Tr\left(  G^{\mu\nu}G^{\lambda\sigma}\right)
\times\int\left\vert \kappa\right\vert ^{5}d\kappa d\lambda B_{+^{\prime
}-^{\prime}}\left(  x,\lambda,\kappa\right)  .$ However, we have shown in the
previous section that such a term is gauge dependent under the
2Tgauge-symmetry. In particular in the reduction to $3+1$ dimensions we have
shown that all $F_{-^{\prime}\mu}\left(  X\right)  =0,\;\;F_{+^{\prime
}-^{\prime}}\left(  X\right)  =0$ vanish not only for $B_{MN}$ but for all
others as well. Hence the topological term vanishes identically.

This resolves the strong CP problem in QCD in the emergent Standard Model in
$3+1$ dimensions.

\section{Mass generation \label{massgen}}

We will mention briefly several mass generation scenarios, namely dynamical
symmetry breaking, Coleman-Weiberg mechanism with only the SU$\left(
2\right)  \times$U$\left(  1\right)  $ doublet field $H$ and no dilaton, new
mechanisms offered by the 2T-physics formulation, and finally discuss in more
detail the dilaton assisted Higgs mechanism.

Dynamical symmetry breaking in the form of extended technicolor is certainly
one of the possibilities for mass generation. This would proceed by adding all
the ingredients of extra techni-matter and techni-interactions in parallel to
what is done in the usual $3+1$ dimensional theories of technicolor. The $4+2$
dimensional theory seems to proceed in the same way and therefore we do not
have new comments on this possibility from the point of view of $4+2$ dimensions.

The Coleman-Weiberg mechanism proceeds through radiative corrections in a
completely massless theory with quartic plus gauge interactions
\cite{colemanweinberg}. This would apply in 3+1 dimensions to the Higgs field
$H$ in interaction with the electroweak gauge bosons, and produce an effective
potential which does lead to spontaneous breakdown. Of course, we would need
to recompute these effects directly in the 4+2 dimensional quantum theory, but
for now let us assume that the result is roughly similar. In the 3+1 theory
this mechanism predicts a definite mass ratio between the massive vector and
the massive Higgs. Using the values of the electroweak coupling constant and
an average of the $W/Z$ masses, the mass of the Higgs comes out in the range
of about 10 GeV, and seems to be ruled out already.

In 2T-physics there are new ways of understanding mass as having a
relationship to some moduli in the embedding of 3+1 dimensions in the higher
space of 4+2 dimensions. This produces the massive relativistic particle
instead of the massless particle, as indicated in Fig.1. This effect, which
has no relation to the Kaluza-Klein type of mass, has been studied in the
worldline formalism for particle dynamics \cite{2tHandAdS}\cite{twistorBP1}.
It has even been suggested as an alternative mechanism to the Higgs
\cite{marnelius} but remained far from being understood. The application of
this approach to understand the various corners of Fig.1 in the context of
field theory is at its infancy \cite{2tfield}, and needs to be studied in the
presence of interactions as formulated in this paper. This has not been
developed so far, but should be mentioned as a new possible source of mass
that remains to be investigated.

Next we turn to the Higgs mechanism, which is the most popular possibility
within the usual Standard Model. We find that there are new twists in the
$4+2$ formulation, and as a result there could be measurable phenomenological
consequences as described below.

The 4+2 action leads to the absence of quadratic mass terms in the Higgs
potential. Therefore, instead of the tachyonic mass, an interaction of the
Higgs $H$ with a dilaton $\Phi$ was introduced in Eq.(\ref{VHphi}) in the
form
\begin{equation}
V\left(  \Phi,H\right)  =\frac{\lambda}{4}\left(  H^{\dagger}H-\alpha^{2}%
\Phi^{2}\right)  ^{2}+V\left(  \Phi\right)  \label{higgsV}%
\end{equation}
to induce the spontaneous breakdown of the electroweak symmetry SU$\left(
2\right)  \times$U$\left(  1\right)  $. The reason that only quartic
interactions were allowed is intimately related to the $b$-symmetry (which
ultimately comes from the underlying Sp$\left(  2,R\right)  $). The
$b$-symmetry removes gauge degrees of freedom from the $4+2$ dimensional
fields and reduce them to the $3+1$ dimensional degrees of freedom.

The reason for the pure quartic interaction can also be understood directly by
studying the equations of motion. As shown in the previous section, the
solution of the kinematic equations of motion for the scalars require that
they must be homogeneous of degree $-1$. This means that their dependence on
the extra dimension $\kappa$ must be of the form
\begin{equation}
H\left(  X\right)  =\frac{1}{\kappa}\hat{H}\left(  x\right)  ,\;\Phi\left(
X\right)  =\frac{1}{\kappa}\hat{\Phi}\left(  x\right)  . \label{homogeneous}%
\end{equation}
The dynamical equations of motion $\partial^{2}\Phi=\partial_{\Phi}V\left(
\Phi,H\right)  +\cdots$ and $\;\partial^{2}H=\partial_{H}V\left(
\Phi,H\right)  +\cdots$ now have the left hand side proportional to
$\kappa^{-3},$ therefore the right hand side must also be proportional to
$\kappa^{-3}.$ Combined with Eq.(\ref{homogeneous}), this requires $V\left(
\Phi,H\right)  $ to be purely quartic. If any additional terms, such as
quadratics are included in $V\left(  \Phi,H\right)  $, they have to vanish on
their own since they will have a different power of $\kappa.$

For this reason the dynamical breakdown of the SU$\left(  2\right)  \times
$U$\left(  1\right)  $ electro-weak symmetry cannot be accomplished with a
tachyonic mass term for the Higgs field $H,$ since this is forbidden in the
$4+2$ formulation. Instead, a coupling to the dilaton as given in
Eq.(\ref{higgsV}) generates the non-trivial vacuum configuration in
Eq.(\ref{vacuum}). The equations of motion for the scalars (assuming all other
fields vanish at the vacuum) are%
\begin{equation}
\partial^{2}H=\lambda H\left(  H^{\dagger}H-\alpha^{2}\Phi^{2}\right)
,\;\partial^{2}\Phi=-2\alpha^{2}\Phi\left(  H^{\dagger}H-\alpha^{2}\Phi
^{2}\right)  +V^{\prime}\left(  \Phi\right)  .
\end{equation}
At the vacuum configuration $\left(  H^{\dagger}H-\alpha^{2}\Phi^{2}\right)
=0$ the Higgs field must satisfy $\partial^{2}H=0$ while $H\left(  X\right)  $
is also homogeneous of the form Eq.(\ref{homogeneous})$.$ We have seen already
that $\partial^{2}H\left(  X\right)  =\kappa^{-3}\partial_{x}^{2}\hat
{H}\left(  x\right)  ,$ therefore $\hat{H}\left(  x\right)  $ must be a
constant at the vacuum, but $H\left(  X\right)  $ still depends on the
$\kappa$ coordinate
\begin{equation}
\langle H\left(  \kappa,\lambda,x^{\mu}\right)  \rangle=\frac{v}{\kappa
}\left(
\genfrac{}{}{0pt}{}{0}{1}%
\right)  ,\;v~\text{= electroweak scale }\sim100~\text{GeV.}%
\end{equation}
The electroweak scale of about 100 $GeV$ is determined by fitting to experiment.

Returning to the equation for $\Phi,$ at the Higgs vacuum, it reduces to the
form $\partial^{2}\Phi=V^{\prime}\left(  \Phi\right)  $. But it must also
satisfy the vacuum value $\left(  H^{\dagger}H-\alpha^{2}\Phi^{2}\right)  =0$
and be homogeneous as in Eq.(\ref{homogeneous})$,$ therefore
\begin{equation}
\langle\Phi\left(  X\right)  \rangle=\pm\frac{v}{\kappa\alpha}.
\label{dilatonvac}%
\end{equation}
Hence $\partial^{2}\langle\Phi\rangle=0$, which requires $V^{\prime}\left(
\langle\Phi\rangle\right)  =0$ at the vacuum. If $V\left(  \Phi\right)  $ is
also taken as a quartic monomial $V\left(  \Phi\right)  =\frac{\lambda
^{\prime}}{4}\Phi^{4}$ (as required by the $b$ symmetry), then the only
solution is $\lambda^{\prime}=0,$ or $V\left(  \Phi\right)  =0$ identically
(at the classical level) to fit phenomenology\footnote{Recall that we can
change the homogeneity degree of $\Phi$ as in Eq.(\ref{s2W}) from $-1$ to
$\left(  -1+\frac{a}{2}\right)  $ by permitting the term $W\left(
\Phi\right)  =\frac{a}{2}\Phi^{2}$ in the action$.$Whatever the new
homogeneity degree of $\Phi$ is, it must appear in the potential $V\left(
\Phi,H\right)  $ with appropriate powers to make every term in $V\left(
\Phi,H\right)  $ have homogeneity degree $-4.$ For example if the degree of
$\Phi$ is $-2$ then every $\Phi^{2}$ we see in $V\left(  \Phi,H\right)  $
should be replaced by $\Phi.$ This gives quadratic terms for $\Phi$ but does
not change the conclusions. More complicated forms of $W\left(  \Phi\right)  $
have not been studied yet.}.

Actually, the above discussion for the dilaton $\Phi$ is incomplete. To fully
understand the interactions of the dilaton one must include the gravitational
fields with which it naturally interacts. For example, in string theory the
vacuum expectation value of the dilaton plays the role of the string coupling
constant that controls all string interactions. There is no full understanding
yet how our four dimensional vacuum (or the $4+2$ upgrade in our case) emerges
from a more fundamental theory that includes gravity. This information will
eventually include the vacuum expectation value of the dilaton. Therefore at
the present we have no theoretical control on how to stabilize the vacuum
expectation value of the dilaton.

For this reason, in our model we simply take the value of $\langle\Phi
\rangle\neq0$ as given by the phenomenology of the Higgs $\langle H\rangle
\neq0$. But we imagine that $\langle\Phi\rangle$ is stabilized by additional
interactions in the gravitational or string theory sectors to have a fixed
value related to $v\sim100~GeV,$ and the dependence on the extra dimension
$\kappa,$ and the coupling $\alpha$ as given above in Eq.(\ref{dilatonvac}).
Then this form of the dilaton plays the same role as the tachyonic mass term
of the Higgs in the usual Standard Model. This is then the source that drives
the electroweak symmetry breaking.

In this way we have given a deeper physical basis for the Higgs vacuum. In the
$4+2$ theory mass generation through a Higgs is not isolated from the
gravitational (or string) interactions. In fact our suggested point of view is
intellectually more satisfactory because a Higgs vacuum fills all space with
the constant $v.$ To imagine that this space-filling vacuum could be achieved
without the cooperation of the gravitational sector or without appealing to
the vacuum selection process in the evolution of the universe since the Big
Bang suggests that something was amiss in the logic of mass generation.

Given the discussion above we suggest that the vacuum value of the dilaton
(effectively the tachyonic Higgs mass) is imposed mainly by the gravitational
or string sector of a more complete theory. But, given that $\langle
\Phi\rangle\neq0$ is not just an isolated constant, but the value of a field,
we should analyze how the small fluctuations of the dilaton around its vacuum
interact with the rest of matter. Therefore, our proposal has phenomenological
consequences as discussed below.

We now discuss the small fluctuations of both the Higgs and the dilaton. We
give the discussion directly in terms of the $3+1$ dimensional fields. We
choose the unitary gauge for the Higgs by absorbing 3 of its degrees of
freedom into the electroweak gauge fields. For the remaining neutral Higgs
field, and the dilaton field we write everywhere the following vacuum shifted
forms
\begin{equation}
H^{0}\left(  X\right)  =\frac{1}{\kappa}\left(  v+h\left(  x\right)  \right)
,\;\Phi\left(  X\right)  =\frac{1}{\alpha\kappa}\left(  v+\alpha\phi\left(
x\right)  \right)  .
\end{equation}
where $h,\phi$ are the small fluctuations. The potential energy in
Eq.(\ref{higgsV}), with $V\left(  \Phi\right)  =0,$ becomes $V\left(
\Phi,H\right)  =\frac{1}{\kappa^{4}}V\left(  h,\phi\right)  $ where $V\left(
h,\phi\right)  $ is the potential energy from the point of view of the $3+1$
dimensional theory. It is given by
\begin{align}
V\left(  h,\phi\right)   &  =\frac{\lambda}{4}\left(  \left(  v+h\right)
^{2}-(v+\alpha\phi)^{2}\right)  ^{2}\\
&  =\frac{\lambda}{4}\left(  h-\alpha\phi\right)  ^{2}\left(  h+\alpha
\phi+2v\right)  ^{2}.
\end{align}
In the limit $\alpha\rightarrow0$ that corresponds to zero coupling to the
dilaton, we recognize the standard theory of the Higgs boson with its usual
potential $V\left(  h\right)  =\lambda v^{2}h^{2}+\lambda vh^{3}+\frac
{\lambda}{4}h^{4}$. If $h$ is observed at the LHC, from its mass given by
$\lambda v^{2}=\frac{1}{2}m^{2},$ the coupling constant $\lambda$ would be
determined, and its self interactions $\lambda vh^{3}+\frac{\lambda}{4}h^{4}$
are then predicted.

If $\alpha$ is very small, then the coupling of the Higgs (and the rest of the
Standard Model) to the dilaton $\phi\left(  x\right)  $ may appear to be well
hidden from measurement in the near future.

However, no matter how small $\alpha$ is, there is an inevitable fact of a
massless Goldstone boson associated with the spontaneous breaking of scale
invariance. We emphasize that the emergent Standard Model is scale invariant
at the classical level because there are no mass terms at all. In fact, the
emergent Standard Model is invariant under the conformal group of
transformations SO$\left(  4,2\right)  $ at the classical level, where the
conformal SO$\left(  4,2\right)  $ is precisely the Lorentz symmetry in the
higher $4+2$ dimensions as explained by 2T-physics. The spontaneous breaking
of the electroweak symmetry simultaneously breaks the global scale symmetry
and generates a Goldstone boson.

To identify the Goldstone boson we define the following orthogonal
combinations of the fields $h,\phi$
\begin{equation}
\tilde{h}=\frac{h-\alpha\phi}{\sqrt{1+\alpha^{2}}},\;\tilde{\phi}=\frac{\alpha
h+\phi}{\sqrt{1+\alpha^{2}}},\;\text{or \ }h=\frac{\tilde{h}+\alpha\tilde
{\phi}}{\sqrt{1+\alpha^{2}}},\;\phi=\frac{-\alpha\tilde{h}+\tilde{\phi}}%
{\sqrt{1+\alpha^{2}}}%
\end{equation}
In terms of $\tilde{h},\tilde{\phi}$ the kinetic terms remain correctly
normalized $\left(  \partial_{\mu}h\right)  ^{2}+\left(  \partial_{\mu}%
\phi\right)  ^{2}=\left(  \partial_{\mu}\tilde{h}\right)  ^{2}+\left(
\partial_{\mu}\tilde{\phi}\right)  ^{2}$ while the potential energy takes the
form%
\begin{equation}
V\left(  \tilde{h},\tilde{\phi}\right)  =\frac{\lambda}{4}\tilde{h}^{2}\left(
\left(  1-\alpha^{2}\right)  \tilde{h}+2\alpha\tilde{\phi}+\sqrt{1+\alpha^{2}%
}2v\right)  ^{2}%
\end{equation}
From this we see that the field $\tilde{h}$ is massive, but the field
$\tilde{\phi}$ is massless since there is no quadratic term proportional to
$\tilde{\phi}^{2}.$

We must emphasize that the analysis of this Goldstone boson is certainly
incomplete. First we must remember that the dilaton couples to the
gravitational or string sector and this can alter its mass. Furthermore, the
scale invariance we mentioned above is known to be broken by quantum
anomalies, at least as it is usually computed in any $3+1$ dimensional theory.
Furthermore, the Coleman-Weinberg mechanism will also add mass-generating
radiative corrections. Any of these or all of these effects would lift the
mass of the dilaton, so the Goldstone boson identified above is not expected
to remain massless. However, being potentially a Goldstone boson, its mass may
not be too large and perhaps it is within the range of possible observations.

Can we expect to see such a dilaton in the coming experiments? Let's try to
estimate its couplings to standard matter by first neglecting the effects
mentioned in the previous paragraph. Evidently, this estimate must go through
the Higgs sector. The dilaton $\tilde{\phi}$ couples to all fermions and
electroweak bosons since the standard coupling of the Higgs $h$ must be
replaced everywhere by $h=\frac{\tilde{h}+\alpha\tilde{\phi}}{\sqrt
{1+\alpha^{2}}}.$ The dimensionless coupling of $h$ is proportional to the
mass of the quarks and leptons in the form $m_{i}/v$ where $m_{i}$ is the mass
of the quark or lepton and $v$ is the electroweak scale. Therefore the
coupling of the dilaton $\tilde{\phi}$ to every quark and lepton is given by
$\approx\left(  m_{i}/v\right)  \times\alpha.$ The strongest coupling is
evidently to the top quark since it has the largest mass. The value of
$\alpha$ will determine whether this coupling is strong enough to be seen in
the coming LHC experiments or in precision measurements that test radiative corrections.

The dilaton-Higgs scenario is certainly among the possible scenarios even in
the usual Standard Model, but it has not been suggested before. The 2T-physics
formulation provides a compelling motivation for favoring this alternative,
especially since the 2T-physics approach solves the strong CP problem.
Therefore, it needs to be taken seriously and studied more thoroughly. The
effects of the dilaton should be incorporated into phenomenological estimates
and it should be included among the experimental searches for new particles
especially as part of understanding the origin of mass.

If the dilaton-Higgs scenario is realized in Nature, it would imply that the
Higgs vacuum expectation value $\langle H^{0}\left(  X\right)  \rangle
=\frac{v}{\kappa}$ or the corresponding dilaton ($\langle\Phi\left(  X\right)
\rangle=\frac{v}{\alpha\kappa}$) is a probe into the extra dimension $\kappa.$

\section{Directions \label{directions}}

In this paper we constructed the principles for field theory in $d+2$
dimensions. Because of the new 2Tgauge-symmetries the interactions are unique
and the physical formulation is ghost free. This produces the physics of
$\left(  d-1\right)  +1$ dimensions but from the vantage of $d+2$ dimensions.
The advantages of formulating physics from the vantage of $d+2$ dimensions are
conveyed by the duality, holography, symmetry and unifying features of
2T-physics illustrated partially in Fig.1.

In this paper we have studied only one of the $3+1$ dimensional holographic
images of the new formulation of the Standard Model in $4+2$ dimensions. This
established first of all that the 2T-physics formalism is completely physical
and capable of correctly describing all we know in physics up to now, as
embodied by the Standard Model of Particles and Forces.

Moreover, the Standard Model in $4+2$ dimensions resolved an outstanding
problem of QCD, namely the strong CP problem. In addition it also provided a
new point of view on mass generation by relating it to a deeper physical basis
for mass. These features indicate that the $4+2$ vantage is capable of leading
our thinking into new fertile territories, and apparently explain more than
what was possible in $3+1$ dimensions.

Indeed a brief look into the ideas conveyed in Fig.1 is sufficient to say that
there is a lot more to explore and explain by using the new formulation of the
Standard Model. Perhaps such ideas will lead to new computational techniques
for analyzing field theory non-perturbatively, and shed more light into
structures such as quantum chromo-dynamics that is still in great need for
technical progress.

Beyond the Standard Model, similar field theory techniques can be used to
discuss grand unification and gravity. Grand unification would proceed through
gauge theories as in the present paper. The equations of motion for gravity
have already been constructed in \cite{2tfield} and these can certainly now be
elevated to an action principle. It would be interesting to explore
gravitational physics, including cosmology, black holes, and the issue of the
cosmological constant by using the 2T-physics formulation.

Supersymmetry should be incorporated by basing it on the formulation of the
superparticle in 2T-physics consistently with the Sp$\left(  2,R\right)  $
symmetry \cite{super2t}-\cite{twistorLect}. The field theory version of this
is likely to have a richer mathematical structure of supersymmetry than $3+1$
dimensions. The twistor formalism \cite{2ttwistor}\cite{2tsuperstring}%
\cite{2ttwistor}\cite{twistorLect}\cite{twistorBP1} that is closely connected
to this approach could also lead to a twistor or supertwistor version of field theory.

Beyond these, we recall that at the basis of what we presented is the very
basic principle of Sp$\left(  2,R\right)  $ gauge symmetry that makes position
and momentum locally indistinguishable. However, the field theory approach we
used here distinguishes position from momentum. We may ask if we can come up
with a more even-handed field theoretic formulation. There is a beginning
along these lines in \cite{2tfieldXP}. When some such approach succeeds to
describe standard physics we will have access to deeper insights.

We have not even began to discuss the quantum theory in this paper. It is
evident that having an action principle was the first concrete step needed to
define the quantum theory consistently through the path integral. This can now
proceed in the standard manner, taking into consideration the gauge
symmetries. Motivated by Dirac's approach to conformal symmetry, there were
some efforts in the past to discuss quantum field theory in $4+2$ dimensions
\cite{adler}\cite{fronsdal}. This was done by using some guesswork and without
the benefit of an action principle, but it could provide some guidance for a
renewed effort to formulate and use quantum field theory directly in $4+2$
dimensions, and then apply it to practical computations.

\section*{Acknowledgments}

I thank Shih-Hung Chen, Yueh-Cheng Kuo, Bora Orcal and Guillaume Quelin at USC
for helpful discussions while the concepts in this paper were developed. The
final stages of this work were accomplished during a visit at the University
of Valencia. I thank the group members there, in particular J. Azcarraga, for
their encouragement and enthusiastic support while this paper was completed.

\section{Appendix A}

In this appendix we give another form of the gauge fixed action for the scalar
field (instead of Eq.(\ref{phiactionFixed})) and the gauge field (instead of
Eq.(\ref{Aaction})) that makes the underlying Sp$\left(  2,R\right)  $
symmetry fully transparent. For fermions we do not need a separate gauge fixed
treatment of the equations of motion to display the underlying Sp$\left(
2,R\right)  $ symmetry, since we have seen that the fully gauge invariant
version can already be written in terms of $L^{MN}$ as in Eq.(\ref{Lpsi}).

We emphasize that the fields $\Phi,A_{M}$ that appear below are the gauge
fixed versions as discussed following Eq.(\ref{phiactionFixed}) and
Eq.(\ref{dsa}) respectively. This means that their remainder proportional to
$X^{2}$ are not the most general, but are restricted to a form allowed by the
remaining 2Tgauge-symmetry.

The only Sp$\left(  2,R\right)  $ invariant in the 2T particle mechanics is
the SO$\left(  d,2\right)  $ orbital angular momentum $L^{MN}=X^{M}P^{N}%
-X^{N}P^{M}.$ In the field theory version we substitute $P^{M}=-i\partial
^{M},$ therefore it is natural for derivatives to appear in the combination
of
\begin{equation}
L^{MN}=-i\left(  X^{M}\partial^{N}-X^{N}\partial^{M}\right)  .
\end{equation}
So, the kinetic term in the action for a scalar field $\Phi$ can be taken as
follows (the potential term is identical to Eq.(\ref{phiactionFixed}) and is
omitted below)
\begin{equation}
S_{0}\left(  \Phi\right)  =-\frac{1}{2}\int d^{d+2}X~\delta^{\prime}\left(
X^{2}\right)  ~\left[  \frac{1}{2}\left(  L^{MN}\Phi\right)  \left(
L_{MN}\Phi\right)  +\left(  1-\frac{d^{4}}{4}\right)  \Phi^{2}\right]  .
\label{s0phi}%
\end{equation}
Note here the $\delta^{\prime}\left(  X^{2}\right)  ~$rather than the
$\delta\left(  X^{2}\right)  $ to implement the Sp$\left(  2,R\right)  $
constraint $X^{2}=0$. The reason for the constant term $\left(  1-\frac{d^{4}%
}{4}\right)  $ is also Sp$\left(  2,R\right)  .$ It is related to the fact
that the SO$\left(  d,2\right)  $ Casimir operator $\frac{1}{2}L^{MN}L_{MN}$
and the Sp$\left(  2,R\right)  $ Casimir operator $C_{2}^{Sp\left(
2,R\right)  }=\frac{1}{2}\left(  Q_{11}Q_{22}+Q_{22}Q_{11}\right)
-Q_{12}Q_{12},$ are related to each other by the following equation
\cite{2treviews}%
\begin{align}
4C_{2}^{Sp\left(  2,R\right)  }  &  =\frac{1}{2}L^{MN}L_{MN}-\left(
1-\frac{d^{2}}{4}\right) \label{casimir2}\\
&  =-X^{2}\partial^{2}+\left(  X\cdot\partial+\frac{d-2}{2}\right)  \left(
X\cdot\partial+\frac{d+2}{2}\right)  . \label{casimir1}%
\end{align}
Thus, after an integration by parts (noting $L^{MN}\delta^{\prime}\left(
X^{2}\right)  =\delta^{\prime}\left(  X^{2}\right)  L^{MN}$) the action can be
written in terms of the Sp$\left(  2,R\right)  $ Casimir operator
\begin{align}
S_{0}\left(  \Phi\right)   &  =\frac{1}{2}\int d^{d+2}X~\delta^{\prime}\left(
X^{2}\right)  ~\Phi\left(  \frac{1}{2}L^{MN}L_{MN}-1+\frac{d^{4}}{4}\right)
\Phi\\
&  =\frac{1}{2}\int d^{d+2}X~\delta^{\prime}\left(  X^{2}\right)  ~\Phi\left(
4C_{2}^{Sp\left(  2,R\right)  }\right)  \Phi.\\
&  =\frac{1}{2}\int d^{d+2}X~\Phi\left\{  \delta\left(  X^{2}\right)
\partial^{2}\Phi+\delta^{\prime}\left(  X^{2}\right)  \left(  X\cdot
\partial+\frac{d-2}{2}\right)  \left(  X\cdot\partial+\frac{d+2}{2}\right)
\right\}  \Phi\label{s0phi2}%
\end{align}
In the last line we inserted the explicit form of the Casimir operator and
used $-X^{2}\delta^{\prime}\left(  X^{2}\right)  =\delta\left(  X^{2}\right)
$ so that the first term has the familiar appearance as the kinetic term of a
Klein-Gordon field in $d+2$ dimensions as in Eq.(\ref{phiactionFixed}), now
with the delta function $\delta\left(  X^{2}\right)  $ instead of
$\delta^{\prime}\left(  X^{2}\right)  .$ We see that with the extra terms
proportional to $\delta^{\prime}\left(  X^{2}\right)  $ we can rebuild the
Sp$\left(  2,R\right)  $ Casimir operator and make it evident that there is an
underlying Sp$\left(  2,R\right)  $ invariance.

The general variation of the action in Eq.(\ref{s0phi}) is $\delta
S_{0}\left(  \Phi\right)  =\int d^{d+2}X~\delta^{\prime}\left(  X^{2}\right)
~\delta\Phi\left(  4C_{2}^{Sp\left(  2,R\right)  }\right)  \Phi$ therefore the
equation of motion is
\begin{equation}
\delta^{\prime}\left(  X^{2}\right)  \left(  4C_{2}^{Sp\left(  2,R\right)
}\right)  \Phi=0. \label{varyPhi1}%
\end{equation}
This can be written in the form%
\begin{equation}
\delta\left(  X^{2}\right)  \partial^{2}\Phi+\delta^{\prime}\left(
X^{2}\right)  \left(  X\cdot\partial+\frac{d-2}{2}\right)  \left(
X\cdot\partial+\frac{d+2}{2}\right)  \Phi=0,
\end{equation}
from which we conclude that $\Phi$ satisfies two equations, not just one.
Provided the remainder of $\Phi=\Phi_{0}+X^{2}\tilde{\Phi}$ is a priori gauge
fixed to be homogeneous\footnote{As emphasized in footnote (\ref{careful}),
the correct two equations emerge only if $\Phi$ is already gauge fixed in this
action. Otherwise the equations include a non-homogeneous remainder
$\tilde{\Phi}$ which completely spoils the dynamical equation.} $\left(
X\cdot\partial+\frac{d+2}{2}\right)  \tilde{\Phi}=0,$ the resulting equations
for the full $\Phi$ are $\left[  \partial^{2}\Phi\right]  _{X^{2}=0}=0$ and
$\left(  X\cdot\partial+\frac{d-2}{2}\right)  \left(  X\cdot\partial
+\frac{d+2}{2}\right)  \Phi=0$. In particular this implies that the Casimir
$C_{2}^{Sp\left(  2,R\right)  }$ vanishes on the free field
\begin{equation}
\left[  C_{2}^{Sp\left(  2,R\right)  }\Phi\right]  _{X^{2}=0}=0.
\end{equation}
Now we see that on $\hat{\Phi}=\delta\left(  X^{2}\right)  \Phi$ the
quantities $X^{2}$ and $\partial^{2}$ vanish, hence their commutator which is
proportional to $\left(  X\cdot\partial+\frac{d-2}{2}\right)  \Phi=0$ must
also vanish. This is the solution we must take consistently with the equations
of motion derived above. This indicates that we have come full circle, and
derived the Sp$\left(  2,R\right)  $ singlet condition on $\left[
\delta\left(  X^{2}\right)  \Phi\right]  $ directly from the field theoretic
action principle. Note that the simpler looking gauge fixed action that we
adopted in the text in Eq.(\ref{phiactionFixed}) gives the identical information.

For gauge bosons we can give an alternative but equivalent form of the gauge
fixed action in Eq.(\ref{Aaction}) to display all derivatives in the form
$L^{MN}.$ Then the action takes the following manifestly Sp$\left(
2,R\right)  $ invariant form%
\begin{equation}
S\left(  A\right)  =\frac{1}{4}\int\left(  d^{d+2}X\right)  ~\delta^{\prime
}\left(  X^{2}\right)  ~\Phi^{\frac{2\left(  d-4\right)  }{d-2}}~Tr\left(
F_{MNK}F^{MNK}\right)  ,
\end{equation}
where
\begin{equation}
F_{MNK}=F_{[MN}X_{K]},\;\text{with ~}F_{MN}=\partial_{M}A_{N}-\partial
_{N}A_{M}-ig_{A}\left[  A_{M},A_{N}\right]
\end{equation}
is invariant under Sp$\left(  2,R\right)  $ since only $L^{MN}$ appears. Note
the$~\delta^{\prime}\left(  X^{2}\right)  $ rather than $\delta\left(
X^{2}\right)  $ in the volume element which is similar to the scalar action in
Eq.(\ref{s0phi}). The$~\delta^{\prime}\left(  X^{2}\right)  $ is just what is
needed to relate to the action in Eq.(\ref{Aaction}). When we compute
$Tr\left(  F_{MNK}F^{MNK}\right)  $ we find%
\begin{align}
\delta^{\prime}\left(  X^{2}\right)  ~Tr\left(  F_{MNK}F^{MNK}\right)   &
=\delta^{\prime}\left(  X^{2}\right)  ~X^{2}Tr\left(  F_{MN}F^{MN}\right)
+\cdots\\
&  =-\delta\left(  X^{2}\right)  ~Tr\left(  F_{MN}F^{MN}\right)  +\cdots
\end{align}
The first term reproduces the other action in Eq.(\ref{Aaction}) while the
extra terms play a role similar to the extra terms in the scalar action above
to complete into the Sp$\left(  2,R\right)  $ invariant. When all the
equations of motion, and gauge symmetries, are taken into account the two
actions give equivalent results in the physical sector. We emphasize that the
remainder $\tilde{A}_{M}$ in $A_{M}=A_{M}^{0}+X^{2}\tilde{A}_{M}$ must be a
priori gauge fixed to satisfy the same properties of the remaining gauge
freedom $a_{M}$ as given in Eq.(\ref{Aa}).

\end{document}